\documentclass[onecolumn]{emulateapj}

\newcommand{\vecn}{{\bf n}}
\newcommand{\vecw}{{\bf w}}
\newcommand{\dln}{\partial\ln\nu}
\newcommand{\dlnp}{\partial\ln\nu^\prime}
\newcommand{\vece}{{\bf e}}
\newcommand{\vecnp}{{\bf n}^\prime}
\newcommand{\nv}{{\bf n}\cdot{\bf v}}

\newcommand{\vc}{\frac{{\bf v}}{c}}
\newcommand{\nvc}{\frac{{\bf n}\cdot{\bf v}}{c}}
\newcommand{\npvc}{\frac{{\bf n}^\prime\cdot{\bf v}}{c}}
\newcommand{\nup}{\nu^\prime}
\newcommand{\dlnsig}{\frac{\partial\ln\sigma_0}{\partial \ln\nu}}

\newcommand{\dlnjl}{\frac{\partial\ln J}{\partial \ln\nu}}
\newcommand{\dlnhl}{\frac{\partial\ln H^j}{\partial \ln\nu}}
\newcommand{\dlnkl}{\frac{\partial\ln K^{jk}}{\partial \ln\nu}}
\newcommand{\dlnkap}{\frac{\partial\ln\kappa_0}{\partial \ln\nu}}
\newcommand{\dlneta}{\frac{\partial\ln\eta_0^{\rm th}}{\partial \ln\nu}}
\newcommand{\ddr}{\frac{1}{r}\frac{\partial}{\partial r}}
\newcommand{\ddz}{\frac{\partial}{\partial z}}

\shortauthors{Hubeny, \& Burrows}
\shorttitle{Neutrino transport in core-collaps supernovae}

\begin{document}

\title{
A New Algorithm for 2-D Transport for Astrophysical
Simulations: I. General Formulation and Tests for the 1-D Spherical Case
}

\author{Ivan Hubeny\altaffilmark{1} \& Adam Burrows\altaffilmark{1}}
%\author{Adam Burrows\altaffilmark{1}}

\altaffiltext{1}{Department of Astronomy and Steward Observatory, The University
of Arizona, Tucson, AZ 85721}

\begin{abstract}

We derive new equations using the {\it mixed-frame} approach for one- 
and two-dimensional (axisymmetric) time-dependent radiation transport 
and the associated couplings with matter.  Our formulation is multi-group 
and multi-angle and includes anisotropic scattering, 
frequency(energy)-dependent scattering and absorption,
complete velocity dependence to order $v/c$, rotation, and 
energy redistribution due to inelastic scattering. Hence, 
the ``2D" realization is actually ``6 1/2"-dimensional.
The effects of radiation viscosity are automatically incorporated.
Moreover, we develop Accelerated-Lambda-Iteration, Krylov subspace (GMRES),
Discontinuous-Finite-Element, and Feautrier numerical methods for solving
the equations and present the results of one-dimensional numerical tests
of the new formalism. The virtues of the mixed-frame approach include simple velocity
dependence with no velocity derivatives, straight characteristics, simple  
physical interpretation, and clear generalization to higher dimensions.
Our treatment can be used for both photon and neutrino transport, but
we focus on neutrino transport and applications to core-collapse
supernova theory in the discussions and examples.

\end{abstract}


\section{Introduction}

Many phenomena in the Universe must be addressed
using the tools of radiation transport 
and radiation hydrodynamics to achieve a
theoretical understanding of their character.  Supernova
explosions, gamma-ray bursts, star formation, planet formation,
nova explosions, X-ray bursts, Luminous-Blue-Variable (LBV) outbursts,
and stellar winds are all time-dependent fluid flow problems 
in which radiation plays a pivotal, often driving, role.
In most circumstances, the radiation is photons, but
in studies of the core-collapse supernova mechanism the radiation
is neutrinos.  In either case, time-dependent techniques to address 
radiation transport and the coupling of radiation with matter are
of central concern to the theorist whose goal is 
explaining the transient, dynamical phenomena of the 
Cosmos.  However, spherically-symmetric algorithms for 
radiation transport and radiation hydrodynamics, generally 
necessary to achieve a first-order understanding, are often not 
sufficient and multi-dimensional approaches are called for.
These are not easy, not only to formalate, but to implement.
Nevertheless, multi-D radiation transport is emerging as 
a necessary tool in the theorist's toolbox and computers
to evolve the associated equations are becoming available 
to a wider cohort of researchers.  

In this paper, we derive new equations using the {\it mixed-frame} approach
for one- and two-dimensional radiation transport and the associated 
coupling with matter, significantly extending the pioneering work of
Mihalas \& Klein (1982) to include anisotropic scattering, 
frequency(energy)-dependent scattering and absorption, 
complete velocity dependence to order $v/c$, rotation, and energy 
redistribution due to inelastic scattering.  Moreover, we develop 
algorithms for solving these equations and present
the results of one-dimensional numerical tests.  Hence, we provide 
both the new formulation and appropriate numerical techniques to solve it.
The mixed-frame approach, in which the radiation quantities are
defined in the laboratory (Eulerian) frame and the matter and coupling
quantities are defined in the comoving frame, has largely been
neglected by the radiation transport and atmospheres communities
because of their focus on line transfer.  The desire to 
include narrow spectral lines and to handle Doppler shifts into 
and out of those lines necessitates many spectral bins and a huge
number of angular bins to ensure the lines are resolved on the 
computational grid.  As a result, most dynamic atmosphere 
and radiation studies are done using the comoving 
(Lagrangian) equations of radiation transport.  For instance, the core-collapse
supernova community, which has been at the forefront of radiation
hydrodynamic developments in astrophysics, has inherited this formulation for their
treatment of neutrino transport.

However, many radiation-hydrodynamic problems do
not require an exquisite treatment of spectral line transport, but 
a good treatment of continuum transport.  The core-collapse supernova
problem is one such case.  The monochromatic opacities and emissivities
of neutrinos are overwhelmingly smooth functions of neutrino energy.
Given this, the mixed-frame formulation is ideally suited for supernova
theory.  Its virtues vis \`a vis the comoving-frame approach 
to the solution of the Boltzmann equation and its related moment 
equations are numerous: 1) Even in one-dimension, instead of requiring 
$\sim$20 velocity-dependent terms (Buras et al. 2006) on the left-hand(streaming) side
of the Boltzmann/transport equation, many of which involve spatial velocity
derivatives, there are no such terms on the left-hand-side in the mixed frame approach
and only {\it one} grouped linear term (to $O(v/c)$) on the right-hand(source) 
side;  2) There are no terms with derivatives 
of the velocity. Therefore, the characteristics of the associated
transport equation are all straight lines.  Furthermore, there
is no need for the monotonicity in the velocity field required
by some implicit solvers; and 3) The mixed-frame method is easily
generalized to two and three dimensions, and the associated solvers
are straightforward (though more expensive) extensions of those employed in
1D. Note that much has been made of the importance of velocity-dependent
terms in the transport equations for the calculation 
of the neutrino energy deposited in the ``net-gain" 
(Wilson 1985) region.  We show that the mixed-frame approach provides the 
most straightforward perspective from which to understand the physics 
of this effect and that
even its sign is frame-dependent.  In particular, in the mixed-frame 
formulation the velocity-dependent term augments the net gain during 
the stalled-shock phase, while in the comoving frame formulation the 
corresponding terms reduce it.  Hence, any statement concerning the 
importance of such terms is very frame- and treatment-dependent.

Though the mixed-frame equations depend simply upon velocity, the Lorentz
transformations that are the core of this formulation introduce frequency(energy)
derivatives of the radiation moments.  These couple adjacent energy groups
and might have compromised implicit algorithms that parallelize in group 
(Livne et al. 2004; Walder et al. 2005; Burrows et al. 2006; 
Dessart et al. 2006ab; Ott et al. 2006ab).  However, we show 
that such terms can be handled semi-implicitly, with the logarithmic derivatives
of the moments with respect to energy handled explicitly during an otherwise
implicit solve. In this way, processors performing updates on only 
a single group are not coupled during the iteration to adversely affect 
parallelization and scalability.  Furthermore, a similar semi-implicit 
tactic works for inelastic scattering terms, since these are sub-dominant 
in the context of core collapse (Thompson, Burrows, \& Pinto 2003).
Note, however, that parallelization in energy groups is viable for
2D calculations, but not for 3D. An entire 2D hydro 
grid can now reside on a single processor, but an entire 
3D hydro grid requires spatial domain decomposition onto many
processors and parallelization in {\it only} energy groups is not yet viable. 
Furthermore, given the seven-dimensional nature of 3D radiation 
transport (3 space + 2 angles + 1 frequency/energy + 1 time),
3D is not yet computationally feasible for astrophysical 
simulations.  Therefore, we focus in this paper on 1D and 2D 
mixed-frame formulations.  

However, a 2D, mixed-frame, azimuthally-symmetric 
formalism is still six-dimensional (lacking 
one spatial dimension and a hemisphere) and is particularly 
straightforward and powerful. Our approach  
includes rotation (and, hence, is actually ``6 1/2"-dimensional).
The Eddington factor becomes an ``Eddington tensor" with five independent entries.
The zeroth- and first-moment equations are accurate approximations 
to the full equations that are closed with Eddington factors that can be
calculated at each timestep to provide the full solution, or every $N$ steps
to provide an excellent, fast, though approximate, solution.  Since  
velocity dependence is included in the multi-D context, the effects
of radiation viscosity (neutrino or photon) are automatically incorporated
into any scheme that includes the hydrodynamics with the radiation force
and energy couplings.

Though our treatment can be used for both photon and neutrino transport,
it was devised with neutrinos in mind. As a result, much of the 
discussion and all the tests we perform are in that context.  
Therefore, a short review of the transport schemes 
employed to date in supernova modeling is in order.  In the realm of 1D 
(spherical) neutrino transport, Bowers \& Wilson (1982), Bruenn (1985),  Wilson (1985),
and Mayle, Wilson, \& Schramm (1987) employed multi-group diffusion codes with flux
limiters
and did not address angle-dependent transport. Later, various groups
achieved a multi-group/multi-angle neutrino Boltzmann capability (Mezzacappa \& Bruenn
1993;
Rampp \& Janka 2000; Mezzacappa et al. 2001; Thompson, Burrows, \& Pinto 2003; 
Liebend\"orfer et al. 2001ab, 2004, 2005).  In 2D simulations, LeBlanc 
\& Wilson (1970) constructed a gray, flux-limited diffusion code, that nevertheless
calculated a two-component vector flux.  Herant et al. (1994) fielded a gray diffusion
code with a simple matching to the free-streaming regime.  Burrows, Hayes, \& Fryxell
(1995) 
developed a gray diffusion code that calculated different solutions for the different
$\theta$ angles (in spherical coordinates) in a 2D hydro 
simulation, but calculated at each angle a spherical model that
employed the matter profiles along that one ray.  They did not calculate lateral
transport in the angular direction.  This is the so-called ``ray-by-ray" approach that 
is still used by some groups today.  Rampp \& Janka (2000,2002), Janka 
et al. (2005ab), and Buras et al. (2006) have updated the 
ray-by-ray method using sophisticated 1D Boltzmann transport along each ray,
but no lateral transport (though they do include in the hydro the effects of lateral
radiation pressure and lepton number transport).  The transport is calculated in the
comoving frame and mapped to the Eulerian frame of their PPM hydrodynamics.  
This prescription is viable only if the
flow maintains rough sphericity and the core does not move off the center of the grid
(as when kicks are imparted to the protoneutron star). However, the ray-by-ray 
approach is not real 2D transport.  The ZEUS-2D radhydro code of Stone, 
Mihalas, and Norman (1992), and its update by Hayes and Norman (2003), solve the
zeroth- and first-moment equations of the gray transport equation and 
use a short-characteristics solution of the
static transfer equation to obtain the second moment for closure.  These codes are
realizations of the variable Eddington factor method and avoid some of the pitfalls of flux
limiters, but since they are gray and not multi-group, they are of limited utility for modern
supernova simulations.

All the above multi-group codes were formulated in 
the comoving frame and none of the 2D variants was 
simultaneously 2D, multi-group, and time-dependent.  A 2D, multi-group, 
flux-limited capability has recently been developed by 
Swesty \& Myra (2005ab, 2006), but they treat the inner core in 1D, not 2D,
and to date have not published long-duration core-collapse and post-bounce simulations.
Cardall, Lentz, \& Mezzacappa (2005) and Cardall \& Mezzacappa (2003) have derived
the general form of the neutrino transport equations including higher-order 
terms in $v/c$ and general relativity, but have not yet developed working
implementations.
The first bonafide 2D multi-group, multi-angle, time-dependent capability was 
achieved by Livne et al. (2004) using the implicit, Arbitrary-Lagrangian-Eulerian 
(ALE) code VULCAN/2D (with remap), for which the radiation field was defined
in the laboratory frame, but this code is not fast, does not include
the Doppler shifts due to the velocity field (though it does include advection), 
and does not include energy redistribution.  However, its multi-group, 
flux-limited diffusion variant is 2D in the entire
computational domain and much faster than its multi-angle version. With it, 
a number of multi-group, fully-2D, radiation/hydrodynamic investigations 
have been possible, with and without rotation (Walder et al. 2005; Burrows 
et al. 2006; Ott et al. 2006ab; Dessart et al. 2006ab).  However, as flops become
cheaper and computer speeds improve, one will need to do better. This is what motivates
the present paper and the development of the new 2D mixed-frame, implicit, multi-group, 
multi-angle algorithm and its two-moment variants. In the context of hydrodynamics,
this code has been christened BETHE\footnote{{\bf B}asic (2-Dimensional), 
{\bf E}xplicit/Implicit, {\bf T}ransport and {\bf H}ydrodynamics
{\bf E}xplosion (Code)}. 

In \S\ref{sec-form}, we present the transport equation and its general formulation, 
but quickly in \S\ref{mixed} derive, using the appropriate Lorentz transformations,
the equations of radiative transfer in the mixed-frame. This section contains
our central analytic results.  In \S\ref{moment}, we follow with derivations
of the associated moment equations.  Then, in \S\ref{hydro} we digress
into a discussion of the frame transformations of the source
terms on the right-hand-sides of the transport equation and of the associated matter 
energy and momentum equations.  Care in these matters is important 
to ensure global energy and momentum conservation to $O(v/c)$ and that one 
employs the correct radiation source terms in the hydro equations. 
Different realizations of the hydro equations are possible and the 
form of the radiation energy source term depends upon the form of the hydro energy
equation.  For instance, the matter energy equation can be in first-law format 
(purely Lagrangian with comoving-frame time derivatives) or can include the 
kinetic energy explicitly (with Eulerian partial time derivatives).
For these two formulations, the radiation source terms are different.
In \S\ref{cylindrical}, we present the mixed-frame formalism in cylindrical
(axisymmetric) coordinates
and in \S\ref{spherical} we present the mixed-frame formalism in spherical coordinates.
Then, in \S\ref{sec-sph}, we derive the mixed-frame equations in spherical symmetry
and explore various solution techniques, including Accelerated-Lambda-Iteration (ALI),
a tridiagonal approximate operator, the use of the Krylov subspace algorithm GMRES,
the Discontinuous-Finite-Element (DFE) method, and the Feautrier scheme.
In \S\ref{moment2}, we derive the associated moment equations,
provide the matrix representation, and introduce sphericity factors.  In
\S\ref{sect-imp},
we derive and discuss procedures for the implicit coupling of matter with radiation,
including an ALI treatment and the linearization of the energy and composition ($Y_e$) 
evolution equations.  We follow this in \S\ref{sec-num} with a series of numerical
tests using the spherical formulation of the mixed-frame transport equations.
These include stationary solutions (\S\ref{sec-num-st}) and a time-dependent cooling 
calculation of an idealized protoneutron star with implicit radiation-matter coupling
(\S\ref{sec-num-imp}).  Section \ref{conclusion} provides a summary and the Appendix
contains
a derivation of the mixed-frame treatment of inelastic energy redistribution.

\section{Formulation} 
\label{formulation}

\subsection{Transport Equation}
\label{sec-form}

The Boltzmann transport equation for the neutrino
occupation probability $f$ is (Mihalas \& Mihalas 1984):
\begin{equation}
\label{rte}
\frac{1}{c}  \frac{\partial f}{\partial t} + (\vecn\cdot\nabla)\, f =
C_{\rm TH}[f] + C_{\rm ES}[f] + C_{\rm NES}[f]\, ,
\end{equation}
where $f$ is the neutrino occupation probability, $\vecn$ the unit vector
in the direction of neutrino propagation,
$C_{\rm TH}$ is the collision integral (net source term)
for ``thermal'' creation and destruction of neutrinos 
(emission and absorption), mostly due to charged-current processes;
$C_{\rm ES}$ is the collisional integral
for elastic scattering of neutrinos; and
$C_{\rm NES}$ is the collision integral
for inelastic scattering (such as neutrino-electron scattering).
We rewrite the transport equation using the specific intensity,
$I$, which can be written in terms of $f$ as:
\begin{equation}
\label{inrel}
I = \frac{\nu^3}{h^3 c^2}\, f\, ,
\end{equation}
where $\nu$ is the neutrino energy, $h$ is Planck's constant, and $c$ is 
the speed of light.

It is customary to formulate the combined $C_{\rm TH} + C_{\rm ES}$ 
contributions in terms of absorption and emission coefficients.
The transfer equation then reads
\begin{equation}
\label{rte1}
\left(\frac{1}{c}\frac{\partial}{\partial t} + \vecn\cdot\nabla \right)
I(\nu,\vecn)=
\eta^{\rm th}(\nu,\vecn) + \eta^{\rm sc}(\nu,\vecn) - 
[\kappa(\nu,\vecn) + \sigma(\nu,\vecn)]\, I(\nu,\vecn)\, ,
\end{equation}
where $\kappa$ is the true absorption coefficient, $\sigma$ is the scattering 
coefficient, $\eta^{\rm th}$ is the thermal emission coefficient, and
$\eta^{\rm sc}$ is the scattering part of the emission coefficient.
This equation has the same form in all frames.

Inelastic scattering is somewhat complicated, but fortunately
it is usually small compared with the thermal and elastic scattering
source terms. We postpone a detailed investigation of inelastic scattering
to a future paper, but provide its formulation in the mixed-frame formalism
in the Appendix.

%--------------------------------------------------------------------

\subsection{Mixed-Frame Formulation}
\label{mixed}

In the mixed-frame approach, the material properties (absorption and emission
coefficients) of the right-hand side of eq. (\ref{rte1})
are expressed using the comoving frame, while the specific intensity and the
left-hand side are expressed in the inertial, observer's frame.
This approach was first suggested in the context of photon transport by
Mihalas \& Klein (1982, hereafter referred to as MK). In this paper, we will
generalize their approach by allowing for energy-dependent anisotropic 
scattering, as well as for non-coherent, inelastic scattering (Appendix).
While the mixed-frame formalism has only a limited applicability for photon
line transport due to a large variation of spectral line opacity as a function
of photon energy, the mixed-frame approach is very well suited to
neutrino transport, in which all the relevant neutrino-matter interaction
cross-sections are smooth functions of neutrino energy.

Denoting by subscript $0$ quantities in the comoving frame, the Lorentz
transforms of the photon/neutrino energy and direction are
\begin{equation}
\label{nutr}
\nu_0 = \nu\gamma\, \left(1 - \nvc\right)\, ,
\end{equation}
and
\begin{equation}
\vecn_0 = (\nu/\nu_0)\left[\vecn - \vc \left(\gamma - 
\frac{\gamma^2}{\gamma+1}\nvc\right)\right]\, .
\end{equation}
To $O(v/c)$, we have the following expressions
\begin{equation}
\label{nutr1}
\nu_0 = \nu \left(1 - \nvc\right)\, ,
\end{equation}
\begin{equation}
\vecn_0 = (\nu/\nu_0)\left(\vecn - \vc\right) = \vecn \left(1+\nvc\right) -
\vc\, ,
\end{equation}
and the absorption ($\kappa$) and scattering ($\sigma$) coefficients transform
as
\begin{equation}
\label{kappatr}
\kappa(\nu,\vecn) = (\nu_0/\nu) \kappa_0(\nu_0)\, ,
\end{equation}
and 
\begin{equation}
\label{sigmatra}
\sigma(\nu,\vecn) = (\nu_0/\nu) \sigma_0(\nu_0)\, .
\end{equation}
The emission coefficient in eq. (\ref{rte1}) transforms as
\begin{equation}
\label{emtr}
\eta(\nu,\vecn) = (\nu/\nu_0)^2 \eta_0(\nu_0, \vecn_0)\, .
\end{equation}

Both absorption coefficients in the inertial frame are expressed through
the comoving frame coefficient and its derivative, exactly as in MK
\begin{equation}
\label{kappa2}
\kappa(\nu,\vecn) =  \kappa_0(\nu) - \nvc \left[\kappa_0(\nu) + \nu 
\frac{\partial\kappa_0}{\partial\nu} \right]\, ,
\end{equation}
which follows from eqs. (\ref{nutr1}) and (\ref{kappatr}) and a Taylor
expansion 
$\kappa_0(\nu_0) = \kappa_0(\nu) +(\nu_0-\nu) \partial\kappa_0/\partial\nu$.
The transformation of $\sigma$ is analogous:
\begin{equation}
\label{sigmatr}
\sigma(\nu,\vecn) =  \sigma_0(\nu) - \nvc \left[\sigma_0(\nu) + \nu \, .
\frac{\partial\sigma_0}{\partial\nu} \right]\, .
\end{equation}
The thermal emission coefficient, as given by MK, is:
\begin{equation}
\eta^{\rm th}(\nu,\vecn) = \eta_0^{\rm th}(\nu) + \nvc \left[2 \eta_0^{\rm
th}(\nu) -
\nu 
\frac{\partial\eta_0^{\rm th}}{\partial\nu} \right]\, ,
\end{equation}
where we assume that the thermal emission coefficient is isotropic in the
comoving frame.

We assume that the comoving-frame elastic scattering emission term is given by
\begin{equation}
\label{etascdef}
\eta_0^{\rm sc}(\nu_0,\vecn_0) = \frac{\sigma_0(\nu_0)}{4\pi}
\oint d\omega_0^\prime\, I_0(\nu_0,\vecn_0^\prime)\, g_0(\vecn_0^\prime,
\vecn_0)\, ,
\end{equation}
where the primed quantities refer to the properties of the absorbed photon/neutrino and
$g_0$ is the scattering phase function in the comoving frame.

In the following, we assume that the scattering phase function is in a simple
form,
\begin{equation}
g_0(\vecn_0^\prime, \vecn_0) = 1 + \delta\, \vecn_0^\prime \cdot \vecn_0\,
.
\end{equation}
For elastic neutrino-matter scattering this is an excellent approximation.
The scattering emission coefficient transforms according to eqs.
(\ref{emtr}) 
and (\ref{etascdef}) as
\begin{equation}
\label{etasc2}
\eta^{\rm sc}(\nu,\vecn) = \left(\frac{\nu}{\nu_0}\right)^2 
\frac{\sigma_0(\nu_0)}{4\pi}\, \cal I\, ,
\end{equation}
where $\cal I$ is the integral term of eq. (\ref{etascdef}).
The specific intensity transforms as
\begin{equation}
I_0(\nup_0,\vecnp_0) = \left(\frac{\nu_0}{\nup}\right)^3
I(\nup,\vecnp)\, ,
\end{equation}
and the element of the solid angle as
\begin{equation}
d\omega_0^\prime\ = \left(\frac{\nu}{\nu_0}\right)^2 d\omega^\prime\, .
\end{equation}
We express the specific intensity at $\nup$ through the Taylor-series expansion
around $\nu$,
\begin{equation}
I(\nup,\vecnp) = I(\nu,\vecnp) + \frac{\partial I}{\partial\nu}(\nup-\nu) =
I(\nu,\vecnp) + \nu\,\frac{\partial I}{\partial\nu} \left(\npvc - \nvc
\right)\, ,
\end{equation}
and the cosine of the scattering angle, again to $O(v/c)$, as
\begin{equation}
\label{nntrans}
\vecnp_0\cdot\vecn_0 = \vecnp\cdot\vecn + (\vecnp\cdot\vecn - 1)\,
\left(\npvc + \nvc\right)\, .
\end{equation}
A final step is the connection between the comoving and laboratory frames, which
consists in accounting for the Taylor expansion of
$\sigma_0$ to
transform the energy from $\nu_0$ to $\nu$:
\begin{equation}
\label{sigma0tr}
\sigma_0(\nu_0) = \sigma_0(\nu) - \frac{\nv}{c}\, \nu\,
\frac{\partial\sigma_0}{\partial\nu}\, .
\end{equation}
To avoid confusion, we stress that we do not need here the transformation
equation (\ref{sigmatr}) that describes the transformation from the
inertial-frame
$\sigma$ to the comoving-frame $\sigma_0$. Here, we already have the
comoving-frame
$\sigma_0$, and all we need is to transform the energy, which is exactly
what eq. (\ref{sigma0tr}) expresses.

After some algebra, we obtain for the transport equation in the
mixed-frame formalism:
\begin{equation}
\label{rtemx}
\left(\frac{1}{c}\frac{\partial}{\partial t} + \vecn\cdot\nabla \right) 
I(\nu, \vecn) =
r_{00}(\nu, \vecn) + r_{01}(\nu, \vecn) + r_{10}(\nu, \vecn) + r_{11}(\nu,
\vecn)\, ,
\end{equation}
where the subscripts refer to terms which are $0$-th or $1$-st order in
$v/c$ and $\delta$, respectively. The terms of the r.h.s of eq. (\ref{rtemx})
are given by
\begin{equation}
\label{r00}
r_{00} = \eta_0^{\rm th} - (\kappa_0 + \sigma_0)\, I(\nu, \vecn) + \sigma_0
J\, ,
\end{equation}
\begin{equation}
r_{01} =\sigma_0 \, \delta \, H^j n_j\, ,
\end{equation}
\begin{eqnarray}
\label{r10}
r_{10} =n_j w^j \left[ \eta_0^{\rm th} \left(2-\dlneta\right) + 
\kappa_0 \left(1+\dlnkap\right) I(\nu, \vecn)\right] + \nonumber \\
+ n_j w^j\left[\sigma_0 \left(1+\dlnsig\right) I(\nu, \vecn) +
\sigma_0 J \left(2 - \dlnsig-\dlnjl\right)\right]
%\nonumber \\
- \sigma_0 w_j H^j \left(1-\dlnhl\right) \, ,
\end{eqnarray}
and
\begin{equation}
\label{r11}
r_{11} = \sigma_0\, \delta \left[-n_j w^j J  
 + n_j w_k \dlnkl\right] 
%\nonumber\\
+ \sigma_0\, \delta\, H^j \left[-w_j +  n_j n_k w^k
\left(3 - \dlnsig -\dlnhl\right)\right]\, ,
\end{equation}
where, again, we omit explicit indication of the energy dependence of most
quantities.
Here we use the convention that one sums over repeated indices, although
our use of subscripts or superscripts is arbitrary and does not necessarily
denote
covariant or contravariant components.
To simplify the notation, we have introduced the normalized velocity,
\begin{equation}
{\bf w} = \frac{\bf v}{c}\, .
\end{equation}
Here, the usual moments of the specific intensity are defined by
\begin{equation}
J = \frac{c E}{4\pi} =\oint I\, d\omega/4\pi\, ,
\end{equation}
\begin{equation}
{\bf H} = \frac{{\bf F}}{4\pi}= \oint I\, \vecn\,d\omega/4\pi\, ,
\end{equation}
\begin{equation}
{\tt K} = \frac{c {\tt P}}{4\pi} =\oint I\, \vecn \vecn\, d\omega/4\pi\, ,
\end{equation}
where $E$, ${\bf F}$, and ${\tt P}$ are the radiation energy density,
flux, and pressure tensor, respectively.

%-------------------------------------------------------------------

\subsection{Radiation Moment Equations}
\label{moment}

Integrating transfer eq. (\ref{rte1}) over angles with the source/sink terms
given by (\ref{r00}) - (\ref{r11}), we obtain the 0th- and 1st-moment equations 
(written here in Cartesian coordinates):
\begin{equation}
\label{momj01}
\frac{1}{c}\frac{\partial J}{\partial t} + \frac{\partial
H^j}{\partial{x^j}} =
\eta_0^{\rm th} - \kappa_0 J+ \Xi_j\, H^j \, ,
\end{equation}
and
\begin{eqnarray}
\label{momh01}
\frac{1}{c}\frac{\partial H^i}{\partial t} + \frac{\partial
K^{ij}}{\partial{x^j}} =
-(\kappa_0+\sigma_{\rm tr}) H^i + w^i \widetilde\eta_0 
 +\frac{w_i}{3} \sigma_0 J \left(2 - \delta -\dlnsig-\dlnjl\right)\nonumber\\
 +w_j K^{ij} \left[ \widetilde\kappa_0 + 
 \sigma_0 \left(1 + \frac{\partial\ln\sigma_0}{\partial\ln\nu} +
\frac{\delta}{3}\dlnkl\right)\right] \, ,
\end{eqnarray}
where
\begin{equation}
\label{bxidef}
\Xi_j = w_j \left[ \widetilde\kappa_0  +
\sigma_{\rm tr} \left(\dlnsig + \dlnhl\right)\right]\, .
\end{equation}
In eqs. (\ref{momj01}) - (\ref{bxidef}), we have introduced the so-called
transport
cross-section,
\begin{equation}
\label{sigtr}
\sigma_{\rm tr} = \sigma_0 \left( 1 - \frac{\delta}{3}\right)\, ,
\end{equation}
and have set
\begin{equation}
\widetilde\kappa_0 = \kappa_0 \left(1+\dlnkap\right)\, ,\quad\quad
\widetilde\sigma_0 = \sigma_0 \left(1+\dlnsig\right)\, ,
\end{equation}
%
%
%\begin{equation}
%\widetilde\sigma_0 = \sigma_0 \left(1+\dlnsig\right)\, ,
%\end{equation}
%
and
\begin{equation}
\widetilde\eta_0 = \frac{1}{3} \eta_0^{\rm th} \left(2-\dlneta\right)\, .
\end{equation}
To close the system of moment equations, we introduce the Eddington tensor:
\begin{equation}
f^{ij} \equiv \frac{K^{ij}}{J}\, ,
\end{equation}
so that the 1st-moment equation is written
\begin{equation}
\frac{1}{c}\frac{\partial H^i}{\partial t} + 
\frac{\partial}{\partial{x^j}} \left(f^{ij}J\right)=
-(\kappa_0+\sigma_{\rm tr}) H^i + w^i \widetilde\eta_0 + \xi_i\, J\, ,
\end{equation}
where
\begin{equation}
\label{xidef}
\xi_i = \frac{w_i}{3} \sigma_0 \left(2 - \delta -\dlnsig-\dlnjl\right) +
w_j f^{ij} \left[ \widetilde\kappa_0 
 + \sigma_0 \left(1 + \dlnsig +
 \frac{\delta}{3}\frac{\partial\ln f^{ij}}{\partial\ln\nu} +
\frac{\delta}{3}\frac{\partial\ln J}{\partial\ln\nu}\right)\right] \, .
\end{equation}
%

% -----------------------------------------------------------------------

\subsection{Hydrodynamical Equations and Energy Coupling}
\label{hydro}

We may write moment equations (\ref{momj01}) and (\ref{momh01}) as equations 
for the energy and momentum of the radiation field, namely
\begin{equation}
\frac{\partial E}{\partial t} + 
\frac{\partial F^i}{\partial x^i} = -cG^0\, ,
\end{equation}
and
\begin{equation}
\frac{1}{c^2}\frac{\partial F^i}{\partial t} + 
\frac{\partial P^{ij}}{\partial x^j} = -G^i\, ,
\end{equation}
where here $E$, $F$, are $P$ are the energy-integrated radiation energy density, flux,
and stress tensor, respectively, and $G^0$ and
$G^i$ are the components of the four-force density vector
(Mihalas \& Mihalas 1984), in the inertial frame,
which are given in our mixed-frame formalism by the trivial
modification of the r.h.s. of eqs. (31) and (32), viz.
\begin{equation}
\label{cg00}
cG^0 = 4\pi\int_0^\infty [\kappa_0\, J - \eta_0^{\rm th} - 
\Xi_j\, H^j]\, d\nu\, ,
\end{equation}
and
\begin{equation}
\label{cgi0}
cG^i = 4\pi\int_0^\infty [(\kappa_0 + \sigma_{\rm tr})\, H^i - 
w^i \widetilde\eta_0 - \xi^i\, J]\, d\nu\, .
\end{equation}

The equation for overall energy conservation of
the radiating fluid (i.e., containing both the matter and neutrino energy), 
correct to $O(v/c)$, is given by (MK;
Mihalas \& Mihalas 1984):
\begin{equation}
\frac{\partial}{\partial t} (\rho e + \frac{1}{2}\rho v^2 + E)
+\frac{\partial}{\partial x^i}
[(\rho e + \frac{1}{2}\rho v^2 +p) v^i + F^i ] = v_i f^i\, ,
\end{equation}
where $e$ is the specific internal energy of the fluid, $p$
is the fluid pressure, and $f^i$ is the external force on the fluid.
This equation can also be written as an total energy for the radiating flow,
\begin{equation}
\label{en_tot_in}
\rho\,\frac{D}{Dt}(e + v^2/2) +\frac{\partial}{\partial x^i}(p v^i)
= v_i f^i - \left( \frac{\partial E}{\partial t} +
\frac{\partial F^i}{\partial x^i}\right) = v_i f^i + cG^0\, ,
\end{equation}
and analogously the momentum equation
\begin{equation}
\label{mom_in}
\rho\,\frac{D v^i}{Dt} = f^i - \frac{\partial p}{\partial x^i}
- \left( \frac{1}{c^2}\frac{\partial F^i}{\partial t} + 
\frac{\partial P^{ij}}{\partial x^j} \right) 
+ \frac{v^i}{c^2}\left(\frac{\partial E}{\partial t} + 
\frac{\partial F^i}{\partial x^i} \right)
= f^i - \frac{\partial p}{\partial x^i}
+ G^i - \frac{v^i}{c}\, G^0\, .
\end{equation}
To obtain the gas-energy equation, one first writes an
equation for mechanical energy which is obtained by multiplying
eq. (\ref{mom_in}) by $v_i$, and subtracts it from the total energy
equation (\ref{en_tot_in}). The equation for mechanical energy,
to $O(v/c)$, reads
\begin{equation}
\rho\,\frac{D(v^2/2)}{Dt} = v_i f^i - v_i\frac{\partial p}{\partial x^i}
- v_i \left( \frac{1}{c^2}\frac{\partial F^i}{\partial t} + 
\frac{\partial P^{ij}}{\partial x^j} \right)\, .
\end{equation}
The resulting comoving-frame gas-energy equation reads
\begin{equation}
\label{gas_en}
\rho\,\left[\frac{De}{Dt} + p\,\frac{D(1/\rho)}{Dt}\right]
= cG^0 - v_i G^i \, .
\end{equation}
This is the appropriate equation for updating the fluid temperature
and is a statement of the first law of thermodynamics. 
If we were to leave density
fixed, we could write the energy equation as an equation for temperature,
\begin{equation}
\label{temp_eq}
\rho\,C_V\,\frac{DT}{Dt} = cG^0 - v_i G^i \, ,
\end{equation}
where $T$ is the temperature, and $C_V$ the specific heat at constant 
volume.

Equation (\ref{temp_eq}) contains the components of the inertial-frame
four-force density vector, which are given by eqs. (\ref{cg00})
and (\ref{cgi0}) that in turn were derived by using the mixed-frame
formalism. However, $cG^0$ and $cG^i$ contain terms that are
proportional to the scattering coefficient $\sigma_0$. 
Since the elastic scattering should not contribute
to the energy balance in either inertial or comoving frames,
one should make sure that such terms cancel exactly. This is very
important in the context of neutrino transport in supernovae,
since in a low-temperature, low-density regions beyond the shock
the scattering coefficient $\sigma_0$ may be larger by many orders of
magnitude than the true absorption coefficient 
$\kappa_0$, and one can, thus, easily introduce spurious terms in the
energy balance because of rounding errors.

It is, therefore, instructive to derive the right-hand-side of
the energy equation in a different way, in which we eliminate the
scattering terms analytically. The idea is first to write down
the four-force density vector in the comoving frame, where it is
easy to formulate, and then to perform a Lorentz transformation 
back to the inertial frame. 

The four-force density vector in the co-moving frame is simply 
written as
\begin{equation}
c G_0^\alpha = \int_0^\infty \oint [\chi_0(\nu_0) I_0(\nu_0,\vecn_0)
- \eta^{\rm th}_0(\nu_0) - \eta^{\rm sc}_0(\nu_0,\vecn_0)]\, 
n_0^\alpha\, d\nu_0\, d\omega_0 \, ,
\end{equation}
where the scattering term in the comoving frame is simply given by
\begin{eqnarray}
\eta^{\rm sc}_0(\nu_0,\vecn_0) = \sigma_0(\nu_0)\int 
\oint I_0(\nu_0,\vecn_0^\prime)\, g_0(\vecn_0^\prime,\vecn_0)\,
(d\omega_0^\prime/4\pi)
\nonumber\\
= \sigma_0(\nu_0)\oint
I_0(\nu_0,\vecn_0^\prime) (1+\delta\, \vecn_0^\prime\cdot\vecn_0)\,
(d\omega_0^\prime/4\pi)
\nonumber\\
= \sigma_0(\nu_0)\, [J_0(\nu_0) + \delta\, {\bf
H}_0(\nu_0)\cdot\vecn_0]\, .
\end{eqnarray}
Therefore,
\begin{equation}
\label{cg000}
cG^0_0 = 4\pi\int_0^\infty [\kappa_0(\nu_0) J_0(\nu_0) -
\eta_0^{\rm th}(\nu_0)]
\, d\nu_0\, ,
\end{equation}
and
\begin{equation}
\label{cgi_com}
cG_0^i = 4\pi\int_0^\infty [\kappa_0(\nu_0) + (1-\delta/3)
\sigma_0(\nu_0)] \, H^i_0(\nu_0)\, d\nu_0
= 4\pi\int_0^\infty [\kappa_0(\nu_0) +\sigma_{\rm tr}(\nu_0)] 
\, H^i_0(\nu_0)\, d\nu_0
\, .
\end{equation}
We can now easily transform the components of the four-force density
to the inertial frame using the standard Lorentz transform 
[c.f., Mihalas \& Mihalas 1984; eqs. (91.22)], which to $O(v/c)$ become:
\begin{equation}
cG^0 = cG^0_0 + v_i G_0^i\, ,
\end{equation}
and
\begin{equation}
cG^i = cG^i_0 + v^i G_0^0\, .
\end{equation}
The inertial-frame four-force density vector is thus (dropping explicit
indication of its dependence on the comoving-frame frequency $\nu_0$)
\begin{equation}
\label{cg0}
cG^0 = 4\pi\int_0^\infty [\kappa_0 J_0 - \eta_0^{\rm th}\,
+\,w_i (\kappa_0 + \sigma_{\rm tr}) H^i_0(\nu_0)]\, d\nu_0\, ,
\end{equation}
and
\begin{equation}
\label{cgi}
cG^i = 4\pi\int_0^\infty [(\kappa_0+\sigma_{\rm tr}) H_0^i
+ w^i (\kappa_0 J_0 - \eta_0^{\rm th} ]\, d\nu_0\, ,
\end{equation}
and, thus, the right-hand-side of the material internal energy equation
reads
\begin{equation}
\label{cg0wgi}
cG^0 - v_i G^i = cG^0 - w_i\, cG^i =
4\pi\int_0^\infty [\kappa_0 J_0(\nu_0) - \eta_0^{\rm th}]\, d\nu_0
\, + \, O(v^2/c^2)\, .
\end{equation}
We see that the second term of eq. (\ref{cg0}) and the first term
of eq. (\ref{cgi}) exactly cancel, as can be expected on physical
grounds, and eq. (\ref{cg000}) is the result.

As a final step, we now have to transform the comoving frame
moments of the specific intensity back to the inertial frame
(while leaving the material quantities in the comoving frame
untouched in keeping with the general philosophy of the mixed-frame
approach).

Specifically,
\begin{eqnarray}
\label{kjtrans}
\int_0^\infty \kappa_0(\nu_0) J_0(\nu_0)\, d\nu_0 =
\int_0^\infty \oint \kappa_0(\nu_0) I_0(\nu_0,\vecn_0)\,
d\nu_0\, d\omega_0 =
\nonumber\\
\int_0^\infty \oint \left[\kappa_0(\nu) + (\nu_0-\nu)
\frac{\partial\kappa_0}{\partial\nu}\right]\,
\left(\frac{\nu_0}{\nu}\right)^3 I(\nu,\vecn)\,
\left(\frac{\nu}{\nu_0}\right)  d\nu\, d\omega =
\nonumber\\
\int_0^\infty \oint \left[\kappa_0(\nu_0) - 
\frac{\partial\kappa_0}{\partial\ln\nu}\, \vecn\cdot{\bf w} \right]
\, \left( 1-2\vecn\cdot{\bf w}\right) I(\nu,\vecn) \,
d\nu\,d\omega =
\nonumber\\
\int_0^\infty\left[\kappa_0(\nu) J(\nu) -
{\bf H}(\nu)\cdot{\bf w}\,\left(2\kappa_0 +
\frac{\partial\kappa_0}{\partial\ln\nu}\right) \right]\, d\nu\, ,
\end{eqnarray}
and, thus, the appropriate right-hand-side of the energy equation
in the inertial frame, written in the mixed-frame formalism
(in which the material quantities are in the comoving frame,
while the radiation quantities and the energy are in
the inertial frame) is:
\begin{equation}
\label{rhecmv}
cG^0 - v_iG^i = 4\pi\int_0^\infty
\left[\kappa_0 J - \eta_0^{\rm th} - w_i\,H^i \left(2 \kappa_0
+\frac{\partial\kappa_0}{\partial\ln\nu}\right) \right]\, d\nu\, .
\end{equation}
We now have to show that we obtain the same expression if we
use $cG^0$ and $cG^i$ given by eqs. (\ref{cg00}) and (\ref{cgi0}).
Using eqs. (\ref{bxidef}) and (\ref{xidef}), we have
\begin{equation}
cG^0 = 4\pi\int_0^\infty\left\{ \kappa_0 J - \eta_0^{\rm th}
- w_i H^i \left[\widetilde\kappa_0 + \sigma_{\rm tr}
\left(\frac{\partial\ln\sigma_0}{\partial\ln\nu} + 
\frac{\partial\ln H^i}{\partial\ln\nu} \right) \right] \right\}\, d\nu \, ,
\end{equation}
and
\begin{equation}
cG^i = 4\pi\int_0^\infty (\kappa_0 + \sigma_{\rm tr} ) H^i\, d\nu\, + O(w^i)\, ,
\end{equation}
and, thus, to $O(v/c)$, we obtain
\begin{eqnarray}
\label{rheine}
cG^0 - w^i\, cG^i = 4\pi\int_0^\infty (\kappa_0 J - \eta_0^{\rm th})\, d\nu -
4\pi \int_0^\infty  w_i H^i \left( 2\kappa_0 + 
\frac{\partial\kappa_0}{\partial\ln\nu} \right)\, d\nu
\nonumber\\
- 4\pi\int_0^\infty  w_i H^i \left[\sigma_{\rm tr} \left(
\frac{\partial\ln\sigma_0}{\partial\ln\nu} + 
\frac{\partial\ln H^i}{\partial\ln\nu} \right) + \sigma_{\rm tr} 
\right]\, d\nu\, .
\end{eqnarray}
The last integral in eq. (\ref{rheine}) is equal to zero, as can be 
easily shown by integrating its first two terms by parts. 
Equation (\ref{rheine}) is thus, indeed, identical to eq. (\ref{rhecmv}).
Here we assume,
consistently with the previous formalism, that the incoherence parameter
$\delta$ is independent of energy. One could easily develop a formalism
in which one can account for the energy dependence of $\delta$
(containing additional terms proportional to 
$w^i \delta\, \partial\ln\delta/\partial\ln\nu$), but we consider
such a complication unnecessary.

Next, we consider the electron fraction equation. In the comoving
frame, it is given by
\begin{equation}
\label{elfrc1}
\rho N_A \frac{D Y_e}{Dt} = -4 \pi \sum_{i} s_i \int_0^\infty \left(
\kappa_0 J_0 - \eta_0^{\rm th} \right) \frac{d\nu_0}{\nu_0}\, ,
\end{equation}
where the sum extends over the neutrino species, and $s_i=-1$
for $\nu_e$ neutrinos, $s_i=1$ for $\bar\nu_e$ neutrinos, and
$s_i=0$ for all other neutrino species; $N_A$ is the Avogadro's
number. The right-hand-side of eq. (\ref{elfrc1}) is easily
expressed in the mixed frame using an analogous procedure as that
used in eq. {\ref{kjtrans}), which differs only by the
occurrence of the term $d\nu_0/\nu_0$ instead of $d\nu_0$.
The resulting equation is:
\begin{equation}
\label{elfrc2}
\rho N_A \frac{D Y_e}{Dt} = -4 \pi \sum_{i} s_i \int_0^\infty \left[
\kappa_0 J - \eta_0^{\rm th}  -  w_i H^i \left( \kappa_0 + 
\frac{\partial\kappa_0}{\partial\ln\nu}\right)\,\right] 
\frac{d\nu}{\nu}\, .
\end{equation}

Next, we consider the momentum equation. It was already given by
eq. (\ref{mom_in}), namely
\begin{equation}
\label{mom_in2}
\rho\,\frac{D v^i}{Dt} 
= f^i - \frac{\partial p}{\partial x^i}
+ G^i - w^i\, G^0\, .
\end{equation}
The radiation-interaction term, which can be viewed in Cartesian coordinates as a
gradient of 
radiation pressure, 
$ -\partial p^{ij}_{\rm rad}/\partial x^j = G^i - w^i G^0$,
is written in eq. (\ref{mom_in2}) in the inertial, Eulerian, frame.
To transform it to the mixed frame, we use the same strategy as
before to obtain the net radiation heating term in the energy balance
equation: we can either use expressions for $G^0$ and $G^i$ given
by eqs. (\ref{cg00}) and (\ref{cgi0}), which are already expressed
in the mixed-frame formalism and perform necessary integrations
over energies, or we can use a simple expression for the comoving-frame
four-force vector, eq. (\ref{cgi_com}), and express the comoving-frame
moment $H_0(\nu_0)$ through the inertial-frame moments. Obviously, both
approaches have to yield the same result, which is:
\begin{equation}
G^i - w^i G^0 = 4\pi\int_0^\infty \left[\left(\kappa_0 + \sigma_{\rm tr}\right)
\left( H^i - w^i J + w_j \frac{\partial(f^{ij} J)}{\partial\ln\nu} \right)
\right]\, d\nu\, .
\end{equation}

Finally, we stress the following feature of our formalism. The material
equation for the conservation of the total energy and momentum,
as well as the electron fraction equation, were written in the
Eulerian frame, where the radiation-interaction is also in the 
Eulerian frame with, however, interaction coefficients (i.e.,
the absorption, emission, and scattering coefficient) in the
comoving (Lagrangian) frame. If the overall hydro scheme is Eulerian,
our present scheme is obviously consistent with it and we would use 
eq. (\ref{cg00}). If the overall hydro
scheme is Lagrangian, the radiation-interaction terms are generally
different. For instance, the right-hand-side of the gas-energy equation
expressing the 1st-law of thermodynamics is given by $cG^0-w_iG^i$ 
in Eulerian frame quantities [see eqs. (\ref{gas_en}) and (\ref{rhecmv})],
while it is given by $cG^0_0$, eq. (\ref{cg000}), in comoving-frame quantities, 
which is formally different. 

%--------------------------------------------------------------------

\section{Mixed-Frame Radiation Equations in Cylindrical Geometry}
\label{cylindrical}

Here, we assume cylindrical geometry with azimuthal symmetry. 
The coordinates are $r$, $z$, and $\phi$. We assume that the
rest-frame material properties (temperature,
density, opacity, emissivity, etc.) depend
only on $r$ and $z$, while the velocity field has a non-zero 
$\phi$-component. In other words, we allow for rotation. Such an
approach is sometimes called the ``2 1/2-D'' case in hydrodynamics.

The unit vector in direction $\vecn$ is
\begin{equation}
\vecn = \sin\theta\,\cos\psi\,\,\vece_r + \cos\theta\,\,\vece_z + 
\sin\theta\,\sin\psi\, \,\vece_\phi \, .
\end{equation}
Here $\theta$ is the polar angle measured from the positive $z$-direction
and $\psi$ is the local azimuthal angle, such that $\psi=0$ is in the local
positive $r$-direction.

In component form, the first and second moments are given by
\begin{equation}
H_r = \oint\sin\theta\, \cos\psi\, I(\nu,\vecn) d\omega/(4\pi)\, ,
\end{equation}
\begin{equation}
H_z = \oint\cos\theta\, I(\nu,\vecn) d\omega/(4\pi)\, ,
\end{equation}
\begin{equation}
H_\phi = \oint\sin\theta\, \sin\psi\, I(\nu,\vecn) d\omega/(4\pi) \, ,
\end{equation}
and 
\begin{equation}
K_{rr} = \oint\sin^2\theta\, \cos^2\psi\, I(\nu,\vecn) d\omega/(4\pi)\, ,
\end{equation}
\begin{equation}
K_{rz}=K_{zr}= \oint\cos\theta\,\sin\theta\, \cos\psi\,I(\nu,\vecn)
d\omega/(4\pi)\, ,
\end{equation}
\begin{equation}
K_{r\phi} = K_{\phi r} =\oint\sin^2\theta\, \sin\psi\,\cos\psi\,
I(\nu,\vecn) d\omega/(4\pi)\, ,
\end{equation}
\begin{equation}
K_{zz}= \oint\cos^2\theta\, I(\nu,\vecn) d\omega/(4\pi)\, ,
\end{equation}
\begin{equation}
K_{\phi z} =K_{z\phi}= \oint\cos\theta\,\sin\theta\, \sin\psi\,
I(\nu,\vecn) d\omega/(4\pi)\, ,
\end{equation}
and
\begin{equation}
K_{\phi\phi} = \oint\sin^2\theta\, \sin^2\psi\, I(\nu,\vecn) d\omega/(4\pi)
\, ,
\end{equation}
which, due to symmetry and the trace condition 
\begin{equation}
\label{trace}
K_{rr} + K_{zz} + K_{\phi\phi} = J \, ,
\end{equation}
leaves five, instead of nine, independent, non-zero components
of the tensor $\tt K$.

The radiative transfer equation, in cylindrical coordinates and with
azimuthal symmetry, written in the conservative form, now becomes
\begin{eqnarray}
\label{rte2}
\frac{1}{c}\frac{\partial I_\nu}{\partial t} 
+\cos\theta\,\frac{\partial I_\nu}{\partial z}
+\frac{\sin\theta\,\cos\psi}{r}\, \frac{\partial}{\partial r}(rI_\nu)
-\frac{\sin\theta}{r}\,\frac{\partial}{\partial \psi}(\sin\psi\, I_\nu)
\nonumber\\
= r_{00}+r_{01}+r_{10}+ r_{11}\, ,
\end{eqnarray}
where the right-hand side is given by eqs. (\ref{r00}) - (\ref{r11}) written
in
component form, viz.
\begin{equation}
\label{r00a}
r_{00} = \eta_0^{\rm th} - (\kappa_0 + \sigma_0)\, I_\nu + \sigma_0 J\, ,
\end{equation}
\begin{equation}
r_{01}=\sigma_0 \, \delta \, 
(H_r \sin\theta\, \cos\psi + H_z\cos\theta  + H_\phi \sin\theta\,\sin\psi )
\, ,
\end{equation}
\begin{equation}
n_j w^j = \sin\theta\, \cos\psi\, w_r + \cos\theta\, w_z +
\sin\theta\,\sin\psi\, w_\phi \, ,
\end{equation}
and
\begin{eqnarray}
 n_j w_k K^{jk} = \sin\theta\, \cos\psi\, 
(w_r K_{rr} + w_z K_{rz} + w_\phi K_{r\phi})\nonumber \\
 + \cos\theta\, (w_r K_{rz} + w_z K_{zz} + w_\phi K_{z\phi}) 
+ \sin\theta\,\sin\psi\, (w_r K_{r\phi} + w_z K_{z\phi} + w_\phi
K_{\phi\phi}).
\end{eqnarray}
The corresponding moment equations read:
\begin{equation}
\label{jmom2}
\frac{1}{c}\frac{\partial J}{\partial t} + \ddr(r H_r) +
\frac{\partial H_z}{\partial z} 
= \eta_0^{\rm th} - \kappa_0 J +
\Xi_r H_r + \Xi_z H_z  + \Xi_\phi H_\phi \, ,
\end{equation}
\begin{equation}
\label{hrmom2}
\frac{1}{c}\frac{\partial H_r}{\partial t} + \ddr(r f_{rr} J) + \ddz(f_{rz}
J)
- \frac{1-f_{rr}-f_{zz}}{r} J =\nonumber\\
-(\kappa_0+\sigma_{\rm tr}) H_r + w_r \widetilde\eta_0 + \xi_r J\, ,
\end{equation}
\begin{equation}
\label{hzmom2}
\frac{1}{c}\frac{\partial H_z}{\partial t} + \ddr(r f_{rz} J) + \ddz(f_{zz}
J)=
%\nonumber\\
 -(\kappa_0+\sigma_{\rm tr}) H_z + w_z \widetilde\eta_0 + \xi_z J\, ,
\end{equation}
and
\begin{equation}
\label{hphmom2}
\frac{1}{c}\frac{\partial H_\phi}{\partial t} + \ddr(r f_{r\phi} J) +
\ddz(f_{z\phi}
J)=
% \nonumber\\
 -(\kappa_0+\sigma_{\rm tr}) H_\phi + w_\phi \widetilde\eta_0 + \xi_\phi J \, ,
\end{equation}
where the individual components of $\Xi$ and $\xi$ are given by eqs.
(\ref{bxidef}) 
and (\ref{xidef}).
%

%--------------------------------------------------------------------

\section{Mixed-Frame Radiation Equations in Spherical Geometry}
\label{spherical}

The spherical coordinates we use are the standard $r$, $\Theta$, and $\phi$. 
Here, $\Theta$ is the polar angle
measured from the $z$-axis and $\phi$ the azimuthal angle.
The unit vector in direction $\vecn$ is
\begin{equation}
\vecn = \cos\theta\,\,\vece_r + \sin\theta\,\cos\psi\,\,\vece_\Theta +  
\sin\theta\,\sin\psi\, \,\vece_\phi \, .
\end{equation}
$\theta$ is the polar angle measured from the $r$-direction,
and $\psi$ is the local azimuthal angle between the projection of $\vecn$
onto the plane perpendicular to vector ${\bf r}$ measured counterclockwise
from $\vece_\Theta$.
The conservative form of the transfer equation in spherical coordinates is then: 
\begin{eqnarray}
\label{rtesph2}
\frac{1}{c}\frac{\partial I}{\partial t} 
+\frac{\cos\theta}{r^2}\,\frac{\partial}{\partial r} \left(r^2 I\right)
+\frac{\sin\theta\cos\psi}{r\sin\Theta}\frac{\partial}{\partial\Theta}
\left(\sin\Theta\, I\right)
-\frac{1}{r \sin\theta}\frac{\partial}{\partial\theta}\left(\sin^2\theta\,
I\right)
\nonumber\\
-\frac{\sin\theta
\cot\Theta}{r}\frac{\partial}{\partial\psi}\left(\sin\psi\, I\right)
= r_{00}+r_{01}+r_{10}+ r_{11}\, ,
\end{eqnarray}
where the individual right-hand side terms are given by eqs. (\ref{r00}) -
(\ref{r11}), specified for spherical coordinates, i.e. with
\begin{equation}
n_j H^j = \cos\theta\, H_r + \sin\theta\cos\psi\, H_\Theta +
\sin\theta\sin\psi\,
H_\phi\, ,
\end{equation}
\begin{equation}
n_j w^j = \cos\theta\, w_r + \sin\theta\cos\psi\, w_\Theta +
\sin\theta\sin\psi\,
w_\phi\, ,
\end{equation}
and 
\begin{eqnarray}
n_j w_k K^{jk} = 
\cos\theta         (w_r K_{rr} + w_\Theta K_{r\Theta} + w_\phi K_{r\phi}) 
\nonumber\\
+ \sin\theta\cos\psi 
(w_r K_{r\Theta} + w_\Theta K_{\Theta\Theta} + w_\phi K_{\Theta\phi}) 
\nonumber\\
+ \sin\theta\sin\psi 
(w_r K_{r\phi} + w_\Theta K_{\phi\Theta} + w_\phi K_{\phi\phi})\, . 
\end{eqnarray}

The corresponding moment equations read
\begin{eqnarray}
\frac{1}{c}\frac{\partial J}{\partial t} 
+\frac{1}{r^2}\,\frac{\partial (r^2 H_r)}{\partial r}
+\frac{1}{r\sin\Theta}\frac{\partial}{\partial\Theta}\left(\sin\Theta\,
H_\Theta\right)\nonumber\\
= \eta_0^{\rm th} - \kappa_0 J +
\Xi_r H_r + \Xi_\Theta H_\Theta  + \Xi_\phi H_\phi \,
,
\end{eqnarray}
\begin{eqnarray}
\frac{1}{c}\frac{\partial H_r}{\partial t} 
+\frac{1}{r^2}\,\frac{\partial (r^2 f_{rr} J)}{\partial r}
+\frac{1}{r\sin\Theta}\frac{\partial}{\partial\Theta}\left(\sin\Theta\,
f_{r\Theta}
J\right)
%+\frac{1}{r\sin\Theta}\frac{\partial K_{r\phi}}{\partial\phi}
\nonumber\\
+\frac{f_{rr}-1}{r} J
= -(\kappa_0+\sigma_{\rm tr}) H_r + w_r \widetilde\eta_0 + \xi_r J\, ,
\end{eqnarray}
\begin{eqnarray}
\frac{1}{c}\frac{\partial H_\Theta}{\partial t} 
+\frac{1}{r^2}\,\frac{\partial (r^2 f_{r\Theta} J)}{\partial r}
+\frac{1}{r\sin\Theta}\frac{\partial}{\partial\Theta}\left(\sin\Theta\,
f_{\Theta\Theta} J\right)
%+\frac{1}{r\sin\Theta}\frac{\partial K_{r\phi}}{\partial\phi}
\nonumber\\
+\frac{f_{r\Theta}-\cot\Theta\, f_{\phi\phi}}{r} J
= -(\kappa_0+\sigma_{\rm tr}) H_\Theta + w_\Theta \widetilde\eta_0 + \xi_\Theta
J\, ,
\end{eqnarray}
and
\begin{eqnarray}
\frac{1}{c}\frac{\partial H_\phi}{\partial t} 
+\frac{1}{r^2}\,\frac{\partial (r^2 f_{r\phi} J)}{\partial r}
+\frac{1}{r\sin\Theta}\frac{\partial}{\partial\Theta}\left(\sin\Theta\,
f_{\phi\Theta}
J\right)
%+\frac{1}{r\sin\Theta}\frac{\partial K_{r\phi}}{\partial\phi}
\nonumber\\
+\frac{f_{r\phi}+\cot\Theta\, f_{\Theta\phi}}{r} J
= -(\kappa_0+\sigma_{\rm tr}) H_\phi + w_\phi \widetilde\eta_0 + \xi_\phi J\, .
\end{eqnarray}
%

%--------------------------------------------------------------------

\section{The Spherically-Symmetric Case: Equations and Solution Techniques}
\label{sec-sph}

For initial tests of the scheme, we focus on the spherically
symmetric case. 
We introduce the usual notation $\mu = \cos\theta$. 
We also assume that the velocity has only the non-zero component, $w_r$,
which we denote as $w$. Due to symmetry, the only non-vanishing components 
of the vector ${\bf H}$ and tensors ${\tt K}$ and $f$ are $H_{r}$, $K_{rr}$, and
$f_{rr}$;
we denote them here as $H$, $K$, and $f$, respectively. 
The transfer equation in the conservative form is written as
\begin{eqnarray}
\label{rtesph}
\frac{1}{c}\frac{\partial I}{\partial t} 
+ \frac{\mu}{r^2} \frac{\partial(r^2 I)}{\partial r} 
+ \frac{1}{r} \frac{\partial[(1-\mu^2) I]}{\partial\mu}\nonumber\\
= \left(\eta_0 + 3\mu w \widetilde\eta_0\right)
-[\kappa_0 + \sigma_0 - \mu w (\widetilde\kappa_0 + \widetilde\sigma_0)]\,
I\nonumber\\
+ \sigma_0 \left[ 1 +\mu w \left(2 - \delta
-\frac{\partial\ln\sigma_0}{\partial\ln\nu} -
\dlnjl\right)\right] \, J \nonumber\\
+ \sigma_0 \left[\delta\mu - w\left(1+ \delta - \frac{\partial\ln
H}{\partial\ln\nu}\right)  
+ \delta w \mu^2\left(3 
-\frac{\partial\ln\sigma_0}{\partial\ln\nu} - 
\frac{\partial\ln H}{\partial\ln\nu}\right)\right] \, H \nonumber\\
+ \sigma_0 \delta \mu w \frac{\partial\ln
K}{\partial\ln\nu} \, K\, .
\end{eqnarray}
The moment equations read
\begin{equation}
\frac{1}{c}\frac{\partial J}{\partial t} + \frac{1}{r^2} \frac{\partial(r^2
H)}{\partial r} =
\eta_0^{\rm th} - \kappa_0 J + \Xi H \, ,
\end{equation}
and 
\begin{equation}
\frac{1}{c}\frac{\partial H}{\partial t} + 
\frac{1}{r^2} \frac{\partial(r^2 f J)}{\partial r} -\frac{1-f}{r}\, J =
-(\kappa_0 +\sigma_{\rm tr}) H + w \widetilde\eta_0 + \xi J\, ,
\end{equation}
where $\Xi = \Xi_r$ and $\xi = \xi_r$.

We rewrite eq. (\ref{rtesph}) in a non-conservative form, and express
the time derivative through backward time differencing, while retaining
the spatial derivatives, viz. 
\begin{eqnarray}
\label{rtesph3}
\mu \frac{\partial I}{\partial r} 
+ \frac{1-\mu^2}{r} \frac{\partial I}{\partial\mu}
= -(\chi_0 - \mu \chi_1) I +
 (\eta_0^T + \mu\eta_1^T) +\nonumber\\
 (\eta_0^J + \mu\eta_1^J) J + 
(\eta_0^H + \mu\eta_1^H + \mu^2\eta_2^H) H
+ \mu\eta_1^K K\, ,
\end{eqnarray}
where
\begin{equation}
\chi_0 = \kappa_0 + \sigma_0 + \frac{1}{c \Delta t}\, ,
\end{equation} 
\begin{equation}
\chi_1 = (\widetilde\kappa_0 + \widetilde\sigma_0) w\, ,
\end{equation} 
\begin{equation}
\eta_0^T = \eta_0^{\rm th} + \frac{I^0}{c \Delta t}\, ,
\end{equation} 
\begin{equation}
\eta_1^T = 3w \widetilde\eta_0\, ,
\end{equation} 
\begin{equation}
\eta_0^J = \sigma_0\, ,
\end{equation} 
\begin{equation}
\eta_1^J = \sigma_0 w \left(2 - \delta
-\frac{\partial\ln\sigma_0}{\partial\ln\nu}
- \frac{\partial\ln J}{\partial \ln\nu}\right)\, ,
\end{equation} 
\begin{equation}
\eta_0^H = -\sigma_0 w
\left(1 + \delta - \frac{\partial\ln H}{\partial\ln\nu}\right)\, ,
\end{equation} 
\begin{equation}
\eta_1^H = \sigma_0 \delta\, ,
\end{equation} 
\begin{equation}
\eta_2^H = \sigma_0 \delta w \left(3 
-\frac{\partial\ln\sigma_0}{\partial\ln\nu} - 
\frac{\partial\ln H}{\partial\ln\nu}\right)\, ,
\end{equation} 
and
\begin{equation}
\eta_1^K = \sigma_0 \delta w 
\frac{\partial\ln K}{\partial\ln\nu}\, .
\end{equation} 
Here, we split the appropriate coefficients into several parts
depending upon which power of $\mu$ it is associated with; subscript 0 
corresponds to the $\mu$-independent part, subscript 1 to\
a linear term in $\mu$, etc. The superscripts $J, H, K$
refer to the parts of emission coefficients that contain moments
$J, H$, and $K$, and $T$ refers to the ``thermal'' emission coefficient.

In the tangent-ray approach, we consider transfer
along the ray specified by a constant impact parameter, $p$. 
The coordinate along $p$ is called $s$ (we do not use the usual notation $z$ to
avoid confusion with $z$-coordinate used in 2-D cylindrical
geometry), where
\begin{equation}
\label{rmu}
r = (p^2 + s^2)^{1/2}\, ,\quad \mu=\frac{s}{r}\, .
\end{equation}
Because the differential operator 
$\mu(\partial/\partial r) + r^{-1}(1-\mu^2)\partial/\partial\mu$
is identically $\partial/\partial s$,
one can integrate along straight lines because the characteristics
of eq. (\ref{rtesph3}) are straight lines.

Because of the symmetry of the problem we consider only positive $s$.
We denote the intensity propagating in the direction of increasing $s$ by
$I^+$ or $I(\mu)$, and that for decreasing $s$ by $I^-$ or $I(-\mu)$.
The transfer equation along the tangent ray in the positive direction
(increasing $s$) is given by
\begin{eqnarray}
\label{rtesph2p}
\frac{\partial I^+}{\partial s} 
= -(\chi_0 - \mu \chi_1) I^+ +
 (\eta_0^T + \mu\eta_1^T) +\nonumber\\
 (\eta_0^J + \mu\eta_1^J) J + 
(\eta_0^H + \mu\eta_1^H + \mu^2\eta_2^H) H
+ \mu\eta_1^K K\, ,
\end{eqnarray}
while for the negative direction (decreasing $s$) it reads
\begin{eqnarray}
\label{rtesph2m}
-\frac{\partial I^-}{\partial s} 
= -(\chi_0 + \mu \chi_1) I^- +
 (\eta_0^T - \mu\eta_1^T) +\nonumber\\
 (\eta_0^J - \mu\eta_1^J) J + 
(\eta_0^H - \mu\eta_1^H + \mu^2\eta_2^H) H
- \mu\eta_1^K K\, .
\end{eqnarray}
The transfer equation along the ray may be written in a compact form:
\begin{equation}
\label{rtepz}
\frac{1}{\chi^\pm(p,s)}\frac{d I^\pm(p,s)}{ds}= I^\pm(p,s) - S^\pm(p,s)\, ,
\end{equation}
where the corresponding total opacities $\chi^\pm$ and source
functions $S^\pm$
easily follow from the above expressions. The only complication
is in the form of the source function, which may be written as
\begin{equation}
\label{sfpz}
S^\pm(p,s) = S_0^\pm(p,s) + a^\pm(p,s) J(r)+ b^\pm(p,s) H(r)+ c^\pm(p,s)
K(r)
\, ,
\end{equation}
where $r= \sqrt{p^2 + s^2}$, and 
\begin{equation}
\chi^+(p,s) = \chi_0(r) - \mu\chi_1(r) = \chi_0(r)-\frac{s}{r}\chi_1(r)\, ,
\end{equation}
\begin{equation}
\chi^-(p,s) = \chi_0(r) + \mu\chi_1(r) = \chi_0(r)+\frac{s}{r}\chi_1(r)\, ,
\end{equation}
\begin{equation}
S_0^+(p,s) = \frac{\eta_0^T + \mu \eta_1^T}{\chi^+(p,s)} =
\frac{\eta_0^T(r) + (s/r)\eta_1^T}{\chi^+(p,s)}\, ,
\end{equation}
\begin{equation}
S_0^-(p,s) = \frac{\eta_0^T - \mu \eta_1^T}{\chi^-(p,s)} =
\frac{\eta_0^T(r) - (s/r)\eta_1^T}{\chi^-(p,s)}\, ,
\end{equation}
\begin{equation}
a^+(p,s) = \frac{\eta_0^J + \mu \eta_1^J}{\chi^+(p,s)} =
\frac{\eta_0^J(r) + (s/r)\eta_1^J}{\chi^+(p,s)}\, ,
\end{equation}
and
\begin{equation}
a^-(p,s) = \frac{\eta_0^J - \mu \eta_1^J}{\chi^-(p,s)} =
\frac{\eta_0^J(r) - (s/r)\eta_1^J}{\chi^-(p,s)}\, ,
\end{equation}
and analogously for the other quantities entering eq. (\ref{sfpz}).

We introduce the following discretization. The radius grid is defined
in terms of depth index $d=1,\ldots,NR$ which increases {\it inward} from
the surface: $r_1 = R > r_2 > \ldots > R_c$, where $R_c$ is the radius
of the ``core.'' The impact parameters are labeled by the same index
as the radii, that is the impact parameter for the $j$-th ray is
$p_j = r_j$. In addition, one introduces $NC$ core rays with
$0 \leq p_{NR+j} \leq R_c,  j=1,\ldots,NC$. The total number of
rays (impact parameters) is, thus, $NI=NR+NC$.

The moments $J,H$, and $K$, which are integrals over angles, may be
expressed as quadratures over the impact parameters, viz.
\begin{equation}
\label{jpz}
J(r_d) = \sum_{j=d}^{NI} w_{jd}\,
[I^+(p_j, s_d) + I^-(p_j,s_d)]\, ,
\end{equation}
\begin{equation}
\label{hpz}
H(r_d) =  \sum_{j=d}^{NI} w_{jd}\, \mu_{jd}\,
[I^+(p_j, s_d) - I^-(p_j,s_d)]\, ,
\end{equation}
and
\begin{equation}
\label{kpz}
K(r_d) = \sum_{j=d}^{NI} w_{jd}\, \mu_{jd}^2\,
[I^+(p_j, s_d) + I^-(p_j,s_d)]\, .
\end{equation}

In the source function (eq. \ref{sfpz}) the parameters $a^\pm$, $b^\pm$,
$c^\pm$ and $S_0^\pm$ are known functions of $r$, while the only
unknowns are the moments $J, H$, and $K$ which have to be solved for 
self-consistently with the transfer equation. 
Using the Feautrier scheme,  
one can in principle obtain an exact (non-iterative) solution, 
as in the case of the standard comoving-frame transfer equation 
in spherical geometry developed by Mihalas, Kunasz, \& Hummer (1975). 
However, in the present mixed-frame approach with anisotropic
scattering, the direct scheme is somewhat cumbersome.
Even if one can solve the problem directly using the Feautrier
formalism, it is nevertheless advantageous to use an iterative
scheme. We shall, thus, outline an iteration scheme in \S\ref{sect-ali}.

\subsection{ALI iteration scheme}
\label{sect-ali}

The transfer eq. (\ref{rtepz}), with the source function given by 
eq. (\ref{sfpz}), is solved by an application of an Accelerated Lambda
Iteration (ALI) scheme. The solution of eq. (\ref{rtepz}) can be written
formally as
\begin{equation}
\label{ilambda1}
I_{j,d}^\pm = \sum_{d^\prime=1}^j \Lambda_{j, d d^\prime}^\pm
S_{j,d^\prime}^\pm\, ,
\end{equation}
where $I_{j,d}^\pm \equiv I^\pm(p_j, s_d)$ and  
$S_{j,d}^\pm \equiv S^\pm(p_j, s_d)$.
In other words, the specific intensity is understood as the result 
of an action of a certain
operator (or matrix, when discretized), $\Lambda$, on the source
function. The iteration scheme adopted here is an application of the
Jacobi preconditioning scheme, for which we write (dropping the
superscripts $\pm$):
\begin{equation}
I_{j,d}^{\rm new} = \Lambda_{j,dd} S_{j,d}^{\rm new} +
\sum_{d^\prime=1}^j \Lambda_{j, d d^\prime} S_{j,d^\prime}^{\rm old}\, .
\end{equation}
In the usual astrophysical language, we use an approximate $\Lambda$
operator
given by the diagonal (local) part of the exact $\Lambda$.
Using this expression, we can express the moments as
\begin{equation}
J_d^{\rm new} = \frac{1}{2} \sum_{j=d}^{NI} w_{jd} \Lambda_{j,dd}^\pm
[S_{jd}^0 + a_{jd}^\pm J_d^{\rm new} + b_{jd}^\pm H_d^{\rm new} + 
c_{jd}^\pm K_d^{\rm new} ] + {\rm ``old"\ terms}\, ,
\end{equation}
\begin{equation}
H_d^{\rm new} = \frac{1}{2} \sum_{j=d}^{NI} w_{jd}\mu_{jd} \Lambda_{j,dd}^\pm
[S_{jd}^0 + a_{jd}^\pm J_d^{\rm new} + b_{jd}^\pm H_d^{\rm new} + 
c_{jd}^\pm K_d^{\rm new} ] + {\rm ``old"\ terms}\, ,
\end{equation}
and
\begin{equation}
K_d^{\rm new} = \frac{1}{2} \sum_{j=d}^{NI} w_{jd}\mu_{jd}^2 \Lambda_{j,dd}^\pm
[S_{jd}^0 + a_{jd}^\pm J_d^{\rm new} + b_{jd}^\pm H_d^{\rm new} + 
c_{jd}^\pm K_d^{\rm new} ] + {\rm ``old"\ terms}\, .
\end{equation}
Here, the superscript
$\pm$ indicates that we sum over both the $+$ and $-$ terms.
After some algebra
we obtain the set of three coupled equations for the new values of
the three moments at the given radius $r_d$:
\begin{equation}
\label{ali1}
\left(\begin{array}{ccc}
1-\Lambda_d^{JJ} & -\Lambda_d^{JH} &-\Lambda_d^{JK} \\
-\Lambda_d^{HJ} & 1-\Lambda_d^{HH} &-\Lambda_d^{HK} \\
-\Lambda_d^{KJ} & -\Lambda_d^{KH} & 1-\Lambda_d^{KK} 
\end{array}\right)
\cdot
\left(\begin{array}{c}
J_d^{\rm new} - J_d^{\rm old} \\
H_d^{\rm new} - H_d^{\rm old} \\
K_d^{\rm new} - K_d^{\rm old} 
\end{array}\right)
=
\left(\begin{array}{c}
J_d^{\rm FS} - J_d^{\rm old} \\
H_d^{\rm FS} - H_d^{\rm old} \\
K_d^{\rm FS} - K_d^{\rm old} 
\end{array}\right)\, ,
\end{equation}
where $J^{\rm FS}$ is given by eq. (\ref{jpz}) with the specific
intensity given by the ``old'' intensity $I_{jd}^{\rm old}$
(and analogously for $H$ and $K$), and the matrix elements are
given by
\begin{equation}
\label{lamjj}
\Lambda_d^{JJ} =  \sum_{j=d}^{NI} w_{jd} \,
(a_{j,d}^+ \Lambda_{j,dd}^+ + a_{j,d}^- \Lambda_{j,dd}^-)\, ,
\end{equation}
\begin{equation}
\Lambda_d^{JH} =  \sum_{j=d}^{NI} w_{jd}  \,
(b_{j,d}^+ \Lambda_{j,dd}^+ + b_{j,d}^- \Lambda_{j,dd}^-)\, ,
\end{equation}
\begin{equation}
\Lambda_d^{JK} = \sum_{j=d}^{NI} w_{jd} \,
(c_{j,d}^+ \Lambda_{j,dd}^+ + c_{j,d}^- \Lambda_{j,dd}^-)\, ,
\end{equation}
\begin{equation}
\Lambda_d^{HJ} = \sum_{j=d}^{NI} w_{jd} \, \mu_{j,d}   \,
(a_{j,d}^+ \Lambda_{j,dd}^+ - a_{j,d}^- \Lambda_{j,dd}^-) \, ,
\end{equation}
\begin{equation}
\Lambda_d^{HH} = \sum_{j=d}^{NI} w_{jd} \, \mu_{j,d} \, 
(b_{j,d}^+ \Lambda_{j,dd}^+ - b_{j,d}^- \Lambda_{j,dd}^-) \, ,
\end{equation}
\begin{equation}
\Lambda_d^{HK} =  \sum_{j=d}^{NI} w_{jd} \, \mu_{j,d} \,  
(c_{j,d}^+ \Lambda_{j,dd}^+ - c_{j,d}^- \Lambda_{j,dd}^-)\, ,
\end{equation}
and
\begin{equation}
\Lambda_d^{KJ} =  \sum_{j=d}^{NI} w_{jd} \, \mu_{j,d}^2 \, 
(a_{j,d}^+ \Lambda_{j,dd}^+ + a_{j,d}^- \Lambda_{j,dd}^-) \, ,
\end{equation}
\begin{equation}
\Lambda_d^{KH} =  \sum_{j=d}^{NI} w_{jd} \, \mu_{j,d}^2  \,
(b_{j,d}^+ \Lambda_{j,dd}^+ + b_{j,d}^- \Lambda_{j,dd}^-) \, ,
\end{equation}
\begin{equation}
\label{lamkk}
\Lambda_d^{KK} = \sum_{j=d}^{NI} w_{jd} \, \mu_{j,d}^2 \, 
(c_{j,d}^+ \Lambda_{j,dd}^+ + c_{j,d}^- \Lambda_{j,dd}^-)\, .
\end{equation}
To evaluate new values of the moments, one has to invert one simple
$3\times 3$ matrix per depth point. The individual matrix elements
$\Lambda_{j,dd}$ are evaluated during the formal solution step.

The iterations procedure is as follows:
\begin{itemize}
\item[(a)] For given moments $J^{(n)}, H^{(n)},K^{(n)}$
(and with a suitable initial estimate of
$J^{(0)}, H^{(0)},K^{(0)}$),
we perform a set of formal solutions for all impact
parameters $p$, so we obtain new specific intensities, which we
denote $I^\pm_{\rm FS}(p,s)$.
\item[(b)] We compute new values for the moments 
$J^{\rm FS}, H^{\rm FS}, K^{\rm FS}$ using eqs. (\ref{jpz}) - (\ref{kpz}),
with the specific intensity $I_{\rm FS}$.
\item[(c)] We solve eq. (\ref{ali1}), radius by radius, to
obtain new values for the three moments.
\item[(d)] As long as the new moments differ from the old moments,
we iterate steps (a) through (c) to convergence.
\end{itemize}

Since the ``acceleration'' (step c) is very simple, the problem is
effectively reduced to a set of formal solutions along the tangent
rays.
There are two main possibilities to solve the transfer equation along the
ray, using either the Feautrier method, or the Discontinuous Finite
Element (DFE) scheme.  One can also use a 1-D
short characteristics scheme; we have tested (\S\ref{sec-num-st})
all three methods for the present study.

We have found that one can use, without any deterioration of
the ALI iteration procedure, a simplified preconditioner (approximate
$\Lambda$ matrix) that is obtained by dropping the off-diagonal
terms of the $3\times 3$ matrix $\Lambda^i_{dd}$, namely by setting
\begin{equation}
\Lambda_d^{JH} = \Lambda_d^{HJ} = \Lambda_d^{JK} = \Lambda_d^{KJ} = 
\Lambda_d^{HK} = \Lambda_d^{KH} = 0\, ,
\end{equation}

\subsubsection{Tri-diagonal Operator}

A better preconditioner than that based on a local
approximation of the exact transport operator is obtained by considering 
not only the local components of the operator $\Lambda$,
but also terms corresponding to the nearest neighbors. In this
case we write for the new moments
\begin{eqnarray}
\label{tridia1}
J_d^{\rm new} = \frac{1}{2} \sum_{j=d}^{NI} w_{jd} \left[
\Lambda^\pm_{d,d-1} \left(S_{j,d-1}^0 + a_{j,d-1}^\pm J_{d-1}^{\rm new} +
b_{j,d-1}^\pm H_{d-1}^{\rm new} +c_{j,d-1}^\pm K_{d-1}^{\rm new}\right)
\right. +
\nonumber\\
\Lambda^\pm_{d,d} \left(S_{j,d}^0 + a_{j,d}^\pm J_{d}^{\rm new} +
b_{j,d}^\pm H_{d}^{\rm new} +c_{j,d}^\pm K_{d}^{\rm new}\right) +
\nonumber\\
\left.
\Lambda^\pm_{d,d+1} \left(S_{j,d+1}^0 + a_{j,d+1}^\pm J_{d+1}^{\rm new} +
b_{j,d+1}^\pm H_{d+1}^{\rm new} +c_{j,d+1}^\pm K_{d+1}^{\rm new}\right)
\right]\ + {\rm ``old\ terms''}\, ,
\end{eqnarray}
and analogously for $H_d^{\rm new}$ and $K_d^{\rm new}$.
Generalizing the procedure described in \S\ref{sect-ali}, we obtain
a block-tridiagonal system (in the physical space) for the components 
of the moments; each block is a $3\times 3$ matrix that couples
all three moments. The diagonal block is the same as the
approximate $\Lambda$ matrix considered in \S\ref{sect-ali}, and
the off-diagonal blocks easily follow from eq. (\ref{tridia1}).

However, following our finding that the off-diagonal terms
corresponding to the coupling of moments can be dropped, we end
up with three separate tridiagonal systems in the physical space
for the three moments $J$, $H$, and $K$.

\subsection{Augmentation of the ALI scheme by GMRES}

The ALI scheme outlined above can be significantly augmented by 
an application of a suitable Krylov subspace method, for instance
the GMRES (Generalized Minimum Residual) method. There are a number
of variants of the scheme; we have implemented the following one.
Let us define the vector $x$ composed of triads $J,H,K$ at all
radii. Its dimension is, thus, $3\times NR$. The general problem can
be formulated as a linear system $A x = b$, where matrix $A$
is a block matrix of $NR\times NR$ blocks, each block being a
$3\times 3$ matrix, analogous to the matrix of eq. (\ref{ali1})
(containing also the off-diagonal elements of the $\Lambda$ matrix).
We define the ``preconditioned residuum'' at the $n$-th iteration,
$R^{(n)}$, as a vector
composed of $J^{(n+1)} - J^{(n)}$, $H^{(n+1)} - H^{(n)}$, and
$K^{(n+1)} - K^{(n)}$ for all radii (i.e., a collection of
the solution vectors of eq. (\ref{ali1}) for all $d$). The GMRES scheme
consists of consecutively finding ``search vectors,'' $P^{(i)}$,
whose products with matrix $A$ are made orthogonal to the subspace
spanned by the previously constructed search vectors, and
which give the new estimate of the solution. 

The adopted algorithm is as follows: We start with $x^{(0)}$ and
set $P^{(0)} = R^{(0)}$. Then, for each $i = 0,1,\ldots,$
we compute
\begin{equation}
\label{gmr55}
\alpha_i = \frac{\left(A P^{(i)}, R^{(i)}\right)} 
                {\left(A P^{(i)}, A P^{(i)}\right)} \, ,
\end{equation}
\begin{equation}
\label{gmr56}
x^{(i+1)} =  x^{(i)} + \alpha_i P^{(i)} \, ,
\end{equation}
\begin{equation}
\label{gmr57}
R^{(i+1)} =  R^{(i)} - \alpha_i\, A P^{(i)} \, ,
\end{equation}
\begin{equation}
\label{gmr58}
\beta_{ij} = -\frac{\left(A P^{(j)}, A R^{(i+1)}\right)} 
                  {\left(A P^{(j)}, A P^{(j)}\right)} \, ,
              \quad\quad j=0,1,\ldots,i\, ,
\end{equation}
and
\begin{equation}
\label{gmr59}
P^{(i+1)} =  R^{(i+1)} + \sum_{j=0}^i\beta_{ij} P^{(j)} \, .
\end{equation}
In practice, one can keep adding newer and newer search vectors.
However, this may be cumbersome or too memory consuming.
If so, one may actually stop and restart the orthogonalization
process, or limit the orthogonalization to the $k$ most recent search
vectors. In this case, the summation in eq. (\ref{gmr58})
is replaced by $\sum_{j=i-k+1}^i\beta_{ij}  P^{(j)}$. Such a method is
sometimes called ORTHOMIN(k) (Klein et al., 1989).

An important point is that one can (and should!) accomplish the procedure
defined by eqs. (\ref{gmr55}) - (\ref{gmr59}) without performing
explicit multiplications with matrix $A$, which in fact is 
never even assembled explicitly. It turns out that one can write
\begin{equation}
\label{gmr60}
A R^{(i+1)} = R^{(i)} - R^{(i+1)}\, ,
\end{equation}
and
\begin{equation}
\label{gmr61}
A P^{(i+1)} =  A R^{(i+1)} + \sum_{j=0}^i\beta_{ij} A P^{(j)} \, ,
\end{equation}
so indeed explicit multiplications with matrix $A$ are not needed.

\subsection{Discontinuous Finite Element (DFE) Scheme}

In 1-D, our implementation of the DFE scheme is a straightforward application of a
method
developed by Castor, Dykema, \& Klein (1992), where the reader is
referred for additional details and derivations. We present
here the final formulae for the specific intensities, together with
a description of how the elements of the approximate $\Lambda$
operator are evaluated in the context of the DFE approach.

In the direction of propagation we have recurrence relations for
the finite elements:
\begin{equation}
I^-_{d+1} = a_{d+1/2} \left( 2\, I^-_d + \Delta\tau_{d+1/2} S_d +
b_{d+1/2} S_{d+1} \right)\, ,
\end{equation}
and
\begin{equation}
I^+_d = a_{d+1/2} \left( c_{d+1/2} I^-_d + b_{d+1/2} S_d -
\Delta\tau_{d+1/2} S_{d+1} \right)\, ,
\end{equation}
where
\begin{equation}
a_{d+1/2} = \left(\Delta\tau_{d+1/2}^2 + 2\Delta\tau_{d+1/2} + 2\right)^{-1}\, ,
\end{equation}
\begin{equation}
b_{d+1/2} = \Delta\tau_{d+1/2} (\Delta\tau_{d+1/2} + 1)\, ,
\end{equation}
and
\begin{equation}
c_{d+1/2} = 2 (\Delta\tau_{d+1/2} + 1)\, .
\end{equation}
The specific intensity at point $d$ is given as a linear combination of
the discontinuous intensities,
\begin{equation}
\label{dfe-lin}
I_d = \frac{I^-_d \Delta\tau_{d+1/2} + I^+_d \Delta\tau_{d-1/2}}
{\Delta\tau_{d+1/2} + \Delta\tau_{d-1/2}}\, .
\end{equation}
It can be shown (Castor et al. 1992) that this choice of linear
combination of the discontinuous intensities makes the scheme second-order
accurate.

As follows from the above expressions, the diagonal and first
off-diagonal elements of the transport matrix $\Lambda$ are given by
linear combinations as in eq. (\ref{dfe-lin}):
\begin{equation}
\Lambda_{d,j} = \frac{\Lambda^-_{d, j} \Delta\tau_{d+1/2} + 
\Lambda^+_{d, j} \Delta\tau_{d-1/2}}
{\Delta\tau_{d+1/2} + \Delta\tau_{d-1/2}}\, ,\quad j=d-1, d, d+1\, ,
\end{equation}
where
%where $j=d-1, d, d+1$, and 
%
\begin{equation}
\Lambda^-_{d+1,d+1} = a_{d+1/2} b_{d+1/2}\, ,
\end{equation}
\begin{equation}
\Lambda^{-}_{d+1,d} = a_{d+1/2}\left( \Delta\tau_{d+1/2}
+ 2 \Lambda^{-}_{d,d} \right)\, ,
\end{equation}
\begin{equation}
\Lambda^{+}_{d,d} = a_{d+1/2} (b_{d+1/2} + c_{d+1/2} \Lambda^{-}_{dd} )\, ,
\end{equation}
and
\begin{equation}
\Lambda^{+}_{d,d+1} = - \Delta\tau_{d+1/2} a_{d+1/2}\, .
\end{equation}

\subsection{Feautrier scheme}

While using the short-characteristics or the DFE schemes represents
a straightforward application of these methods (for a review, see, e.g., 
Hubeny 2003), an application of the Feautrier scheme requires a 
generalization of the Mihalas, Kunasz, \& Hummer (1975) formalism, 
which we describe in this section.
 
To use the Feautrier scheme to solve the transfer equation,
we introduce the usual symmetric (mean-intensity-like) and antisymmetric
(flux-like) Feautrier variables
\begin{equation}
\label{udef}
%U(\mu) = \frac{1}{2}\left[ I(\mu) + I(-\mu)\right]\, ,
U = \frac{1}{2}\left(I^+ + I^-\right)\, ,
\end{equation}
and
\begin{equation}
\label{vdef}
%V(\mu) = \frac{1}{2}\left[ I(\mu) - I(-\mu)\right]\, ,
V = \frac{1}{2}\left(I^+ - I^-\right)\, .
\end{equation}
Eqs. (\ref{rtesph2p}) and (\ref{rtesph2m}) can then
be written as a set of two differential equations 
for $U$ and $V$:
\begin{equation}
\label{feauv}
\frac{\partial V}{\partial s} =  - \chi_0 U + \mu \chi_1 V +
\eta_0^{T+} + \eta_0^J J +\left( \eta_0^H +\mu^2\eta_2^H \right) H\, ,
\end{equation}
and
\begin{equation}
\label{feauu}
\frac{\partial U}{\partial s} =  - \chi_0 V + \mu\chi_1 U\ +\eta_0^{T-} +
\mu\left(\eta_1^T + \eta_1^J J +\eta_1^H H + \eta_1^K K\right)\, ,
\end{equation}
where
\begin{equation}
\eta_0^{T+} = \eta_0^T + 
\frac {1}{2 c\Delta t}\left[ I^0(\mu) + I^0(-\mu) \right]\, ,
\end{equation}
and
\begin{equation}
\label{etatmi}
\eta_0^{T-} = 
\frac {1}{2 c\Delta t}\left[ I^0(\mu) - I^0(-\mu) \right]\, .
\end{equation}
Here, we take into account the fact that $\eta_0^T$ depends on $\mu$
only through the intensity at the previous time step, $I^0(\mu)$.
The symmetric and antisymmetric averages are, thus, denoted as
$\eta_0^{T+}$ and $\eta_0^{T-}$, respectively.

Furthermore, we introduce a modified optical depth along the tangent ray,
\begin{equation}
d\tau = -\chi_0 ds\, ,
\end{equation}
and the ``source terms'' $S^+$ and $S^-$, where
\begin{equation}
\label{splus}
S^+ = S_0^{T+} + S_0^J J + (S_0 + \mu^2 S_2) H\, ,
\end{equation}
\begin{equation}
\label{sminus}
S^- = S_0^{T-} + \mu(S_1^T + S_1^J J + S_1^H H + S_1^K K)\, ,
\end{equation}
and 
\begin{equation}
S_i^X = \frac{\eta_i^X}{\chi_0}\, ,\quad {\rm for}\quad i=0,1,2\quad 
{\rm and} \quad X=T,J,H,K\, .
\end{equation}
The equations for $U$ and $V$ can then be written simply as
\begin{equation}
\label{feauv2}
\frac{\partial V}{\partial\tau} = U - \alpha\, V - S^+\, ,
\end{equation}
and
\begin{equation}
\label{feauu2}
\frac{\partial U}{\partial\tau} = V - \alpha\, U - S^-\, ,
\end{equation}
where
\begin{equation}
\alpha = \mu\,\frac{\chi_1}{\chi_0}\, .
\end{equation}

To reflect the mean-energy character of $U$ and the flux-like
character of $V$, we stagger the $U$ and $V$ meshes by half a 
zone. That is, $U$ is taken with integer indices $d$, while
$V$ is taken with half-integer indices $d \pm 1/2$. 
The discretized eqs. (\ref{feauv2}) and (\ref{feauu2}) read
\begin{equation}
\label{feauv3}
\frac{V_{d+1/2} - V_{d-1/2}}{\Delta\tau_d} =  U_d -
\frac{\alpha_d}{2} ( V_{d+1/2} + V_{d-1/2}) - S^+_d\, ,
\end{equation}
and
\begin{equation}
\label{feauu3}
\frac{U_{d+1} - U_{d}}{\Delta\tau_{d+1/2}} = V_{d+1/2} -
\frac{\alpha_{d+1/2}}{2} (U_{d+1} + U_{d}) - S^-_{d+1/2}\, ,
\end{equation}
where
\begin{equation}
\Delta\tau_{d+1/2} = \frac{1}{2} (s_d - s_{d+1})
[(\chi_0)_{d+1} + (\chi_0)_d]\, ,
\end{equation}
\begin{equation}
\Delta\tau_{d} = \frac{1}{2}(\Delta\tau_{d+1/2} + \Delta\tau_{d-1/2})\, ,
\end{equation}
and
\begin{equation}
\label{mui}
\mu_d= \frac{s_d}{\sqrt{p^2 + s_d^2}}\, .
\end{equation}
The half-integer $\mu_{d+1/2}$ is given by an equation analogous
to eq. (\ref{mui}), where the corresponding radial points are
given by
\begin{equation}
r_{d+1/2}^3 = \frac{1}{2}(r_{d+1}^3 + r_d^3)\, .
\end{equation}
That is, the cell center is defined in such a way that
the volume at $r_{d+1/2}$ is half of the volume
between $r_d$ and $r_{d+1}$.

Discretized equations may be written in a more compact way by
introducing
\begin{equation}
b_d^{\pm} = \frac{1}{\Delta\tau_d} \pm \frac{\alpha_d}{2} =
\frac{1}{\Delta\tau_d} \pm \frac{\mu_d(\chi_1)_d}
{2(\chi_0)_d}\, ,
\end{equation}
where $d$ can assume integer or half-integer values.
Equation (\ref{feauu3}) can then be solved for $V_{d+1/2}$ to read
\begin{equation}
V_{d+1/2} = U_{d+1} b_{d+1/2}^{+} - U_d b_{d+1/2}^{-} + S_{d+1/2}^-\, .
\end{equation}
Analogously for $V_{d-1/2}$, we have
\begin{equation}
V_{d-1/2} = U_d b_{d-1/2}^{+} - U_{d-1} b_{d-1/2}^{-} + S_{d-1/2}^-\, .
\end{equation}
Using these equations, we can eliminate $V_{d\pm 1/2}$ from
eq. (\ref{feauv3}) to obtain
\begin{eqnarray}
\label{fe3a}
-U_{d-1} b_{d-1/2}^{-} b_d^{-} 
+ U_d \left(1 + b_{d-1/2}^{+} b_d^{-} + b_{d+1/2}^{-} b_d^{+}\right)
- U_{d+1} b_{d+1/2}^{+} b_d^{+} = \nonumber\\
S_d^+ + b_d^{+} S_{d+1/2}^- - b_d^{-} S_{d-1/2}^- \, .
\end{eqnarray}

The inner and outer boundary conditions can be handled in straightforward
fashion.
For the inner boundary condition, $d=D$, we have
$I^+=I^-$ and $\mu=0$. We, thus, have $V=0$ and $S^- = 0$,
the latter equality following from eqs. (\ref{sminus}) and (\ref{etatmi}).
The two Feautrier equations are written
\begin{equation}
\label{lbca}
\left(\frac{\partial V}{\partial\tau}\right)_D = U_D - S^+_D\, ,
\end{equation}
\begin{equation}
\label{lbcb}
\left(\frac{\partial U}{\partial\tau}\right)_D = V_D 
- \frac{\partial S^-}{\partial\tau}\, ,
%= - \frac{\partial S^-}{\partial\tau}\, ,
\end{equation}
and, thus,
\begin{equation}
\label{lbcc}
\left(\frac{\partial^2 U}{\partial\tau^2}\right)_D = U_D -  S^+_D
- \frac{\partial S^-}{\partial\tau}\, .
\end{equation}
The 2nd-order form of the boundary condition follows from a
Taylor expansion of $U_{D-1}$ around $D$,
\begin{equation}
U_{D-1} = U_D - 
\Delta\tau_{D-1/2} \left(\frac{\partial U}{\partial\tau}\right)_D +
\frac{1}{2}
\Delta\tau_{D-1/2}^2 \left(\frac{\partial^2 U}{\partial\tau^2}\right)_D 
\, .
\end{equation}
Using eqs. (\ref{lbca}) - (\ref{lbcc}), and expressing
$\partial S^-/\partial\tau$ as a difference, we obtain
\begin{equation}
\label{fe3b}
U_D \left( 1 + \frac{2}{\Delta\tau_{D-1/2}^2} \right) - 
U_{D-1}\frac{2}{\Delta\tau_{D-1/2}^2} = 
S_D^+  + \left(\frac{2}{\Delta\tau_{D-1/2}^2} - 
\frac{1}{\Delta\tau_{D-1/2}} \right) S_{D-1}^-\, .
\end{equation}

For the outer boundary condition at $d=1$,
we take $I^- = 0$, and, thus, $U_1 = V_1$. In order to write
the 2nd-order form of the boundary condition, we first
write a general 2nd-order equation that is derived from eqs. 
(\ref{feauv2}) and (\ref{feauu2}):
\begin{equation}
\label{d2u}
\frac{\partial^2 U}{\partial\tau^2} = U \left(1 + \alpha^2 -
\frac{\partial \alpha}{\partial\tau} \right) - 2\alpha V -
S^{+} +\alpha S^{-} - \frac{\partial S^-}{\partial\tau} \, .
\end{equation}
The 2nd-order form of the boundary condition follows from
the Taylor expansion of $U$ around $d=1$,
%
%\begin{eqnarray}
\begin{equation}
\label{lbc2}
U_2=U_1+\Delta\tau_{3/2} \left.\frac{\partial U}{\partial\tau}\right|_1 +
\frac{1}{2}\Delta\tau_{3/2}^2 
\left.\frac{\partial^2 U}{\partial\tau^2}\right|_1  \, .
\end{equation}
Substituting eqs. (\ref{feauv2}) and (\ref{d2u}) into eq. (\ref{lbc2}),
we obtain
\begin{eqnarray}
U_1 \left[(1-\alpha_1)^2 - \frac{\partial \alpha}{\partial\tau}
-\frac{2(1-\alpha_1)}{\Delta\tau_{3/2}} + 
\frac{2}{\Delta\tau_{3/2}^2} \right]
- U_2\, \frac{2}{\Delta\tau_{3/2}^2} = \nonumber\\
S_1^{+} + \left( \frac{2}{\Delta\tau_{3/2}} - \alpha_1\right) S_1^{-}
+ \frac {\partial S^{-}}{\partial\tau} \, .
\end{eqnarray}
Finally, expanding the derivatives of $\alpha$ and $S^-$ we obtain
\begin{eqnarray}
\label{fe3c}
U_1 \left[(1-\alpha_1)^2 - \frac{\alpha_1 + \alpha_2}{\Delta\tau_{3/2}} +
\frac{2}{\Delta\tau_{3/2}} + \frac{2}{\Delta\tau_{3/2}^2} \right]
- U_2\, \frac{2}{\Delta\tau_{3/2}^2} =\nonumber\\
S_1^{+} + \frac{S_1^- + S_2^-}{\Delta\tau_{3/2}}\, -\, \alpha_1 S_1^-\, .
\end{eqnarray}
Equations (\ref{fe3a}), (\ref{fe3b}), and (\ref{fe3c}) form a tridiagonal
system that is solved by the standard elimination method.

\medskip

An alternative, and in fact more accurate, way of formulating
the Feautrier scheme is by introducing the integration factor, $q$, 
defined by
\begin{equation}
\frac{1}{q}\, \frac{\partial q}{\partial\tau} = \alpha\, ,
\end{equation}
with which the Feautrier eqs. (\ref{feauv2}) and (\ref{feauu2})
are rewritten as
\begin{equation}
\frac{1}{q}\, \frac{\partial(q V)}{\partial\tau} = U - S^+\, ,
\end{equation}
and 
\begin{equation}
\frac{1}{q}\, \frac{\partial(q U)}{\partial\tau} = V - S^-\, .
\end{equation}
Using these equations, we can eliminate $V$, to end up with
a single second-order equation for $U$:
\begin{equation}
\frac{\partial^2 (qU)}{\partial\tau^2} = qU - qS^+ - \frac{d(qS^-)}{d\tau}\, .
\end{equation}
This equation is discretized in the standard way, namely for $d=2,\ldots,D-1$
we have
\begin{equation}
-\frac{q_{d-1}U_{d-1}}{\Delta\tau_d \Delta\tau_{d-1/2}}
+ \frac{q_d U_d}{\Delta\tau_d}\, 
  \left(\frac{1}{\Delta\tau_{d-1/2}} + \frac{1}{\Delta\tau_{d+1/2}} \right)
-\frac{q_{d+1} U_{d+1}}{\Delta\tau_d \Delta\tau_{d+1/2}}
= q_d S^+_d + \frac{q_{d+1} S^-_{d+1} - q_{d-1} S^-_{d-1}}{2 \Delta\tau_d}\, ,
\end{equation}
where we represented the derivative $d(qS^-)/d\tau$ by a centered difference.

The discretized boundary conditions are obtained similarly as above, and
are given by
\begin{equation}
q_1 U_1 \left(1 + \frac{2}{\Delta\tau_{3/2}} + \frac{2}{\Delta\tau_{3/2}^2}\right) -
q_2 U_2 \frac{2}{\Delta\tau_{3/2}^2} =
q_1 S^+_1 + \frac{q_2U_2 + q_1U_1}{\Delta_{3/2}}\, ,
\end{equation}
and
\begin{equation}
q_D U_D \left(1 + \frac{2}{\Delta\tau_{D-1/2}^2}\right) -
q_{D-1} U_{D-1} \frac{2}{\Delta\tau_{D-1/2}^2} =
q_D S^+_D - \frac{q_{D-1} U_{D-1}}{\Delta_{D-1/2}}\, .
\end{equation}
%

%------------------------------------------------------------------

\subsection{Moment equations}
\label{moment2}

At first sight it may seem unnecessary to deal with the moment
equations because all the necessary information about the radiation
field is provied by the specific intensity, whose evaluation was
described in the previous parts of \S\ref{sec-sph}. However, there are 
several reasons for considering the moment equations:

i) The hydrodynamical equations, and in particular the energy and momentum
balance equations, and the electron fraction equation (for core-collapse supernovae),
contain only moments
of the specific intensity, not the intensity itself. It is, thus, natural
to work in terms of the radiation moments.

ii) It is actually more consistent to work in terms of moments, since in
the moment equations the angular integrations are performed analytically,
while when using the specific intensities the radiation moments are
evaluated by a numerical integration. Generally, these two ways give somewhat
different results, and one should carefully assure the consistency between
the radiation moments used in the formal solution of the transfer equation,
and those used in the material equations.

iii) There is a practical aspect. Solving moment equations is considerably
faster than solving the full angle-dependent transfer equation. In fact,
the full angle-dependent transfer solver may be viewed as a tool to provide 
just the Eddington factor, while the moments are obtained by a numerical
solution of the moment equation. In many cases the Eddington factor changes
only slowly with time, and therefore one does not actually have to update the Eddington
factor in each timestep; instead the Eddington factor may be held fixed for
several timesteps. This may obviously lead to a significant reduction of the
computer time required. We shall return to this point in \S\ref{sec-num-imp}.

The moment equations in conservative form are written as
\begin{equation}
\frac{1}{c}\frac{\partial J}{\partial t} + \frac{1}{r^2} \frac{\partial(r^2
H)}{\partial r} =
\eta_0 - \kappa_0 J + \Xi H \, ,
\end{equation}
and
\begin{equation}
\frac{1}{c}\frac{\partial H}{\partial t} + 
\frac{1}{r^2} \frac{\partial(r^2 f J)}{\partial r} -\frac{1-f}{r}\, J =
-(\kappa_0 +\sigma_{\rm tr}) H + w \widetilde\eta_0 + \xi J\, ,
\end{equation}
where
\begin{equation}
\Xi = \left[\widetilde\kappa_0 + \sigma_{\rm tr}\left(
\frac{\partial\ln\sigma_0}{\partial\ln\nu}  +
\frac{\partial\ln H}{\partial\ln\nu}\right)\right] w\, ,
\end{equation}
and
\begin{equation}
\label{sr2}
\xi = \left[\frac{\sigma_0}{3} \left(2 - \delta -
\frac{\partial\ln\sigma_0}{\partial\ln\nu}  -
\frac{\partial\ln J}{\partial\ln\nu} 
\right) +
f \widetilde\kappa_0 + 
 f \sigma_0 \left(
1 + \frac{\partial\ln\sigma_0}{\partial\ln\nu} 
+ \frac{\delta}{3} \frac{\partial\ln K}{\partial\ln\nu} \right)
\right] w\, .
\end{equation}
We rewrite these equations in a more compact and useful form, again
expressing the time derivative through backward time differencing:
\begin{equation}
\label{dhdt}
\frac{d h}{d\tau} = k j - \beta h - S_J\, ,
\end{equation}
and
\begin{equation}
\label{djdt}
\frac{d j}{d\tau} = h - \alpha j - S_H\, ,
\end{equation}
where the modified moments are given by
\begin{equation}
j = r^2 f J\, , 
\end{equation}
and
\begin{equation}
h = r^2 H\, ,
\end{equation}
and where the auxiliary quantities are given by
\begin{equation}
S^J = \frac{\eta_J}{\kappa_H}\, ,\quad S^H = \frac{\eta_H}{\kappa_H}\, ,
\end{equation}
\begin{equation}
\eta_J = r^2 \eta_0^{\rm th} + \frac{r^2 J_0}{c\Delta t}\, , \quad
\eta_H = r^2 w \widetilde\eta_0 + \frac{r^2 H_0}{c\Delta t}\, ,
\end{equation}
\begin{equation}
\kappa_J = \left(\kappa_0 + \frac{1}{c\Delta t}\right) \frac{1}{f}\, ,\quad
\kappa_H = \kappa_0 + \sigma_{\rm tr} + \frac{1}{c\Delta t}\, ,
\end{equation}
\begin{equation}
\alpha = \frac{\xi + (1-f)/r}{f \kappa_H}\, ,\quad
\beta = \frac{\Xi}{\kappa_H}\, ,\quad
k = \frac{\kappa_J}{\kappa_H}\, .
\end{equation}
and
\begin{equation}
d\tau = -\kappa_H dr\, .
\end{equation}
Discretization of eqs. (\ref{dhdt}) and (\ref{djdt}) yields
\begin{equation}
\label{dhdt2}
\frac{h_{d+1/2} - h_{d-1/2}}{\Delta\tau_d} =
k_d j_d - \frac{\beta_d}{2} (h_{d+1/2} + h_{d-1/2}) - S^J_d\, ,
\end{equation}
and
\begin{equation}
\label{djdt2}
\frac{j_{d+1} - j_d}{\Delta\tau_{d+1/2}} =
h_{d+1/2} - \frac{\alpha_{d+1/2}}{2} (j_{d+1} + j_d) - S^H_{d+1/2}\, .
\end{equation}
We eliminate $h$ from eq. (\ref{djdt2}) to obtain 
\begin{equation}
\label{hdp12}
h_{d+1/2} = j_{d+1} \gamma_{d+1/2}^{+} - j_d \gamma_{d+1/2}^{-} +
S_H^{d+1/2}\, .
\end{equation}
Performing the same operation for $h_{d-1/2}$ and substituting into 
eq. (\ref{dhdt2}), we obtain 
\begin{eqnarray}
\label{jmomtri}
- j_{d-1}\, c_d^- \gamma_{d-1/2}^-
+ j_d \left( k_d + c_d^+ \gamma_{d+1/2}^- + c_d^- \gamma_{d-1/2}^+ \right)
- j_{d+1}\, c_d^+ \gamma_{d+1/2}^+ = \nonumber\\
S_J + c_d^+ S_H^{d+1/2} - c_d^- S_H^{d-1/2}\, ,
\end{eqnarray}
where
\begin{equation}
c_d^{\pm} = \frac{1}{\Delta\tau_d} \pm \frac{\beta_d}{2}\, ,
\end{equation} 
and
\begin{equation}
\label{gam1}
\gamma_d^{\pm} = \frac{1}{\Delta\tau_d} \pm \frac{\alpha_d}{2}\, ,
\end{equation}
and where we understand that the depth indices in eq. (\ref{gam1})
have half-integer values.

The boundary conditions are expressed using the second-order form
of the moment equation. Eliminating $h$ from eq. (\ref{djdt}) and using
eq. (\ref{dhdt}), we obtain
\begin{equation}
\label{jsec}
\frac{d^2 j}{d\tau^2} = 
\left( k - \frac{d\alpha}{d\tau} + \alpha^2 \right) j -
(\alpha+\beta) h - 
\left( S_J - \alpha S_H + \frac{d S_H}{d\tau}\right)\, .
\end{equation}
The outer boundary condition (at $d=1)$ follows from expanding $j$ in a
Taylor series:
\begin{equation}
j_2 = j_1 + \Delta\tau \left.\frac{dj}{d\tau}\right|_1 +
\frac{1}{2} \Delta\tau^2 \left.\frac{d^2j}{d\tau^2}\right|_1\, .
\end{equation}
Substituting for derivatives from eqs. (\ref{dhdt}) and (\ref{jsec}),
we obtain
\begin{eqnarray}
\label{umbc}
j_1 \left[\frac{2}{\Delta\tau_{3/2}^2} + f_H
\left(\frac{2}{\Delta\tau_{3/2}}-\beta_1 \right) -
\frac{\alpha_1 + \alpha_2}{\Delta\tau_{3/2}} + k_1 + \alpha_1^2 \right]
- j_2 \frac{2}{\Delta\tau_{3/2}^2} = \nonumber\\
S_J^1 - \alpha_1 S_H^1 + \frac{S_H^1 + S_H^2}{\Delta\tau_{3/2}}\, ,
\end{eqnarray}
where we have also used the 1-st order form of the derivatives of $\alpha$ and
$S_H$, [for instance, 
$(d\alpha/d\tau)_1 = (\alpha_2 - \alpha_1)/\Delta\tau_{3/2}$].
Finally,
\begin{equation}
f_H \equiv \frac{h_1}{j_1} = \frac{H_1}{f_1 J_1}\, ,
\end{equation}
is the ``flux Eddington factor'' which is evaluated in the full
(angle-dependent) transfer solution.

The inner boundary condition ($d=D$) is analogous. There, we use
eq. (\ref{jsec}) and the symmetry condition $h_D = 0$ to obtain
\begin{eqnarray}
\label{lmbc}
j_D \left[\frac{2}{\Delta\tau_{D-1/2}^2} + 
\frac{\alpha_D + \alpha_{D-1}}{\Delta\tau_{D-1/2}} + k_D + \alpha_D^2
\right]
- j_{D-1} \frac{2}{\Delta\tau_{D-1/2}^2} = \nonumber\\
S_J^D - \alpha_D S_H^D - \frac{S_H^{D}+S_H^{D-1}}{\Delta\tau_{D-1/2}}\, .
\end{eqnarray}
Equation (\ref{jmomtri}), together with boundary eqs. (\ref{umbc}) and
(\ref{lmbc}), form a tridiagonal system which is solved by a standard
Gaussian elimination (also called a forward-backward sweep).

\subsubsection{Matrix representation}

To prepare for the formalism used in \S\ref{sec-num}, we note
that we can write down the tridiagonal system for $j$ defined by eqs. 
(\ref{jmomtri}), (\ref{umbc}), and (\ref{lmbc}) in a matrix form
\begin{equation}
T \cdot {\bf j} = {\bf S}^J + U \cdot {\bf S}^H\, ,
\end{equation}
where ${\bf j} = (j_1, j_2, \ldots, j_D)^T$ is a column vector of moments $j_d$
at all depth points.
There is an analogous expression for the source function vectors $S^J$ and
$S^H$.
Matrices $T$ and $U$ are tridiagonal matrices. The matrix elements of $T$ are
given by 
\begin{equation}
T_{d,d-1} = -c_d^-\gamma_{d-1/2}^-\, ,\quad
T_{d,d} = k_d + c_d^+\gamma_{d+1/2}^- + c_d^-\gamma_{d-1/2}^+\, ,\quad
T_{d,d+1} = -c_d^+\gamma_{d+1/2}^+\, ,
\end{equation}
for $2 \leq d \leq D-1$. Expressions for $d=1$ and $d=D$ easily follow from eqs.
(\ref{umbc}) and (\ref{lmbc}). The matrix elements of $U$ are given by
\begin{equation}
U_{d,d-1} = -\frac{c_d^-}{2}\, ,\quad
U_{d,d} = \frac{c_d^+}{2} -\frac{c_d^-}{2}\, ,\quad
U_{d,d+1} = \frac{c_d^+}{2}\, ,\quad 2 \leq d \leq D-1 ,
\end{equation}
\begin{equation}
U_{11} = \frac{1}{\Delta\tau_{3/2}} - \alpha_1\, ,\quad
U_{12} = \frac{1}{\Delta\tau_{3/2}}\, ,
\end{equation}
and
\begin{equation}
U_{DD} = -\frac{1}{\Delta\tau_{D-1/2}} - \alpha_D\, ,\quad
U_{D,D-1} = -\frac{1}{\Delta\tau_{1-D/2}}\, .
\end{equation}
One can formally write a solution for $j$ as
\begin{equation}
\label{jl01}
{\bf j} = T^{-1} \cdot {\bf S}^J + T^{-1} U \cdot {\bf S}^H \equiv
\Lambda_0  \cdot {\bf S}^J + \Lambda_1  \cdot {\bf S}^H\, ,
\end{equation}
with new matrices $\Lambda_0 = T^{-1}$ and
$\Lambda_1 = T^{-1} U$.

We can write a similar equation for vector ${\bf h} = (h_1, h_2, \ldots, h_D)^T$
using eqs. (\ref{hdp12}), (\ref{jl01}), and expressions
$h_d = (h_{d-1/2} + h_{d+1/2})/2$, $h_1 = f_H j_1$, and $h_d = 0$:
\begin{equation}
\label{hl012}
{\bf h} = V \cdot {\bf j} + W \cdot {\bf S}^H =
(V \Lambda_0) \cdot {\bf S}^J + (V \Lambda_1 + W) \cdot {\bf S}^H\, ,
\end{equation}
where $V$ and $W$ are tridiagonal matrices whose elements are given by
\begin{equation}
V_{d,d-1} = - \frac{\gamma_{d-1/2}^-}{2}\, , \quad
V_{dd} = \frac{\gamma_{d-1/2}^+}{2} - \frac{\gamma_{d+1/2}^-}{2}\, , \quad
V_{d,d+1} = \frac{\gamma_{d+1/2}^+}{2}\, ,
\quad 2 \leq d \leq D-1\, ,
\end{equation}
\begin{equation}
V_{11} = f_H\, ,\quad V_{12} = V_{DD} = V_{D,D-1} = 0\, .
\end{equation}
and
\begin{equation}
W_{dd} = \frac{1}{2}\, \quad W_{d,d\pm 1} = \frac{1}{4}\, ,\quad d\leq D-1\, ,
\end{equation}
\begin{equation}
W_{D,D-1} = W_{DD} = 0\, .
\end{equation}

We write the expression for ${\bf h}$ as
\begin{equation}
\label{hl23}
{\bf h} = \Lambda_2 \cdot {\bf S}^J + \Lambda_3 \cdot {\bf S}^H\, ,
\end{equation}
where $\Lambda_2 =  V \Lambda_0$ and $\Lambda_3 = V \Lambda_1 + W$.

Finally, the true moments $J$ and $H$ are given as
\begin{equation}
\label{jl01t}
{\bf J} = {\rm diag}(1/r_d^2 f_d)\cdot {\bf j}\, , \quad {\rm i.e.} \quad
J_d = \frac{j_d}{r_d^2 f_d}\, ,
\end{equation}
and
\begin{equation}
\label{hl012t}
{\bf H} = {\rm diag}(1/r_d^2)\cdot {\bf h}\, , \quad {\rm i.e.} \quad
H_d = r_d^{-2} h_d\, .
\end{equation}

\subsubsection{Generalized Sphericity Factors}

It turns out that the discretization of moment equations
represented by eqs. (\ref{jmomtri}), (\ref{umbc}) and
(\ref{lmbc}), may lead to significant inaccuracies in the
evaluated moments $j$ and $h$. This is essentially due to
a first-order representation of the $\beta h$ and $\alpha j$ terms.
An improved numerical technique easily follows from introducing
integration factors of eqs. (\ref{dhdt}) and (\ref{djdt}).
In the classical photon transport, such an approach was pioneered
by Auer (1971), who coined the term ``sphericity factors.''

We rewrite eqs. \ref{dhdt}) and (\ref{djdt}) as
\begin{equation}
\label{momqh}
\frac{1}{q_h}\,\frac{d(q_h h)}{d\tau} = k j - S_J\, ,
\end{equation}
and
\begin{equation}
\label{momqj}
\frac{1}{q_j}\,\frac{d(q_j j)}{d\tau} = h - S_H\, ,
\end{equation}
where the integration factors are defined by
\begin{equation}
\label{defqh}
\frac{d(\ln q_h)}{d\tau} = \beta\,
\end{equation}
and
\begin{equation}
\label{defqj}
\frac{d(\ln q_j)}{d\tau} = \alpha\, .
\end{equation}
We use eq. (\ref{momqh}) to eliminate $h$ from differenced eq. (\ref{momqj}), 
which leads to a second-order equation for $q_j j$:
\begin{equation}
\label{momqfac}
\frac{d^2(q_j j)}{d x^2} = \frac{q_h^2}{q_j}\, k\, j -
\left[\frac{q_h^2}{q_j}\, S_J + \frac{d(q_h S_H)}{dx}\right]\, ,
\end{equation}
where
\begin{equation}
d x = \frac{q_j}{q_h}\, d \tau = - \frac{q_j}{q_h}\,\kappa_H\, dr\, ,
\end{equation}
represents a modified optical depth increment. Equation (\ref{momqfac})
is second-order accurate. Discretization of this equation, and
the second-order boundary conditions, are quite analogous to the classical
Feautrier scheme.

%------------------------------------------------------------------

\section{Implicit Coupling of Radiation to Matter}
\label{sect-imp}

We outline here a procedure for the implicit coupling of
radiation and matter that is based on an application of the
Accelerated Lambda Iteration (ALI). The procedure is a generalization
of a treatment developed by Burrows et al. (2000); an analogous
procedure was developed in the context of static stellar atmospheres
by Hubeny \& Lanz (1995), who coined it the ``hybrid CL/ALI''
method (CL stands for Complete Linearization).

\subsection{ALI treatment of radiation}

The specific intensity can be written as
\begin{equation}
I_{\mu\nu} = \Lambda_{\mu\nu}  S_{\mu\nu}\, ,
\end{equation}
which is just another way of expressing eq. (\ref{ilambda1}).
Here, we do not discuss energy redistribution and inelastic scattering, but
provide our formalism for the coupling of energy groups in the Appendix.
With independent energy groups, we drop the energy subscript, $\nu$, 
but remember that all quantities still depend on energy. 

We assume that at the end of a given time step we know the
source function and the corresponding specific intensity,
which we denote as ``old'' quantities,
$I_\mu^{\rm old} = \Lambda_\mu S_\mu^{\rm old}$.
We express the ``new'' specific intensity as
\begin{equation}
\label{inew1}
I_{\mu}^{\rm new} = \Lambda_{\mu}(S_{\mu}^{\rm old} + \delta S_{\mu})\, .
\end{equation}
We will also denote 
$I_{\mu}^{\rm new} = I_{\mu}^{\rm old} + \delta I_{\mu} $, and express
$\Lambda_{\mu} = \Lambda_{\mu}^{\ast} + \delta \Lambda_{\mu}$.
Substituting these into eq. (\ref{inew1}), and neglecting the
2-nd order term $\delta S\, \delta I$ we obtain
\begin{equation}
\label{inew2}
\delta I_{\mu} = \Lambda_{\mu}^\ast\, \delta S_{\mu}\, .
\end{equation}
This states that the correction to the specific intensity is given by the
action of an approximate operator on the correction to the
source function.

The source function is given by eqs. (\ref{splus}) and (\ref{sminus}),
which we rewrite here as
\begin{equation}
\label{smu1}
S_{\mu} = S_0 + a_\mu J + b_\mu H + c_\mu K\, ,
\end{equation}
where $S_0, a_\mu, b_\mu$, and $c_\mu$ are known functions of the
material state parameters $T$, $Y_e$, and $\rho$. The correction
to the source function can, thus, be written
\begin{equation}
\label{scorr1}
\delta S_\mu = \frac{\partial S_\mu}{\partial T} \delta T +
\frac{\partial S_\mu}{\partial Y_e} \delta Y_e +
\frac{\partial S_\mu}{\partial \rho} \delta \rho +
a_\mu \delta J + b_\mu \delta H + c_\mu \delta K\, , 
\end{equation}
where, for instance,
\begin{equation}
\label{sder1}
\frac{\partial S_\mu}{\partial T} =
\frac{\partial S_0}{\partial T} + 
\frac{\partial a_\mu}{\partial T} J +
\frac{\partial b_\mu}{\partial T} H +
\frac{\partial c_\mu}{\partial T} K\, ,
\end{equation}
and analogously for the other derivatives. The expression for
$\delta I_\mu$ easily follows from eqs. (\ref{inew2}) and (\ref{sder1}).
Integrating over angles, we obtain equations for the corrections to
the moments $\delta J, \delta H$, and $\delta K$. For instance,
for $\delta J$ we obtain
\begin{eqnarray}
\label{jcor1}
\delta J = 
\delta T\, \int_{-1}^1 \Lambda_\mu^\ast \frac{\partial S_\mu}{\partial T} 
d\mu/2 +
\delta Y_e\, \int_{-1}^1 \Lambda_\mu^\ast \frac{\partial S_\mu}{\partial
Y_e}
d\mu/2 +
\delta \rho\, \int_{-1}^1 \Lambda_\mu^\ast \frac{\partial S_\mu}{\partial
\rho} d\mu/2 \nonumber\\
+ \delta J \, \int_{-1}^1 \Lambda_\mu^\ast a_\mu d\mu/2
+ \delta H \, \int_{-1}^1 \Lambda_\mu^\ast b_\mu d\mu/2
+ \delta K \, \int_{-1}^1 \Lambda_\mu^\ast c_\mu d\mu/2\, ,
\end{eqnarray}
and analogously for $\delta H$ and $\delta K$, the only difference
being that the angular integrals are modified to $\int \cdots \mu d\mu$ and
$\int \cdots \mu^2 d\mu$, respectively.

The simplest choice for the $\Lambda^\ast$ operator is the diagonal
(local) part of the exact $\Lambda$, as discussed in \S\ref{sect-ali}. 
Notice that Burrows et al. (2000) used a tridiagonal part of the
exact $\Lambda$, which generally gives faster convergence.  
A great advantage of the diagonal operator is that
it yields relatively simple expressions in 1-D, and is very easily
generalized to multi-D, where anything but a diagonal operator would
lead to somewhat cumbersome expressions that involve a number of
neighboring cells.
We will consider both cases, a diagonal as well as a tridiagonal 
approximate operator.

With the choice of diagonal operator, eq. (\ref{jcor1}), and analogous
equations for $\delta H$ and $\delta K$, can be written as separate 
systems for each spatial zone (depth point):
\begin{eqnarray}
\label{jhkcor1}
\left(\begin{array}{c}
\delta J_d \\
\delta H_d \\
\delta K_d
\end{array}\right) =
\left(\begin{array}{ccc}
1-\Lambda_d^{JJ} & -\Lambda_d^{JH} &-\Lambda_d^{JK} \\
-\Lambda_d^{HJ} & 1-\Lambda_d^{HH} &-\Lambda_d^{HK} \\
-\Lambda_d^{KJ} & -\Lambda_d^{KH} & 1-\Lambda_d^{KK} 
\end{array}\right)^{-1}
\cdot \nonumber\\
\left[
\left(\begin{array}{c}
\Sigma_d^{T,J} \\
\Sigma_d^{T,H} \\
\Sigma_d^{T,K} 
\end{array}\right) \cdot \delta T +
\left(\begin{array}{c}
\Sigma_d^{Y_e,J} \\
\Sigma_d^{Y_e,H} \\
\Sigma_d^{Y_e,K} 
\end{array}\right) \cdot \delta Y_e +
\left(\begin{array}{c}
\Sigma_d^{\rho,J} \\
\Sigma_d^{\rho,H} \\
\Sigma_d^{\rho,K} 
\end{array}\right) \cdot \delta\rho \right]\, ,
\end{eqnarray}
where we observe that the elements of the moment-coupling matrix
are exactly those defined by eqs. (\ref{lamjj}) - (\ref{lamkk})
needed in the formal solution of the transfer equation, and where
the elements of the vectors representing the appropriately
angle-averaged derivatives of the source function are given by
\begin{equation}
\Sigma_d^{T,J} = \frac{1}{2}
\int_{-1}^1 \Lambda_\mu^\ast \frac{\partial S_\mu}{\partial T} d\mu\, ,
\end{equation}
\begin{equation}
\Sigma_d^{T,H} = \frac{1}{2}
\int_{-1}^1 \mu\Lambda_\mu^\ast \frac{\partial S_\mu}{\partial T} d\mu\, ,
\end{equation}
\begin{equation}
\Sigma_d^{T,K} = \frac{1}{2}
\int_{-1}^1 \mu^2\Lambda_\mu^\ast \frac{\partial S_\mu}{\partial T} d\mu\,
,
\end{equation}
and analogously for $\Sigma^{Y_e,J}$, etc. The essential point here is that
we are able to express $\delta J$ ($\delta H$, $\delta K$) as functions
of $\delta T$, $\delta Y_e$, and $\delta\rho$, through a simple 
$3\times 3$ matrix inversion.

These equations are quite general. In practice, we perform an implicit
update for the temperature and $Y_e$ only, because the density is 
most naturally updated in the operator-split fashion
in the explicit hydro step. In this case, we formally set $\delta\rho = 0$,
and write (dropping the depth index $d$):
\begin{equation}
\label{jdelta}
\delta J = \Psi^{T, J} \delta T + \Psi^{Y_e, J} \delta Y_e\, ,
\end{equation}
(and analogously for $\delta H$ and $\delta K$) where
\begin{eqnarray}
\left(\begin{array}{c}
\Psi^{T, J}\\
\Psi^{T, H}\\
\Psi^{T, K} \end{array} \right) =
\left(\begin{array}{ccc}
1-\Lambda_d^{JJ} & -\Lambda_d^{JH} &-\Lambda_d^{JK} \\
-\Lambda_d^{HJ} & 1-\Lambda_d^{HH} &-\Lambda_d^{HK} \\
-\Lambda_d^{KJ} & -\Lambda_d^{KH} & 1-\Lambda_d^{KK} 
\end{array}\right)^{-1}
\cdot 
\left(\begin{array}{c}
\Sigma_d^{T,J} \\
\Sigma_d^{T,H} \\
\Sigma_d^{T,K} 
\end{array}\right) \, .
\end{eqnarray}
There are analogous expressions for $\Psi^{Y, J}$, etc.

There is an alternative way of formulating the ALI-based corrections
$\delta J$ and $\delta H$, namely those based directly on the moment
equations (\ref{jl01}) and (\ref{hl012}) or (\ref{hl23}), together with eqs.
(\ref{jl01t}) and (\ref{hl012t}). Since the material equations are
written in a form completely consistent with these moment equations,
this approach seems to be more suitable.
We write for the correction $\delta J_d$, again using for the approximate
operator the diagonal (local) part of the exact operator:
\begin{eqnarray}
r_d^2\, f_d\, \delta J_d\, =\, (\Lambda_0)_{dd}\, \delta S^J_d + 
(\Lambda_1)_{dd}\, \delta S^H_d\, =\, 
\nonumber\\
(\Lambda_0)_{dd} \left( 
\frac{\partial S^J}{\partial T}\, \delta T_d +
\frac{\partial S^J}{\partial Y_e}\, \delta Y_{e,d}\right) +
%\nonumber\\
(\Lambda_1)_{dd} \left( 
\frac{\partial S^H}{\partial T}\, \delta T_d +
\frac{\partial S^H}{\partial Y_e} \delta Y_{e,d}
\right) \, .
\end{eqnarray}
Thus, the auxiliary quantities $\Psi$ can be written
\begin{equation}
\Psi^{T,J}_d =  \frac{1}{r_d^2 f_d} \left[
(\Lambda_0)_{dd} \left( \frac{\partial S^J}{\partial T} \right)_{\! d}  + 
(\Lambda_1)_{dd} \left( \frac{\partial S^H}{\partial T} \right)_{\! d} 
\right]\, ,
\end{equation}
and
\begin{equation}
\Psi^{Y,J}_d =  \frac{1}{r_d^2 f_d} \left[
(\Lambda_0)_{dd} \left( \frac{\partial S^J}{\partial Y_e} \right)_{\! d} +
(\Lambda_1)_{dd} \left( \frac{\partial S^H}{\partial Y_e} \right)_{\! d} 
\right]\, .
\end{equation}
Here, the integrations over angle have been performed analytically, as opposed
to the numerical integration of the previous approach.

Analogously, for $\delta H$ one obtains
\begin{eqnarray}
r_d^2\, \delta H_d\, =\, (\Lambda_2)_{dd}\, \delta S^J_d + 
(\Lambda_3)_{dd}\, \delta S^H_d\, =\,
\nonumber\\
(\Lambda_2)_{dd} \left( 
\frac{\partial S^J}{\partial T}\, \delta T_d +
\frac{\partial S^J}{\partial Y_e}\, \delta Y_{e,d}\right) +
%\nonumber\\
(\Lambda_3)_{dd} \left( 
\frac{\partial S^H}{\partial T}\, \delta T_d +
\frac{\partial S^H}{\partial Y_e}\, \delta Y_{e,d}\right) \, ,
\end{eqnarray}
and, thus,
\begin{equation}
\Psi^{T,H}_d =  \frac{1}{r_d^2} \left[
(\Lambda_2)_{dd} \left( \frac{\partial S^J}{\partial T} \right)_{\! d}  + 
(\Lambda_3)_{dd} \left( \frac{\partial S^H}{\partial T} \right)_{\! d} 
\right]\, ,
\end{equation}
and
\begin{equation}
\Psi^{Y,H}_d =  \frac{1}{r_d^2} \left[
(\Lambda_2)_{dd} \left( \frac{\partial S^J}{\partial Y_e} \right)_{\! d} +
(\Lambda_3)_{dd} \left( \frac{\partial S^H}{\partial Y_e} \right)_{\! d} 
\right]\, .
\end{equation}
In order to evaluate these coefficients, one needs first to compute
the diagonal and a few off-diagonal elements of the inverse
of the tridiagonal matrix $T$. This is done very efficiently using the
procedure suggested by Rybicki \& Hummer (1991). These elements are
evaluated during the solution of the original set of moment equations
and comes at almost no additional cost. We, thus, take elements
$T^{-1}_{dd}, T^{-1}_{d,d \pm 1}$, and $T^{-1}_{d,d\pm 2}$ as known,
and write
\begin{equation}
(\Lambda_0)_{dd} = T^{-1}_{dd}\, 
\end{equation}
\begin{equation}
(\Lambda_1)_{dd} = T^{-1}_{d,d-1} U_{d-1,d}\, + \,T^{-1}_{dd} U_{dd}\, +\,
                   T^{-1}_{d,d+1} U_{d+1,d}\,  
\end{equation}
\begin{equation}
(\Lambda_2)_{dd} = V_{d,d-1} T^{-1}_{d-1,d}\, +\, V_{dd} T^{-1}_{dd}\, +\,
V_{d,d+1} T^{-1}_{d+1,d}\, 
\end{equation}
and
\begin{eqnarray}
(\Lambda_3)_{dd} = V_{dd}\, (\Lambda_1)_{dd}\, +\,
V_{d,d-1} \left(T^{-1}_{d-1,d-1} U_{d-1,d} + T^{-1}_{d-1,d} U_{dd}\right) +
\nonumber\\
V_{d,d+1} \left(T^{-1}_{d+1,d} U_{dd} + T^{-1}_{d+1,d+1} U_{d+1,d}\right)\, +\,
W_{dd}\, .
\end{eqnarray}
Numerical experience has shown that while the first way of evaluating the
moment ``derivatives'' $\Psi$ is fast and simple, the second way is more
consistent and accurate. This is because the material equations (energy
balance and electron fraction equations) are based on the equations for 
the radiation moments, in which the integrations over angles are done
analytically. Thus, the approximate operator based on the radiation
moment equations
can naturally be used in the material equations, because in the first
way of computing the approximate operator the necessary angular integrations 
are performed numerically.

%---------------------------------------------------------------------------------

\subsection{Linearization of the energy and the electron fraction
equations}

\label{sect-imp-lin}

As follows from the analysis presented in \S\ref{hydro} (eqs. \ref{temp_eq} and
\ref{rhecmv}), the energy balance 
equation is written as
\begin{equation}
\rho C_V \frac{D T}{D t} = - 4\pi \sum_i \int_0^\infty\left[
\eta_0^i(\nu, T, Y_e) - \kappa_0^i(\nu, T, Y_e) J_\nu^i +
\lambda^i_T(\nu, T, Y_e)  H_\nu^i
\right] d\nu\, ,
\end{equation}
where the summation extends over all neutrino species $i$ ($\eta_0^i$ is the
corresponding $\eta_0^{\rm th}$), and
where
\begin{equation}
\lambda_T = w\left(2\kappa_0 + \frac{\partial\kappa_0}{\partial\ln\nu}\right) = 
w\left(\kappa_0 + \widetilde\kappa_0 \right)\, ,
\end{equation}
The electron fraction equation is
\begin{equation}
\rho N_A \frac{D Y_e}{D t} = 4\pi \sum_i s^i \int_0^\infty
\left[
\eta^i_0(\nu, T, Y_e) - \kappa^i_0(\nu, T, Y_e) J_\nu^i + \lambda^i_Y(\nu, T, Y_e)
H_\nu^i \right] \frac{d\nu}{\nu} \, ,
\end{equation}
where $s^i = -1$ for $\nu_e$ neutrinos, $s^i = 1$ for $\bar\nu_e$
neutrinos, and $s^i = 0$ for other neutrino species, 
$N_A$ is the Avogadro's number, and
\begin{equation}
\lambda_Y =  w\, \widetilde\kappa_0\, .
\end{equation}

The implicit (backward time differencing) forms of these equations are
\begin{equation}
\label{t1}
\frac{\rho C_V}{4\pi} \frac{T-T_0}{\Delta t} +  
\sum_i \int_0^\infty\left[
\eta_0^i(\nu, T, Y_e) - \kappa_0^i(\nu, T, Y_e) J_\nu^i +
\lambda^i_T(\nu, T, Y_e)  H_\nu^i
\right] d\nu + {\rm ADV}(T) = 0\, ,
\end{equation}
and 
\begin{equation}
\label{y1}
\frac{\rho N_A}{4\pi} \frac{Y_e-Y_e^0}{\Delta t} -  
\sum_i s^i \int_0^\infty\left[
\eta_0^i(\nu, T, Y_e) - \kappa_0^i(\nu, T, Y_e) J_\nu^i +
\lambda^i_Y(\nu, T, Y_e)  H_\nu^i
\right] \frac{d\nu}{\nu}  + {\rm ADV}(Y_e)= 0\, .
\end{equation}
Here, ${\rm ADV}(T)$ and ${\rm ADV}(Y_e)$ are the formal advection terms.
Depending on the overall hydro scheme to which the present formalism
is being implemented, the actual advection terms may be already considered in
the hydro step, and not in the implicit update of $T$ and $Y_e$ due to
radiation, which we consider here. In this case, we have to set
${\rm ADV}(T) = {\rm ADV}(Y_e) = 0$. For certain testing purposes, 
we may consider the advection terms as a part of the present implicit update.
Therefore, we write the advection term at radial zone $d$ (at radius $r_d$), 
using a second-order representation of the spatial derivative, as:
\begin{equation}
{\rm ADV}(T)_d = Z\, \frac{\rho c w C_V}{4\pi}\,
[e_{d,d-1} T(r_{d-1}) - (e_{d,d-1} + e_{d,d+1}) T(r_{d}) + 
e_{d,d+1} T(r_{d+1})]\, ,
\end{equation}
and analogously for ${\rm ADV}(Y_e)$.  Here, $Z=0$ if the advection was
already treated in the hydro step, and $Z=1$ if the advection is being treated
as a part of the implicit update.
The coefficients $e$ depend on an adopted form of differentiation formula.
For a simple centered difference,
\begin{equation}
e_{d,d+1} = -e_{d,d-1} = 1/(r_{d+1} - r_{d-1})\, .
\end{equation}
One can also use a truly 2nd-order formula, where
\begin{equation}
e_{d,d+1} = \frac{r_d-r_{d-1}}{(r_{d+1}-r_d)(r_{d+1} - r_{d-1})}\, ,\quad\quad
e_{d,d-1} = -\frac{r_{d+1}-r_{d}}{(r_{d}-r_{d-1})(r_{d+1} - r_{d-1})}\, .
\end{equation}

They can be treated either explicitly or implicitly. If they are treated 
explicitly, and if we use a diagonal $\Lambda^\ast$ approximate operator to treat
radiation quantities, then the problem can be formulated as a {\em local}
problem in space. We shall first consider the local formulation, while we
will consider the non-local formulation, with an implicit treatment of the
advection term, later.
 
We introduce the state vector $\psi = (T, Y_e)^T$, and write the
system of material equations (\ref{t1}) and (\ref{y1}) as
\begin{equation}
P(\psi) = 0\, ,
\end{equation}
which we linearize using the standard Newton-Raphson technique,
\begin{equation}
\label{nr1}
P^\prime(\psi^{(n)}) \cdot \delta\psi^{(n)} = - P(\psi^{(n)})\, ,
\end{equation}
where $P^\prime$ is the Jacobian of the system;
\begin{equation}
P^\prime_{ij} = \frac{\partial P_i}{\partial \psi_j}\, ,
\end{equation}
i.e., the $i,j$ element of the Jacobian is the partial derivative of
the $i$-th equation with respect to the $j$-th variable.
In expressing the Jacobian, we use the expression  
$\delta J = (\partial J/\partial T) \delta T +(\partial J/\partial Y_e)
\delta Y_e$, together with eq. (\ref{jdelta}),
namely $\delta J = \psi^{T,J} \delta T +
\psi^{Y_e, J} \delta Y_e$, which allows us to
associate $\partial J/\partial T = \Psi^{T,J}$ and
$\partial J/\partial Y_e = \Psi^{Y_e,J}$. We can write analogous expressions 
for $\partial H/\partial T$ and $\partial H/\partial Y_e$. 
The energy and electron fraction equations do not couple the
adjacent depths, and, thus, the Jacobian has a simple block-diagonal
structure; each block is a $2\times 2$ matrix which we denote $A$
(we drop the index indicating the radial zone). The matrix elements
are then given by
\begin{equation}
A_{11} = \frac{\rho C_V}{4\pi\Delta t} + \sum_i \int_0^\infty\left(
\frac{\partial\eta_0^i}{\partial T} - 
\frac{\partial\kappa_0^i}{\partial T} J_0^i
+ \frac{\partial\lambda^i_T}{\partial T} H_0^i -
\kappa_0^i \Psi^{T, J, i} + \lambda^i_T \Psi^{T, H, i}
\right)  d\nu\, ,
\end{equation}
\begin{equation}
A_{12} =  \sum_i  \int_0^\infty\left(
\frac{\partial\eta_0^i}{\partial Y_e} - 
\frac{\partial\kappa_0^i}{\partial Y_e} J_0^i
+ \frac{\partial\lambda^i_T}{\partial Y_e} H_0^i -
\kappa_0^i \Psi^{Y_e, J, i} + \lambda^i_T \Psi^{Y_e, H, i}
\right) d\nu\, ,
\end{equation}
\begin{equation}
A_{21} =  -\sum_i s^i \int_0^\infty\left(
\frac{\partial\eta_0^i}{\partial T} - 
\frac{\partial\kappa_0^i}{\partial T} J_0^i
+ \frac{\partial\lambda^i_Y}{\partial T} H_0^i -
\kappa_0^i \Psi^{T, J, i} + \lambda^i_Y \Psi^{T, H, i}
\right)\frac{d\nu}{\nu}\, ,
\end{equation}
and
\begin{equation}
A_{22} = \frac{\rho N_A}{4\pi\Delta t} - \sum_i s^i \int_0^\infty\left(
\frac{\partial\eta_0^i}{\partial Y_e} - 
\frac{\partial\kappa_0^i}{\partial Y_e} J_0^i
+ \frac{\partial\lambda^i_Y}{\partial Y_e} H_0^i -
\kappa_0^i \Psi^{Y_e, J, i} + \lambda^i_Y \Psi^{Y_e, H, i}
\right)\frac{d\nu}{\nu}\, .
\end{equation}
The quantities that depend on $T$ and $Y_e$, such as $\kappa^i$,
$\eta^i$, and $\Xi^i$, are to be evaluated at the current iterate
$T^{(n)}$ and $Y_e^{(n)}$. The starting estimate is obviously
given by $T^{(0)} = T_0$ and $Y_e^{(0)} = Y_e^0$, that is by the values
at the end of previous time step.

The individual blocks of the right-hand-side vector, $P(\psi^{(n)})$,
which we denote as $b$, are given by
\begin{equation}
b_1 = -\frac{\rho C_V}{4\pi}\frac{T^{(n)} - T_0}{\Delta t} -
\sum_i \int_0^\infty\left(
\eta_{0}^i - \kappa_{0}^i J^i + \lambda_T^i H^i \right) d\nu -{\rm ADV}(T_0)\, ,
\end{equation}
and
\begin{equation}
b_2 = -\frac{\rho N_A}{4\pi}\frac{Y_e^{(n)} - Y_e^0}{\Delta t} + 
\sum_i s^i \int_0^\infty\left(
\eta_{0}^i - \kappa_{0}^i J^i + \lambda_Y^i H^i\
\right) \frac{d\nu}{\nu}-{\rm ADV}(Y_{e, 0})\, .
\end{equation}
Notice that we have to treat here the advection terms (if they
are taken into account, $Z=1$) explicitly. 
%We shall show how they
%can be treated implicitly in \S\ref{coupl_tri}.

The problem is, thus, reduced to solving, for each radial zone, the
system $A \delta\psi = b$. However, the moments, as well as their
``derivatives,'' $\Psi$, are evaluated at the end of the previous time
step. 

In fact, one can avoid inverting a $2\times 2$ matrix $A$ by
setting its off-diagonal elements to zero: $A_{12} = A_{21} = 0$.
We have verified that this does not decrease the convergence
speed in any appreciable way. We can also introduce a slightly
modified notation and write $A_{11} = B^T_{dd}$,
$A_{22} = B^Y_{dd}$, $b_1 = R^T_d$, and $b_2 = R^Y_d$ (where
all the state parameters and radiation moments are taken at depth $d$),
and write the global linearization scheme for the corrections as
\begin{equation}
\label{blin}
B^T \delta T = R^T\, , \quad  B^Y \delta Y_e = R^Y\, ,
\end{equation}
where
\begin{equation}
(B^T)_{d,d^\prime} = B^T_{dd} \delta_{d,d^\prime} \, ,\quad 
(B^Y)_{d,d^\prime} = B^Y_{dd} \delta_{d,d^\prime}\, 
\end{equation}
are diagonal matrices and $R^T = \{R^T_1, R^T_2, \ldots , R^T_D\}$,
and $\delta T = \{\delta T_1, \delta T_2, \ldots, \delta T_D\}$ 
(and analogously for $Y_e$) are the appropriate column vectors.

The iterations proceed as follows:
%Therefore, we have to iterate as follows:

\medskip

 (a) Taking current estimates of $J^i$ and $H^i$
(and taking $J^i_0$ and $H^i_0$ as starting values), we perform
the inner Newton-Raphson iteration loop, eq. (\ref{nr1}) to determine 
estimates of $T$ and $Y_e$ at the end of the given time step.

 (b) We perform a formal solution of the transfer equation
using newly computed values of $T$ and $Y_e$ to compute new
estimates of $J^i$ and $H^i$, and possibly also new $\Psi^{x,y}$.

 (c) Return to step (a) and iterate. This is an outer
iteration loop.

\medskip

In this nested iteration loop, one can in principle obtain a fully
self-consistent implicit solution for $J$, $H$, $T$, and $Y_e$,
i.e. for the radiation moments and the material quantities.
However, this may be too time-consuming (although possible in 1-D), so
we usually resort to approximate schemes. The most
useful approximations (roughly in the order of decreasing overall
importance, i.e. their expected influence in saving computer time)
are:
\medskip

 (1) In the outer iteration loop, one does not perform
the full transfer solution for specific intensities. Instead one
only updates the radiation {\em moments}, keeping the Eddington
factor fixed. This means that the Eddington factor is going to be
treated explicitly, while the rest of the quantities (radiation moments
and material quantities) are treated implicitly. In this case, the 
Eddington factor may be updated at the end of the nested iteration
loop (at the end of the given timestep). It may even be kept fixed
for several time steps (which will almost certainly be necessary
in the 2-D case).

(2) We may reduce the number of iterations in the outer
loop, perhaps even to 1 (only recalculating the radiation moments
after the first implicit update of $T$ and $Y_e$ is done.

(3) In the inner loop, we may hold some quantities fixed.
For instance, we may update $\eta$, $\kappa$, $\Xi$, and their
derivatives, but keep the $\Psi$ factors, that depend on the 
approximate $\Lambda$ operator, fixed.

(4) The extreme variant of the above strategy is to use 
the so-called Kantorovich variant of the Newton-Raphson procedure 
that consists in keeping the Jacobian fixed altogether. Since setting
up the Jacobian is the most time-consuming part of the inner loop
(its inversion is easy, since it is only a $2\times 2$ matrix, or just
two divisions, if we use diagonalized matrices), this
may lead to considerable savings.
\medskip

The essential point behind the two last simplifications is
that one needs solve only the material equations 
(\ref{t1}) and (\ref{y1})
(together with the transfer equation) exactly. Since the Jacobian
is only one possible means to obtain the exact solution, there
is no need to compute the Jacobian ``exactly'' from its mathematical
definition. In other words, the basic linearization equation (\ref{nr1})
computes only {\em corrections} to quantities, not quantities
themselves, and, thus, one can afford approximations. Here, any matrix
will do, as long as the process converges sufficiently quickly.

\subsection{Linearization using tridiagonal operator}
\label{coupl_tri}

A useful generalization of the previous formalism is to replace the
diagonal representation of $\Lambda^\ast$ by a tridiagonal one.
Although it would be easy to retain the coupling of $\delta T$ and
$\delta Y_e$, which would lead to a block-tridiagonal system for the
corrections, we consider here the case of uncoupled $\delta T$ and
$\delta Y_e$, as in eq. (\ref{blin}). Moreover, we can now treat
the advection terms implicitly at no cost, because the matrices already
consider the radial zone coupling. Specifically, eq. (\ref{blin})
remains valid, but matrices $B^T$ and $B^Y$ become tridiagonal,
where the diagonal elements are the same as before, and the off-diagonal
elements are given by
\begin{equation}
\label{matbt}
B^T_{d, d\pm 1} = - \sum_i \int_0^\infty\left(
\kappa^i_{0, d \pm 1}, \Psi^{T, J, i}_{d,d\pm 1} - 
\lambda^i_{T,d\pm 1} \Psi^{T, H, i}_{d,d\pm 1} \right)  d\nu
+ Z \rho_d c w_d C_V e_{d,d\pm 1}/(4\pi)\, ,
\end{equation}
and
\begin{equation}
\label{matby}
B^Y_{d, d\pm 1} = - \sum_i s^i \int_0^\infty\left(
\kappa^i_{0, d \pm 1}, \Psi^{Y_e, J, i}_{d,d\pm 1} - 
\lambda^i_{Y,d\pm 1} \Psi^{Y_e, H, i}_{d,d\pm 1} \right)  d\nu
+ Z \rho_d c w_d N_A e_{d,d\pm 1}/(4\pi)\, ,
\end{equation}
where 
\begin{equation}
\Psi^{X,J}_{d,d\pm 1} =  \frac{1}{r_{d\pm 1}^2 f_{d\pm 1}} \left[
(\Lambda_0)_{d,d\pm 1} \left( \frac{\partial S^J}{\partial X} \right)_{\! d\pm
1}  + 
(\Lambda_1)_{d,d\pm 1} \left( \frac{\partial S^H}{\partial X} \right)_{\! d\pm
1} 
\right]\, ,
\end{equation}
and
\begin{equation}
\Psi^{X,H}_{d,d\pm 1} =  \frac{1}{r_{d\pm 1}^2} \left[
(\Lambda_2)_{d,d\pm 1} \left( \frac{\partial S^J}{\partial X} \right)_{\! d\pm
1}  + 
(\Lambda_3)_{d,d\pm 1} \left( \frac{\partial S^H}{\partial X} \right)_{\! d\pm
1} 
\right]\, ,
\end{equation}
and where $X$ stands for $T$ or $Y_e$.
Obviously, if we adopt the explicit treatment of the advection terms, the
last terms of eqs. (\ref{matbt}) and (\ref{matby}) disappear, and appear
in the right-hand-side vectors $R^T$ and $R^Y$.

\subsection{Treatment of the optically thick region}

Since the radial optical depths close to the center may be very large
for most neutrino species, we must assure that we recover the diffusion
limit exactly. This must apply not only for the solution of the moments
themselves (which we are doing by construction since we are using 
methods which are second-order accurate), but also in the implicit
update. To this end, we found it best to replace the above outlined
approximate operator by that based on the diffusion approximation.
Such an approach is often used in neutron transport theory where it
is known by the name Diffusion Synthetic Acceleration (DSA).
In other words, we employ the approximate operator, not as a diagonal
or tridiagonal part of the exact transport operator, but rather
as the corresponding expression in the diffusion approximation.

In this case we use the approximation
\begin{equation}
J = S + \frac{1}{3}\, \frac{d^2 S}{d\tau^2}\, ,\quad {\rm and}\quad
H = \frac{1}{3}\, \frac{d S}{d\tau}\, ,
\end{equation}
where $S=\eta_0^{\rm th}/\kappa_0$, and we set $\tau = \tau_H$.
In the discretized form (for $1 < d < NR$)
\begin{equation}
J_d = S_d \left[1 - \frac{1}{3\Delta\tau_d}\left(
\frac{1}{\Delta\tau_{d-1/2}} + \frac{1}{\Delta\tau_{d+1/2}}\right)
\right]
+ \frac{S_{d-1}}{3\Delta\tau_d\Delta\tau_{d-1/2}}
+ \frac{S_{d+1}}{3\Delta\tau_d\Delta\tau_{d+1/2}}\, ,
\end{equation}
and (using a centered numerical differentiation) 
\begin{equation}
H_d = \frac{2}{3}\, \frac{S_{d+1} - S_{d-1}}{\Delta\tau_d}\, .
\end{equation}
We stress that we do not use these expression for evaluating
the moments themselves (this is done by exact solution
of the moment equation), but only to evaluate the derivatives
$\partial J/\partial T$, $\partial J/\partial Y_e$ (and
analogous ones for $H$) to be used in the implicit update.
In our notation, we write:
\begin{equation}
\Psi^{T,J}_{d,d} = \left[1 - \frac{1}{3\Delta\tau_d}\left(
\frac{1}{\Delta\tau_{d-1/2}} + \frac{1}{\Delta\tau_{d+1/2}}\right)
\right] \frac{\partial S_d}{\partial T_d}\, ,
\end{equation}
and
\begin{equation}
\Psi^{T,J}_{d,d\pm 1} = \frac{1}{3\Delta\tau_d\Delta\tau_{d\pm 1/2}}
\frac{\partial S_{d\pm 1}}{\partial T_{d\pm 1}}\, ,
\end{equation}
and
\begin{equation}
\Psi^{T,H}_{d,d\pm 1} = \pm\frac{2}{3\Delta\tau_d}
\frac{\partial S_{d\pm 1}}{\partial T_{d\pm 1}}\, ,\quad
\Psi^{T,H}_{d,d} = 0\, ,
\end{equation}
and analogously for $\Psi^{Y,T}$ and $\Psi^{Y,H}$.

To assure a smooth transition between the previous formalism,
which should be used in the optically thin regime, and the
present one, we set the ``derivatives'' $\Psi$ equal to a linear combination
of the original quantities $\Psi$, which we denote by $\Psi_0$,
and the present ones, which we denote by $\Psi_{\rm DSA}$
(for Diffusion Synthetic Acceleration):
\begin{equation}
\Psi^{T,J} = \exp(-\tau) \Psi^{T,J}_0 + [1 - \exp(-\tau)] \Psi^{T,J}_{\rm DSA}\, .
\end{equation}
This choice of the linear combination is not unique, and some other
choice might be better,  but we found it to be quite robust.

%--------------------------------------------------------------------

\section{Numerical Test and Results} 
\label{sec-num}

\subsection{Stationary solutions}
\label{sec-num-st}

We first study the convergence pattern and the behavior
of the solution in the stationary case. We consider a fictitious
structure, based on an angle-averaged snapshot
of a VULCAN/2D simulation taken from Dessart et al. (2006)
at 200 ms after bounce. 
The original 2-D structure was appropriately averaged to yield a
1-D, spherically symmetric structure. Figure \ref{fig1} displays
the temperature ($T$), density ($\rho$), and the electron
fraction ($Y_e$) as a function of radius. 

In the present tests, we consider three neutrino species, $\nu_e$,
$\bar\nu_e$, and a composite of $\nu_\mu$, $\nu_\tau$,
$\bar\nu_\mu$, and $\bar\nu_\tau$, which we denote as ``$\nu_\mu$.''
For each species, we consider 16 energy groups logarithmically
equidistant between 1 and $E_{\rm max}$, where $E_{\rm max}=300$ MeV
for $\nu_e$ neutrinos, and $E_{\rm max}=100$ MeV for the other two species.

First, we test the global ALI iteration scheme.
We consider five cases: diagonal and tridiagonal operator,
with or without the GMRES augmentation, and
another acceleration scheme related to the GMRES, namely
Ng acceleration (Ng 1984; Auer 1991; Hubeny \& Lanz 1992),
which is widely used in astrophysical photon transport work.
The solver for the formal solution of the transfer
equation used in these tests is the DFE scheme. 
We stress that we consider here the most stringent test case of
the ALI iteration scheme, the one for which the initial value
of the radiation intensity is set to zero, and we compute the
stationary solution directly. In a realistic time-dependent solution,
one starts with the radiation intensities at the previous timestep,
which are already relatively close to the solution at the given
timestep, so this generally requires far fewer iterations.

Figure \ref{fig2} displays the convergence pattern for $\nu_e$ and 
$\nu_\mu$ neutrinos with $E=8.6$ MeV, for the 5 different setups of 
the transport solver. Convergence is pretty fast for $\nu_e$
neutrinos; an application of a tri-diagonal approximate operator
leads to a somewhat faster convergence, as does the application of
the GMRES scheme. The case of $\nu_\mu$ neutrinos is more interesting.
Here, we see that the tridiagonal operator yields a substantial
speed up; however, the biggest gain is achieved by using the
GMRES scheme. Interestingly, when the GMRES scheme is applied,
the advantages of using a tridiagonal operator are relatively
modest compared to the diagonal (local) operator, which is 
a very encouraging result because in 2-D or 3-D cases an application 
of anything but a local operator is rather cumbersome.
Ng acceleration provides a speed-up comparable to the case
of no acceleration, but the GMRES scheme is clearly superior.

To fully appreciate the speed of convergence for all species and
energy groups, we present in Figs. \ref{fig3} $-$ \ref{fig4a} the
number of iterations required as a function of neutrino energy.
The convergence criterion is  $\max(\delta J/J) < 10^{-5}$.
For electron neutrinos, the convergence is pretty rapid for
all energy groups, while it is somewhat slower for the lowest
energies for the standard ALI scheme without GMRES. With GMRES,
the solver converges for most energies in 5$-$8 iterations, for
both diagonal and tridiagonal operators. For $\bar\nu_e$ neutrinos,
the convergence is again slower for low energies. For $\nu_\mu$
neutrinos, the convergence is slowest for higher energies.
This behavior is directly related to the proportion of scattering
in the deep layers. As is customary in photon transport we 
define the parameter $\epsilon$ as
\begin{equation}
\epsilon = \frac{\kappa}{\kappa + \sigma}\, .
\end{equation}
Quantity $1-\epsilon$ is sometimes called a single-scattering
albedo. The lower the $\epsilon$, the higher the contribution
of scattering, and, hence, the transport is more
non-local and, thus, numerically more difficult.

In Fig. \ref{fig_eps},
we display $\epsilon$ for the three species (full lines for
$\nu_e$, dotted lines for $\bar\nu_e$, and dashed lines for
$\nu_\mu$) and for three energies (the thickest line for $E=1$ MeV,
the intermediate for 10 MeV and thin line for 100 MeV).
For electron neutrinos, $\epsilon$ is relatively large for all
energies, and thus the scheme converges well. For the lowest energies
for $\bar\nu_e$ neutrinos $\epsilon$ becomes smaller, which
is reflected in slower convergence. For $\nu_\mu$ neutrinos,
the lower energies exhibit the largest $\epsilon$ in the deep
layers (close to unity), and consequently the scheme converges
very rapidly. For higher energies, $\epsilon$ becomes very small,
which combined with a sharp drop of the source function toward
the center, leads to slower convergence.
Notice that $\epsilon$ exhibits a sharp drop beyond 150 -- 200 km;
this, however, does not lead to any significant deterioration 
of the iteration scheme because these layers are already optically
thin.

Next, we compare the results using three different formal solvers,
namely the Discontinuous Finite Element (DFE) scheme, a first-order
short characteristics (SC) scheme, and the Feautrier scheme.
In Fig. \ref{fig_sol2}, the mean intensity $J$ is
plotted as a function of radius for four selected energy groups,
$E = 4.6$ MeV, 11.7 MeV, 40 MeV, and 74 MeV.
Full lines display the DFE results, dotted lines the SC results,
and dashed lines the results obtained by the Feautrier scheme.
Since $J$ spans many orders of magnitude, the lines are almost
always indistinguishable on the plots, with the exception of $J$
for $\nu_\mu$ neutrinos at higher energies. We, therefore, plot
also the relative difference of the mean intensities with those computed
by the DFE scheme, namely $J({\rm solver})/J({\rm DFE}) - 1$.
The results are displayed on the right panels of Fig. \ref{fig_sol2}
by dashed (for Feautrier) or dotted (for SC) lines.

For $\nu_e$ neutrinos,
all three solvers produce mean intensities
which are generally within 1\% of each other, with the exception
of the highest energy, where the difference reaches about 9\%.
However, it should be realized that this difference occurs at
layers where the value of the mean intensity is some {\em 10 or more orders
of magnitude} lower that the peak value; in view of this fact
the overall accuracy is remarkably high.

For the $\bar\nu_e$ neutrinos the accuracy
is also pretty high, although somewhat lower than for electron
neutrinos (around 2\%); the accuracy decreases to about 10\% for
the three lower energies around 10 km, and for the highest energy 
around 40 km, where $J$ is very low anyway. 
The differences between the solvers is largest in the case of
$\nu_\mu$ neutrinos, particularly for the highest 
energies. Interestingly, the Feautrier solver
produces larger differences with the DFE than SC at lower energies,
while for the highest energies the SC solver is very inaccurate and
the Feautrier solver stays at the same level of accuracy as at the 
lower energies.

In the following tests and production runs, we adopt the DFE solver 
as our default solver. This is based partly on the results displayed
above, and partly on the fact that we will adopt DFE for 
future 2-D simulations, which is essentially the only viable choice for
irregular or unstructured grids.

Next, we examine the accuracy of the moment equation solver.
We display in Fig. \ref{fig_mom1} a comparison of the mean intensity
computed by the angle-dependent solver (solid lines), and
from the moment equation solver. The dotted lines represent the
original solver, without the sphericity factors and the dashed lines
represent the moment solver with sphericity factors. As expected, the
sphericity factors improve the agreement considerably.  In particular,
for lowest energies the moment equation solver results in a difference
of about 10\% or more. Using sphericity factors, the agreement
is within a few percent (except, again, at the highest energy), which is
quite reasonable.

Finally, we compare in Fig. \ref{fig_vel1} the full
solution with that setting $\delta=0$ -- i.e., assuming isotropic
scattering (solid lines), and with that setting $w=0$ (dashed lines).
The differences between the solutions assuming anisotropic and
isotropic scattering are rather small at small radii, which is
quite understandable in view of the large optical depth. Farther
from the center, the differences become larger, but are still
quite modest (typically around 5\% for most energy groups).
Neglecting velocities has a larger effect, reaching about 15\% for 
the lowest neutrino energies. 

In Fig. \ref{fig10}, we plot a comparison of the total net heating rates 
for the models displayed in Fig \ref{fig_vel1}.  This quantity is
germane to the neutrino mechanism of core-collapse supernova explosions.
The net heating rate differs relatively little, reaching about 8\%. 
In the so-called 
``gain" region just behind the shock, unlike in the comoving case (Buras et 
al. 2006), adding the velocity-dependent term in the mixed-frame 
formalism, for which the radiation field is calculated in the laboratory 
frame, leads to additional heating.  This is explained simply as the 
inclusion of the blue-shift of the laboratory-frame radiation
(both the monochromatic energy and the specific intensity) into the frame
of the matter due to inward accretion against the outward laboratory-frame flux.
This is not to say that the Buras et al (2006) analysis is incorrect; the two
approaches
should give the same results.  Rather, it shows that the velocity
corrections depend upon the frame in which one calculates the radiation
quantities.  If done in the comoving frame, the numerous velocity
corrections result in a slight diminution in the heating rate behind the 
shock, whereas if done in the laboratory frame the velocity correction is 
slightly positive. This has a bearing on the consequences of including 
the Doppler term in VULCAN/2D simulations and we see that doing so will 
have the opposite effect to that suggested by Buras et al. (2006).

\subsection{Implicit coupling: Protoneutron star
cooling and deleptonization for fixed density}
\label{sec-num-imp}

We have made several tests of temporal evolution of the initial structure
described in \S\ref{sec-num-st}.
At each timestep, we perform an implicit update of $T$ and $Y_e$,
as described in \S\ref{sect-imp}. Since we do not do any hydro in the
present tests, we keep density and velocity fixed during the evolution.
We start with an initial
timestep (typically 1 $\mu$s), and the next timesteps are set
dynamically depending on the value of the maximum relative change
of $T$ and $Y_e$. Specifically, the timestep $\Delta t$ is set to
$\Delta t = \Delta t_0 (\delta_0/|\delta_{\rm max}|)^p$, where
$\Delta t_0$ is the previous timestep, $\delta_{\rm max}$ is the
maximum relative change of $T$ or $Y_e$ (whichever is larger),
and $\delta_0$ and $p$ are pre-set parameters. We set 
$\delta_0 = 10^{-3}$ and $p=1/2$ in the following tests.

As discussed in \S\ref{sect-imp-lin}, the implicit update of $T$
and $Y_e$ is in principle obtained by Newton-Raphson iterations.
We found that our scheme converges very fast. When we
forced the Newton-Raphson loop to determine new $T$ and $Y_e$
to perform just one single iteration, we found that this 
procedure leads to results which are indistinguishable from
letting the iterations proceed until the convergence criterion
is reached (typically, the maximum relative change of $\delta T$
or $\delta Y_e$ is less that $10^{-3}$). The reason is that the second
and subsequent iterations typically lead to very small changes in the new
$T$ and $Y_e$ (and, moreover, only in one or a few radial zones),
so this tiny (or questionable) improvement in accuracy does not 
warrant an accompanied increase in computer time. Consequently,
in the following tests we have forced the number of total
iterations of the implicit update of $T$ and $Y_e$ to be 1.

As in \S\ref{sec-num-st}, our standard model considers 16 energy groups 
for each neutrino species, with logarithmically equidistant energies between
$E=1$ MeV and 300 MeV for $\nu_e$ neutrinos and 100 MeV for other
neutrino species. We test this setting by considering two models, calculated in serial mode,
with 8 groups per species and 32 groups per species, with the
same energy limits. The results are displayed in Fig. \ref{fig_e1},
where we show the temperature, and Fig. \ref{fig_e2}, where we show $Y_e$,
as a function of radius, for the initial configuration, and
for 0.3 seconds and 1 second of evolution. To reach 1 second,
the code required between 3500
to 6000 timesteps, with the larger number of timesteps for lower
number of energy groups. This was offset by a lower computer time
per timestep, so finally it required on a 1.6 GHz AMD Linux box
roughly 590, 730, and 1700 seconds of computer time to reach 1 second of evolution
for 8, 16, and 32 energy groups per species, respectively.

Figures \ref{fig_e1} and \ref{fig_e2} show 
that the models with 16 and 32 energy groups per species differ
only a little, thus validating our choice of 16 groups/species as a
default model. However, after 1 s of evolution one already sees some
differences in the temperature (mostly in the core between 10 and
20 km, where the relative difference reaches some 5\%), and in $Y_e$,
where we found differences of also about 5\% in the whole range between 70
and 200 km. On the other hand, considering only 8 groups per species
leads to larger inaccuracies, reaching 40\% in $T$ and 10-15\% in
$Y_e$. 

An important question is whether one has to perform a full angle-dependent
radiation transport solution, and thus to update the Eddington factor,
in every timestep, or whether one may instead update the Eddington
factor each $n$ timesteps. In Figs. \ref{fig_e3} and \ref{fig_e4}, 
we show the results of such a test. As before, we display the temperature and $Y_e$ after
0.3 seconds and 1 second of evolution, updating the Eddington factor every 10, 100,
and 1000 timesteps. The results are extremely encouraging: they
show that updating the Eddington factor even every 1000 timesteps
is quite accurate! This is not so crucial in the case of 1-D models,
where, thanks to a large efficiency of the transfer solver, the timestep
with the full transfer solution takes only about 2 to 3 times more
computer time than the timestep with only the moment equation solver.
However, this will be very important in 2 or 3 dimensions, where the
full angle-dependent solver will take proportionally larger chucks 
of computer time. 

In the case of the models of Figs. \ref{fig_e3} and \ref{fig_e4} 
it took 930 seconds, 760 seconds, and
730 seconds, to reach 1 second of evolution when an update of the
Eddington factor was done every 10, 100, and 1000 timesteps,
respectively. However, we stress that that the code is not fully
optimized for 1-D models since we view those models as mere tests
for choosing the best setups for future 2-D models.

%--------------------------------------------------------------------

\section{Discussion and Conclusions}
\label{conclusion}
In this paper, we develop the mixed-frame formalism for radiation
transport in both one and two dimensions and provide various 
solution methodologies and 1D numerical tests.  Velocity-dependence to O($v/c$),
anisotropic 
scattering, energy-dependent cross sections, energy redistribution
due to inelastic scattering, and radiation-matter coupling terms are derived 
and incorporated into the algorithm.  The equations in cylindrical, 
spherical, and planar coordinates are provided and the effects of 
bulk and shear radiation viscosity are automatically embedded into the 
approach. In two dimensions, rotation is consistently 
included, extending the Eddington factor to an Eddington tensor with five
components and introducing an azimuthal component to the radiation flux.  
The zeroth- and first-moment equations are derived and their
roles in the transport solution and the radiation-matter coupling formalism 
are fully explored.  Various solution philosophies, such as  
Discontinuous Finite Element (DFE), Feautrier, and short-characteristics,
and various convergence approaches, such as the Accelerated-Lambda-Iteration (ALI),
Ng acceleration, and GMRES, are also implemented and compared.

Many radiation-hydrodynamic problems do not require an 
exquisite treatment of spectral line transport, but a 
good treatment of continuum transport. When such is the case, the mixed-frame
approach is clearly superior to comoving-frame formalisms for which 
the velocity-dependence is complicated. The virtues of the 
mixed-frame perspective are many and include simple velocity
dependence with no velocity derivatives, straight characteristics, simple
physical interpretation, and clear generalization to higher dimensions.
Moreover, since multi-dimensional radiation-hydrodynamic simulations
frequently employ Eulerian grids and hydrodynamics, calculation
of the radiation quantities in the laboratory (Eulerian) frame 
would seem to be natural. 

We also stress that one can still use the mixed-frame formalism
for treating spectral lines, provided one works with high enough
frequency resolution. In the past, this was considered an
excessive requirement. However, modern stationary atmosphere models in
1-D already achieve essentially ``full'' frequency
resolution, e.g., with frequency spacing of 0.75 fiducial Doppler widths
throughout the whole spectrum, that employs some 2 to 3$\times 10^5$
frequency points (Lanz \& Hubeny 2003). In this spirit, one can use
to advantage the mixed-frame approach as well to for such models 
and for general mass motions.

The equations and algorithms we have developed can be used for the 
transport of any radiation, in particular photons and neutrinos. 
In addition, the two-moment closure approach to approximate transport that 
springs naturally from our fully angle-dependent formalism is superior to the
incorporation
of ad hoc flux limiters into the zeroth-moment (energy) radiation equation.
Not only is the finite ``speed of light" effect automatically included 
in a two-moment closure, but the solution can be made arbitrarily accurate,
depending upon how often the correct Eddington tensor is updated.

Radiation transport methods are core tools of the astrophysicist.  
Time-dependent techniques to address transport and the coupling 
of radiation with matter are of central concern to the theorist 
trying to explain the dynamical phenomena of the Universe. 
Therefore, along with the ongoing advance of the computational arts,
the mixed-frame formalism and numerical techniques we have derived 
and tested in this paper should prove of great value for the 
detailed future investigation of complex astrophysical problems.

%--------------------------------------------------------------------

\acknowledgments

We thank Luc Dessart, Eli Livne, Jeremiah Murphy, and Todd Thompson
for fruitful discussions during the course of this work
and acknowledge support from both the Scientific Discovery 
through Advanced Computing
(SciDAC) program of the DOE, grant number DE-FC02-01ER41184,
and from the NSF under grant AST-0504947.

\begin{appendix}

\section{Mixed-frame treatment of neutrino inelastic scattering}
\label{appendix}

In order to derive the proper expressions for the neutrino 
inelastic scattering in the mixed frame we need to
consider very carefully the transformation  of the individual
components of the transfer equation.
Let us first repeat the Boltzmann transport equation for the neutrino
occupation probability $f$:
\begin{equation}
\label{bte}
\frac{1}{c}  \frac{\partial f}{\partial t} + (\vecn\cdot\nabla)\, f =
C[f] \, ,
\end{equation}
where $C = C_{\rm NES}$ is the collisional integral (net source term)
for inelastic neutrino scattering (the dominant contribution is typically that of 
neutrino-electron scattering). Here, we assume that only the inelastic scattering
term is present, since we have already treated the other terms in
\S\ref{sec-form}. 

The relation between the occupation probability and the specific intensity
is given by eq. (\ref{inrel}),
\begin{equation}
I = b(\nu)\, f\, .
\end{equation}
where the conversion factor $b(\nu)$ is given by
\begin{equation}
b(\nu) = \frac{\nu^3}{h^3 c^2}\, .
\end{equation}
The transport equation (\ref{bte}) is written as
\begin{equation}
\label{nte}
\frac{1}{c}  \frac{\partial I_\nu}{\partial t} + 
(\vecn\cdot\nabla)\, I_\nu = b(\nu) C \, .
\end{equation}
So far, the equations have been written without any reference to a coordinate
frame. Transfer equations (\ref{nte}) and (\ref{bte}) have the same form
in all frames. Since the collision term is evaluated in the frame
in which the matter is at rest (comoving frame), we will now write
the transfer equation explicitly in the comoving frame, viz.
\begin{equation}
\label{rtecom}
\frac{1}{c}  \frac{\partial I_0(\nu_0, \vecn_0)}{\partial t_0} +
(\vecn_0\cdot\nabla_0) \,
I_0(\nu_0, \vecn_0) = b(\nu_0) C_0 \, ,
\end{equation}
where, as usual, subscript $0$ indicates quantities in the comoving
frame.
The left-hand side of this equation transforms as
\begin{equation}
\left(\frac{\nu_0}{\nu}\right) \left[
\frac{1}{c}  \frac{\partial}{\partial t_0} + (\vecn_0\cdot\nabla_0)\right]
\left(\frac{\nu}{\nu_0}\right)^3  I_0(\nu_0, \vecn_0) =
\left[ \frac{1}{c}  \frac{\partial}{\partial t} + (\vecn\cdot\nabla)
\right]
I(\nu, \vecn)\, ,
\end{equation}
so that
\begin{equation}
\left(\frac{1}{c}  \frac{\partial}{\partial t} + \vecn\cdot\nabla \right) 
I(\nu, \vecn)
= \left(\frac{\nu}{\nu_0}\right)^2 b(\nu_0)\, C_0(\nu_0, \vecn_0) \, .
\end{equation}
We write this equation as
\begin{equation}
\left(\frac{1}{c}  \frac{\partial}{\partial t} + \vecn\cdot\nabla \right) 
I(\nu, \vecn) = C(\nu, \vecn) \, ,
\end{equation}
where
\begin{equation}
C(\nu, \vecn)
= \left(\frac{\nu}{\nu_0}\right)^2 b(\nu_0)\, C_0(\nu_0, \vecn_0) \, .
\end{equation}
We now have to express $C(\nu,\vecn)$ through the comoving-frame
cross sections and redistribution functions and the inertial-frame
specific intensities.

The source term $C_0$ is given by Bruenn (1985):
\begin{eqnarray}
C_0(\nu_0, \vecn_0) = [1 - f_0(\nu_0,\vecn_0)] \oint\int d\omega_0^\prime
\frac{(\nu_0^\prime)^2\, d\nu_0^\prime}{c(2\pi \hbar c)^3}\,
f_0(\nu_0^\prime, \vecn_0^\prime)\,
R_0^{\rm in}(\nu_0, \nu_0^\prime, \vecn_0\cdot\vecn_0^\prime) \nonumber\\
- f_0(\nu_0,\vecn_0) \oint\int d\omega_0^\prime
\frac{(\nu_0^\prime)^2\, d\nu_0^\prime}{c(2\pi \hbar c)^3}\,
[1 - f_0(\nu_0^\prime, \vecn_0^\prime)]\,
R_0^{\rm out}(\nu_0, \nu_0^\prime, \vecn_0\cdot\vecn_0^\prime) \, ,
\end{eqnarray}
where $R_0^{\rm out,in}$ are the scattering kernels.
This equation is easily rewritten using the specific intensity:
\begin{eqnarray}
\label{b0}
c^2 C_0(\nu_0, \vecn_0) =
\left[1 - \frac{I_0(\nu_0,\vecn_0)}{b(\nu_0)} \right] \oint\int
d\omega_0^\prime
\frac{d\nu_0^\prime}{\nu_0^\prime}\, I_0(\nu_0^\prime, \vecn_0^\prime)\,
R_0^{\rm in}(\nu_0, \nu_0^\prime, \vecn_0\cdot\vecn_0^\prime) \nonumber\\
- \frac{I_0(\nu_0,\vecn_0)}{b(\nu_0)} \oint\int d\omega_0^\prime
\frac{d\nu_0^\prime)}{\nu_0^\prime}\,
[b(\nu_0^\prime) - I_0(\nu_0^\prime, \vecn_0^\prime)]\,
R_0^{\rm out}(\nu_0, \nu_0^\prime, \vecn_0\cdot\vecn_0^\prime) \, .
\end{eqnarray}
Because $I_\nu/b_\nu$ is invariant, i.e.
$I(\nu_0,\vecn_0)/b(\nu_0) = I(\nu, \vecn)/b(\nu)$,
we have
\begin{eqnarray}
\left(\frac{\nu}{\nu_0}\right)^2 b(\nu_0)\, 
\left[1 - \frac{I_0(\nu_0,\vecn_0)}{b(\nu_0)} \right] =
\left(\frac{\nu}{\nu_0}\right)^2 \left(\frac{\nu_0}{\nu}\right)^3\,
b(\nu)  \left[1 - \frac{I(\nu,\vecn)}{b(\nu)} \right]\nonumber\\ =
\left(\frac{\nu_0}{\nu}\right) \left[b(\nu) - I(\nu,\vecn)\right]\, ,
\end{eqnarray}
and, thus,
\begin{equation}
C(\nu, \vecn) = [b(\nu) - I(\nu,\vecn)]\, \eta^{\rm sc}(\nu,\vecn) -
I(\nu,\vecn)\, \left[ \kappa^{\rm sc}(\nu, \vecn) - 
\kappa^{\rm stim}(\nu, \vecn) \right]\, ,
\end{equation}
where
\begin{equation}
\label{etasc}
\eta^{\rm sc}(\nu,\vecn) = \frac{1}{c^2}
\left(\frac{\nu_0}{\nu}\right)\oint\int d\omega_0^\prime
\frac{d\nu_0^\prime}{\nu_0^\prime}\, I_0(\nu_0^\prime, \vecn_0^\prime)\,
R_0^{\rm in}(\nu_0, \nu_0^\prime, \vecn_0\cdot\vecn_0^\prime)\, ,
\end{equation}
\begin{equation}
\label{kappasc}
\kappa^{\rm sc}(\nu, \vecn) = \frac{1}{c^2}\left(\frac{\nu_0}{\nu}\right)
\oint\int d\omega_0^\prime
\frac{d\nu_0^\prime}{\nu_0^\prime}\, b(\nu_0^\prime)\,
R_0^{\rm out}(\nu_0, \nu_0^\prime, \vecn_0\cdot\vecn_0^\prime) \, ,
\end{equation}
and 
\begin{equation}
\label{kappasti}
\kappa^{\rm stim}(\nu,\vecn) = \frac{1}{c^2}
\left(\frac{\nu_0}{\nu}\right)\oint\int d\omega_0^\prime
\frac{d\nu_0^\prime}{\nu_0^\prime}\, I_0(\nu_0^\prime, \vecn_0^\prime)\,
R_0^{\rm out}(\nu_0, \nu_0^\prime, \vecn_0\cdot\vecn_0^\prime)\, .
\end{equation}
In evaluating eqs. (\ref{etasc}) - (\ref{kappasti}), 
we use the following transformation properties
\begin{equation}
\frac{d\nu_0^\prime}{\nu_0^\prime} = \frac{d\nu^\prime}{\nu^\prime}\, 
\end{equation}
and 
\begin{equation}
I_0(\nu_0^\prime,\vecn_0^\prime)\, d\omega_0^\prime = 
\left(\frac{\nu_0^\prime}{\nu^\prime}\right)^3 I(\nu^\prime,\vecn^\prime)
\left(\frac{\nu^\prime}{\nu_0^\prime}\right)^2 d\omega^\prime =
\left(\frac{\nu_0^\prime}{\nu^\prime}\right) I(\nu^\prime,\vecn^\prime)\,
d\omega^\prime\, ,
\end{equation}
where we use the primed quantities ($\nu_0^\prime, \vecn_0^\prime,
\nu^\prime,
\vecn^\prime$) to refer to incoming (``absorbed'') neutrinos, while
unprimed quantities ($\nu_0, \vecn_0, \nu, \vecn$) refer to outgoing
(``emitted'') neutrinos. Finally, we assume that the scattering kernels $R$ are
expanded in Legendre polynomials, and only the first two terms are retained
(Bruenn 1985; Thompson et al. 2003):
\begin{equation}
R_0^{\rm out} = \frac{1}{2} \widetilde\Phi_0^{\rm out}(\nu_0,\nu_0^\prime)
+ \frac{3}{2} \widetilde\Phi_1^{\rm out}(\nu_0,\nu_0^\prime)\, 
\vecn_0\cdot\vecn_0^\prime\, .
\end{equation}
There is an analogous expression for $R_0^{\rm in}$. 
Although in the comoving frame one has a simple relation between the kernels, 
$R_0^{\rm in} = \exp[\nu_0^\prime - \nu_0)/T] R_0^{\rm out}$,
we keep both kernels separate because a transformation 
of the exponential to the inertial frame leads to very 
cumbersome expressions.

We now absorb the factor $1/c^2$, as well as the corresponding
Legendre factors $1/2$ and $3/2$, into the kernels and write
\begin{equation}
\frac{1}{c^2} R_0^{\rm in} = \Phi_0^{\rm in}(\nu_0,\nu_0^\prime)
+ \Phi_1^{\rm in}(\nu_0,\nu_0^\prime)\, \vecn_0\cdot\vecn_0^\prime\, ,
\end{equation}
i.e.
\begin{equation}
\Phi_0^{\rm in} = \frac{1}{2 c^2} \widetilde\Phi_0^{\rm in}\,\quad , \quad
\Phi_1^{\rm in} = \frac{3}{2 c^2} \widetilde\Phi_1^{\rm in}\, ,
\end{equation}
and analogously for $R^{\rm out}$. In the following, we use only the
comoving-frame $\Phi_0$ and $\Phi_1$ so we do not need to use a more
cumbersome, though more consistent, notation such as $(\Phi_i)_0$,
where the first subscript ($i=0, 1$) would
mean the order of the corresponding expansion
term, and the second subscript ($0$) would indicate the comoving-frame
quantity. 

In order to evaluate eq. (\ref{etasc}), we have to expand $\Phi_0$
and $\Phi_1$, 
\begin{equation}
\Phi_0^{\rm x}(\nu_0, \nu_0^\prime) = \Phi_0^{\rm x}(\nu, \nu^\prime) 
- \vecn\cdot\vecw\, \frac{\partial\Phi_0^{\rm x}}{\dln} 
- \vecn^\prime\cdot\vecw\, \frac{\partial\Phi_0^{\rm x}}{\dlnp} \, ,
\end{equation}
and
\begin{equation}
\Phi_1^{\rm x}(\nu_0, \nu_0^\prime) = \Phi_1^{\rm x}(\nu, \nu^\prime) 
- \vecn\cdot\vecw\, \frac{\partial\Phi_1^{\rm x}}{\dln} 
- \vecn^\prime\cdot\vecw\, \frac{\partial\Phi_1^{\rm x}}{\dlnp} \, ,
\end{equation}
where superscripts ``x'' stands for ``in'' or ``out''.
Equation (\ref{etasc}) then reads
\begin{eqnarray}
\eta^{\rm sc}(\nu,\vecn) =\left(\frac{\nu_0}{\nu}\right)
\int \frac{d\nu^\prime}{\nu^\prime} \oint d\omega^\prime  
(1 - \vecn^\prime \cdot \vecw) I(\nu^\prime, \vecn^\prime)\nonumber\\
\left\{\Phi_0(\nu, \nu^\prime) 
- \vecn\cdot\vecw \frac{\partial\Phi_0}{\dln} 
- \vecn^\prime\cdot\vecw \frac{\partial\Phi_0}{\dlnp}+ \right. \nonumber\\
\left. \left[ \Phi_1(\nu, \nu^\prime) 
- \vecn\cdot\vecw \frac{\partial\Phi_1}{\dln} 
- \vecn^\prime\cdot\vecw \frac{\partial\Phi_1}{\dlnp} \right]
[\vecn\cdot\vecn^\prime - (1+\vecn\cdot\vecn^\prime)\,
(\vecn\cdot\vecw + \vecn^\prime\cdot\vecw)] 
\right\} \, ,
\end{eqnarray}
where we used eq. (\ref{nntrans}) for transforming 
$\vecn_0\cdot\vecn_0^\prime$. We express
\begin{equation}
\label{edef}
\eta^{\rm sc}(\nu, \vecn) = \left(\frac{\nu_0}{\nu}\right)
\int \frac{d\nu^\prime}{\nu^\prime}\, e(\nu, \nu^\prime, \vecn)\, ,
\end{equation}
where for $e$ we obtain after some algebra that is completely
analogous to that used in deriving eqs. (\ref{rtemx}) - (\ref{r11}):
\begin{eqnarray}
\label{esc}
e(\nu, \nu^\prime, \vecn) = 
J(\nu^\prime) \left[ \Phi_0^{\rm in} - n^j w_j 
\left( \frac{\partial\Phi_0^{\rm in}}{\dln} + \Phi_1^{\rm in} \right)
\right] 
+ \nonumber\\
H^j(\nu^\prime) 
\left[ n_j \Phi_1^{\rm in} - w_j \left(\Phi_0^{\rm in} + \Phi_1^{\rm in} + 
\frac{\partial\Phi_0^{\rm in}}{\dlnp} \right)
- n_j n^k w_k \left( \Phi_1^{\rm in} + \frac{\partial\Phi_1^{\rm in}}{\dln} 
\right) \right] - \nonumber\\
K^{jk}(\nu^\prime) n_j w_k 
\left( 2 \Phi_1^{\rm in} + \frac{\partial\Phi_1^{\rm in}}{\dlnp}\right) \, .
\end{eqnarray}
According to eqs. (\ref{edef}) and (\ref{esc}), 
\begin{equation}
\eta^{\rm sc}(\nu, \vecn) = \left(\frac{\nu_0}{\nu}\right)
 \int \frac{d\nu^\prime}{\nu^\prime}\,
 e(\nu, \nu^\prime, \vecn)
= (1 -\vecn\cdot\vecw) 
\int \frac{d\nu^\prime}{\nu^\prime}\, e(\nu, \nu^\prime, \vecn)\, ,
\end{equation}
and, thus,
\begin{eqnarray}
\label{etasci}
\eta^{\rm sc}(\nu, \vecn) = 
\int \frac{d\nu^\prime}{\nu^\prime}\, 
J(\nu^\prime) \left[ \Phi_0^{\rm in} - n^j w_j 
\left(\Phi_0^{\rm in} + \frac{\partial\Phi_0^{\rm in}}{\dln} + 
\Phi_1^{\rm in} \right) 
\right] \nonumber\\
+ \int \frac{d\nu^\prime}{\nu^\prime} \,
H^j(\nu^\prime) \left[ n_j \Phi_1^{\rm in}
- w_j \left(\Phi_0^{\rm in} + \frac{\partial\Phi_0^{\rm in}}{\dlnp} +
\Phi_1^{\rm in} \right)
- n_j n^k w_k \left(2 \Phi_1^{\rm in} + \frac{\partial\Phi_1^{\rm
in}}{\dln} \right) \right] \nonumber\\
- \int \frac{d\nu^\prime}{\nu^\prime}\, K^{jk}(\nu^\prime)\, n_j w_k 
\left( 2 \Phi_1^{\rm in} + \frac{\partial\Phi_1^{\rm in}}{\dlnp}\right) \,
.
\end{eqnarray}
Analogously,
\begin{eqnarray}
\label{kapstii}
\kappa^{\rm stim}(\nu, \vecn) = 
\int \frac{d\nu^\prime}{\nu^\prime}\, 
J(\nu^\prime) \left[ \Phi_0^{\rm out} - n^j w_j 
\left(\Phi_0^{\rm out} + \frac{\partial\Phi_0^{\rm out}}{\dln} + 
\Phi_1^{\rm out} \right) 
\right] \nonumber\\
+ \int \frac{d\nu^\prime}{\nu^\prime} \,
H^j(\nu^\prime) \left[ n_j \Phi_1^{\rm out}
- w_j \left(\Phi_0^{\rm out} + \frac{\partial\Phi_0^{\rm out}}{\dlnp} +
\Phi_1^{\rm out} \right)
- n_j n^k w_k \left(2 \Phi_1^{\rm out} + \frac{\partial\Phi_1^{\rm out}}{\dln}
\right) \right] \nonumber\\
- \int \frac{d\nu^\prime}{\nu^\prime}\, K^{jk}(\nu^\prime)\, n_j w_k 
\left( 2 \Phi_1^{\rm out} + \frac{\partial\Phi_1^{\rm out}}{\dlnp}\right) \,
.
\end{eqnarray}
Finally, the ``absorption'' coefficient is given by eq. (\ref{kappasc}),
\begin{eqnarray}
\label{kappasc2}
\kappa^{\rm sc}(\nu, \vecn) = \left(\frac{\nu_0}{\nu}\right) 
\kappa_0^{\rm sc}(\nu_0, \vecn_0) = \nonumber\\
\left(\frac{\nu_0}{\nu}\right)
\oint\int d\omega_0^\prime
\frac{d\nu_0^\prime}{\nu_0^\prime}\, b(\nu_0^\prime)\,
\left[\Phi_0^{\rm out}(\nu_0,\nu_0^\prime) + \Phi_1^{\rm
out}((\nu_0,\nu_0^\prime)
\, \vecn_0\cdot\vecn_0^\prime) \right] \, ,
\end{eqnarray}
where $\kappa_0^{\rm sc}$ represents the comoving-frame coefficient.
Since they do not depend on specific intensity,
the integrations can be done in advance and the resulting coefficients
tabulated.

\end{appendix}

\newpage

%%%%%%%%%%%%%%%%%%%%%%%%%%%%%%%%%%%%%%%%%%%%%%%%%%%

\begin{figure}
\epsscale{1.0}
\plotone{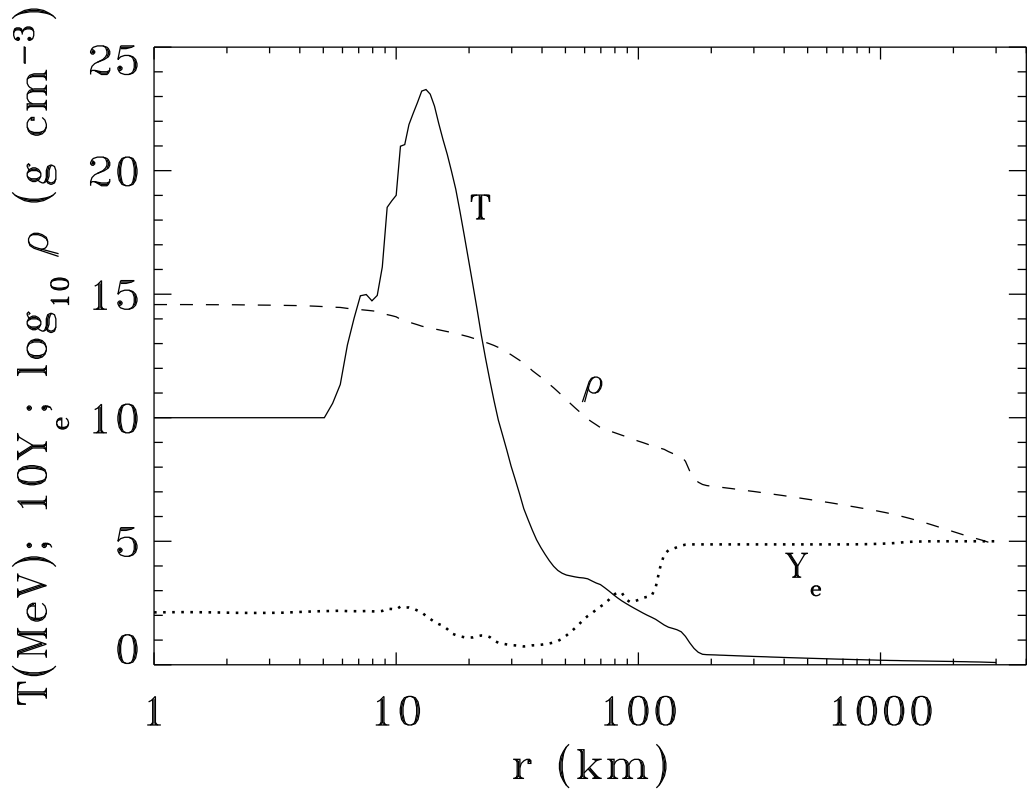}
\caption{The temperature ($T$), density ($\rho$), and the electron
fraction ($Y_e$) as a function of radius, for the fictitious model
used in our numerical tests.
}
\label{fig1}
\end{figure}

%%%%%%%%%%%%%%%%%%%%%%%%%%%%%%%%%%%%%%%%%%%%%%%%%%%

\begin{figure}
\epsscale{1.0}
\plotone{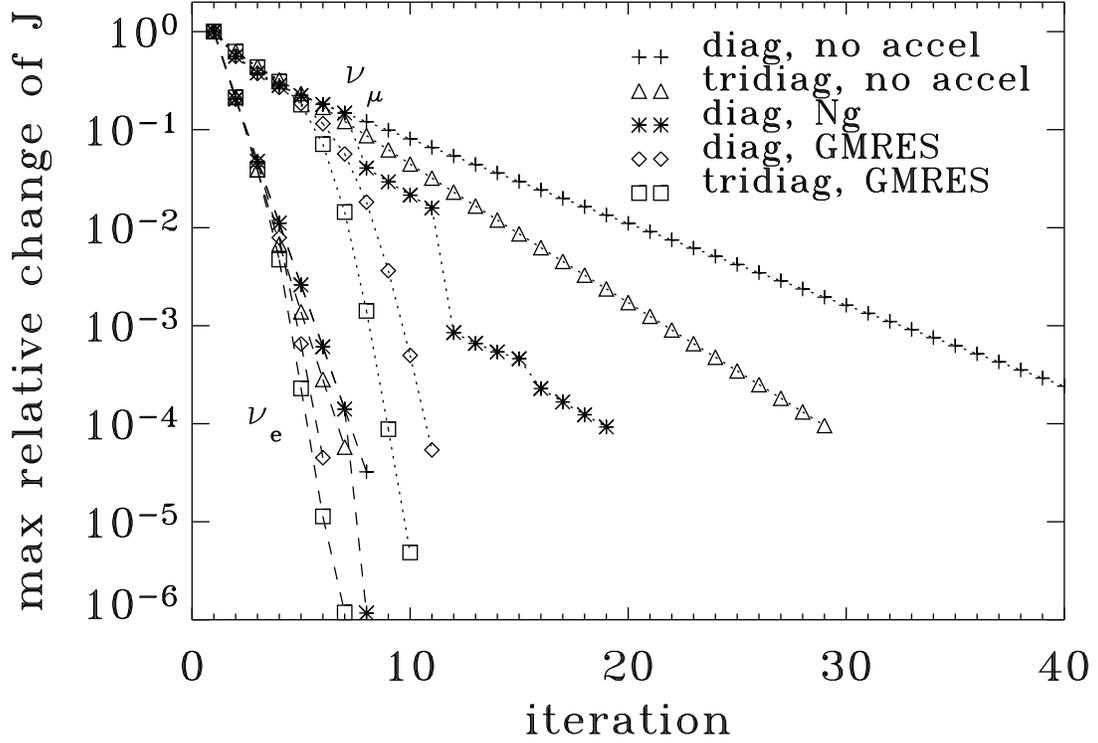}
\caption{Convergence pattern for $\nu_e$ and $\nu_\mu$ neutrinos
for $E=8.6$ MeV, and for 5 different setups of the 
iterative transport solver. See text for a discussion.
}
\label{fig2}
\end{figure}

%%%%%%%%%%%%%%%%%%%%%%%%%%%%%%%%%%%%%%%%%%%%%%%%%%%

\begin{figure}
\epsscale{1.0}
\plotone{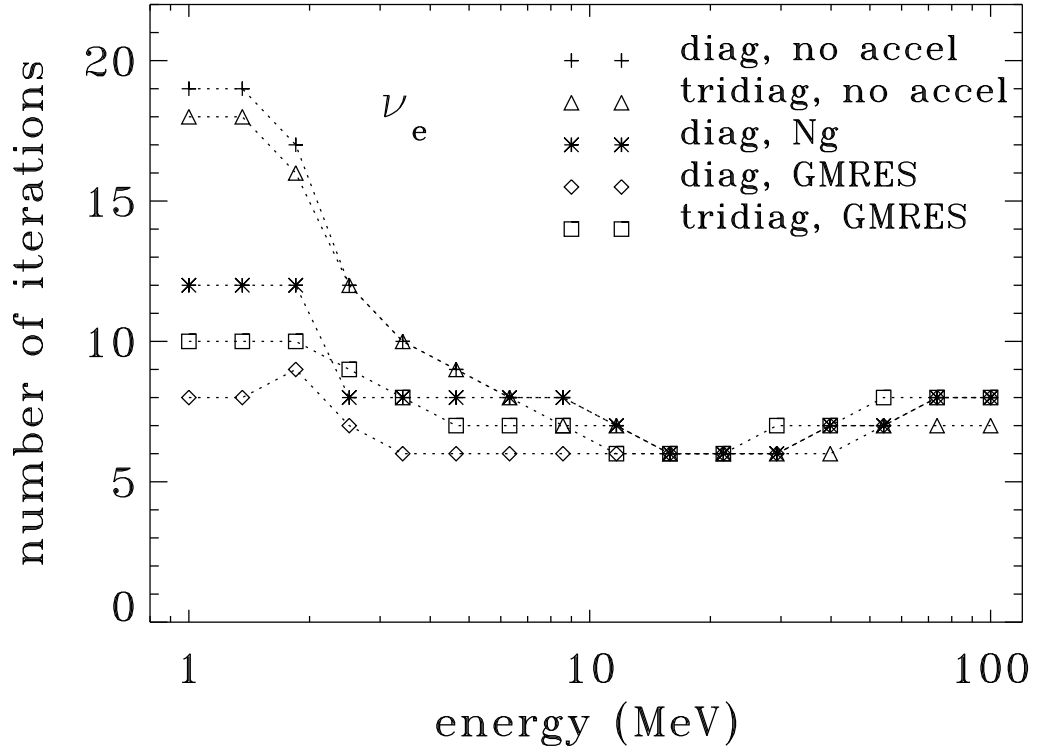}
\caption{The number of iterations to reach a converged
model, defined by the convergence criterion 
$\max(\delta J/J) < 10^{-5}$, for the individual setups of
the iterative transfer solver, for all energy groups of
the $\nu_e$ neutrino.
}
\label{fig3}
\end{figure}
%%%%%%%%%%%%%%%%%%%%%%%%%%%%%%%%%%%%%%%%%%%%%%%%%%%

\begin{figure}
\epsscale{1.0}
\plotone{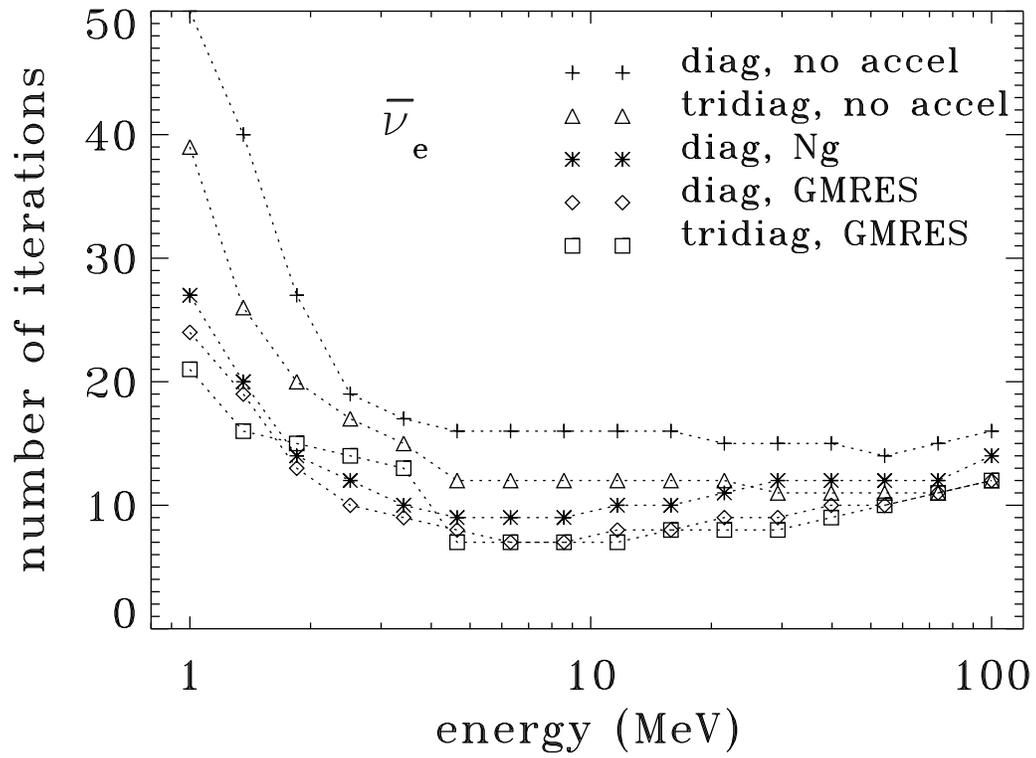}
\caption{The same as Fig. \ref{fig3}, but for $\bar\nu_e$
neutrinos.
}
\label{fig4}
\end{figure}

%%%%%%%%%%%%%%%%%%%%%%%%%%%%%%%%%%%%%%%%%%%%%%%%%%%

\begin{figure}
\epsscale{1.0}
\plotone{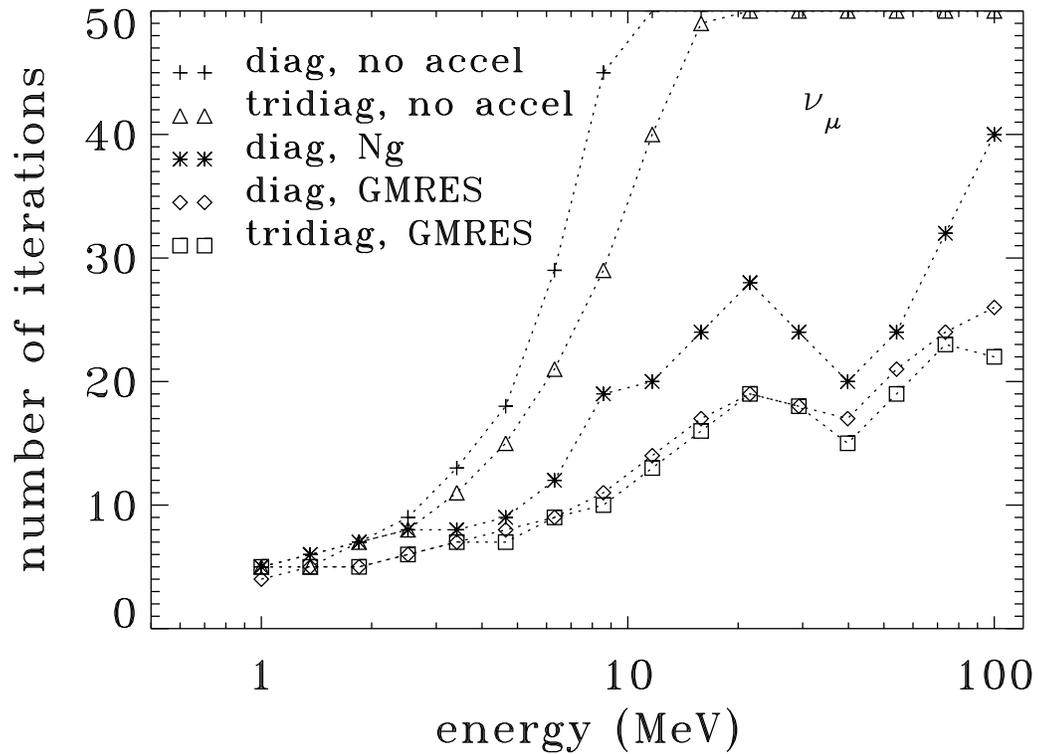}
\caption{The same as Fig. \ref{fig3}, but for
$\nu_\mu$ neutrinos.
}
\label{fig4a}
\end{figure}

%%%%%%%%%%%%%%%%%%%%%%%%%%%%%%%%%%%%%%%%%%%%%%%%%%%

\begin{figure}
\epsscale{1.0}
\plotone{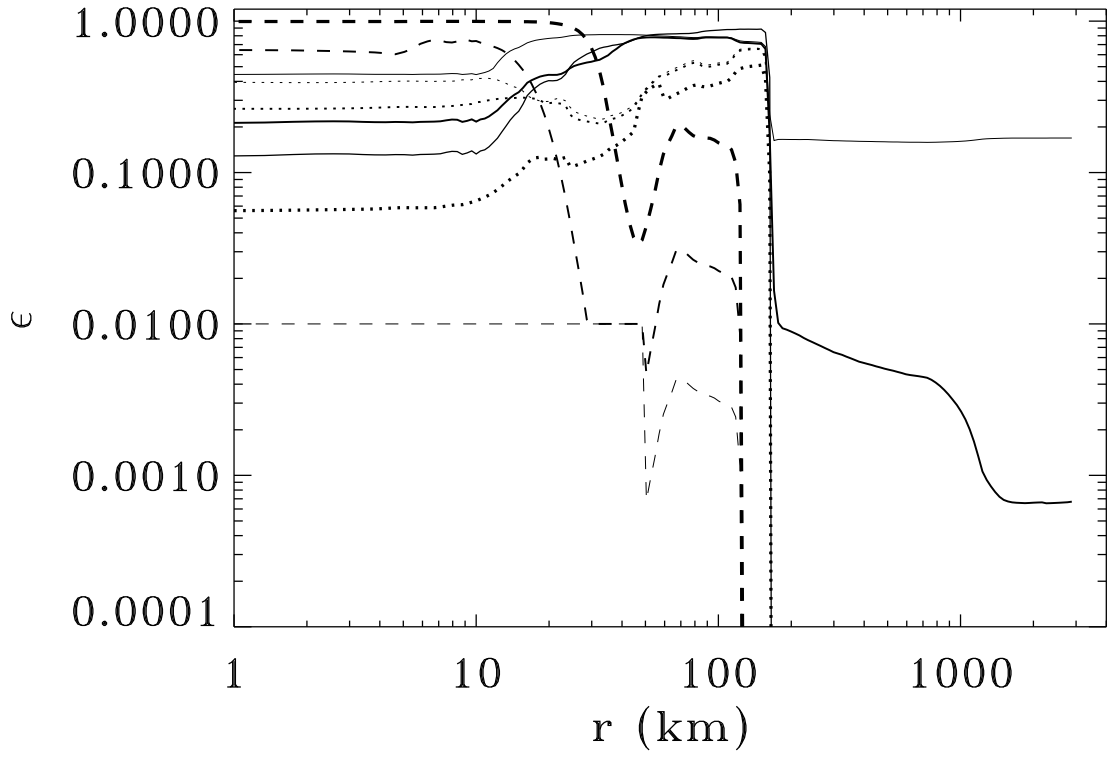}
\caption{Parameter $\epsilon = \kappa/(\kappa+\sigma)$
as a function of radius for the three energy groups:
$E=1$ MeV - the thickest lines; $E=10$ MeV - thinner
lines; and $E=100$ MeV - thin lines), for three neutrino
species: $\nu_e$ neutrinos -- solid lines, $\bar\nu_e$ --
dotted lines, and $\nu_\mu$ neutrinos --
dashed lines.
}
\label{fig_eps}
\end{figure}

%%%%%%%%%%%%%%%%%%%%%%%%%%%%%%%%%%%%%%%%%%%%%%%%%%%

\begin{figure}
\epsscale{0.77}
\plotone{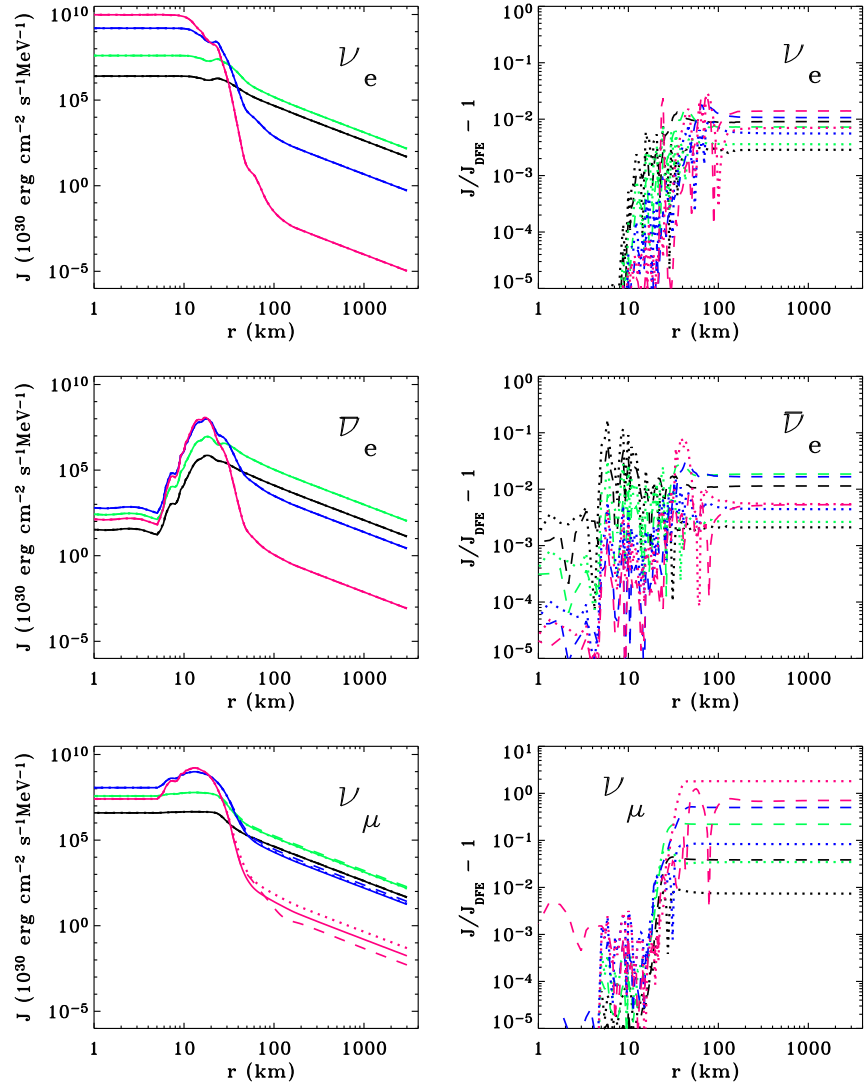}
\caption{A comparison of the three different formal solvers,
namely the Discontinuous Finite Element (DFE) scheme, the first-order
short characteristics (SC) scheme, and the Feautrier scheme.
On the left-hand panels, the mean intensity $J$ is
plotted as a function of radius for four selected energy groups,
with $E = 4.6$ MeV (black), 11.7 MeV (green), 40 MeV (blue), 
and 74 MeV (red).
Full lines display the DFE results, dotted lines the SC results,
and dashed lines the results obtained by the Feautrier scheme.
The right-hand panels display the corresponding
relative differences of the mean intensities with those computed
by the DFE scheme, namely $J({\rm solver})/J({\rm DFE}) - 1$:
dashed lines -- Feautrier; dotted lines -- SC.
}
\label{fig_sol2}
\end{figure}

%%%%%%%%%%%%%%%%%%%%%%%%%%%%%%%%%%%%%%%%%%%%%%%%%%%

\begin{figure}
\epsscale{1.0}
\plotone{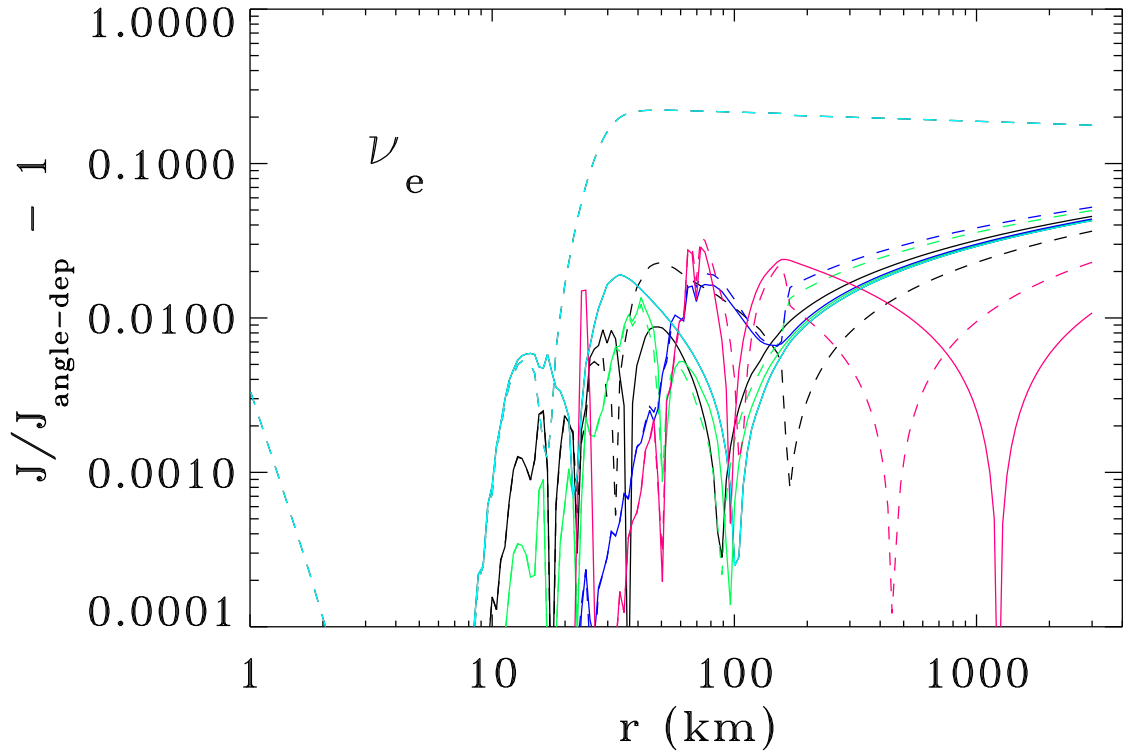}
\caption{
A comparison of the relative differences of the mean intensity 
computed by the angle-dependent solver and by
the moment equation solver, 
$J({\rm angle\ dep.\ solver})/J({\rm moment\ solver}) - 1$.
%, for the same four four energy groups
%of $\nu_e$ neutrino as in Fig. \ref{fig_sol2}.
The dashed lines represent the
original solver, without the sphericity factors; solid lines
the moment solver with sphericity factors.
The color pattern is analogous to that used in Fig. \ref{fig_sol2};
the only difference is an added energy group with $E=1$ MeV, represented
by the light-blue [turquoise] lines.
}
\label{fig_mom1}
\end{figure}

%%%%%%%%%%%%%%%%%%%%%%%%%%%%%%%%%%%%%%%%%%%%%%%%%%%

\begin{figure}
\epsscale{1.0}
\plotone{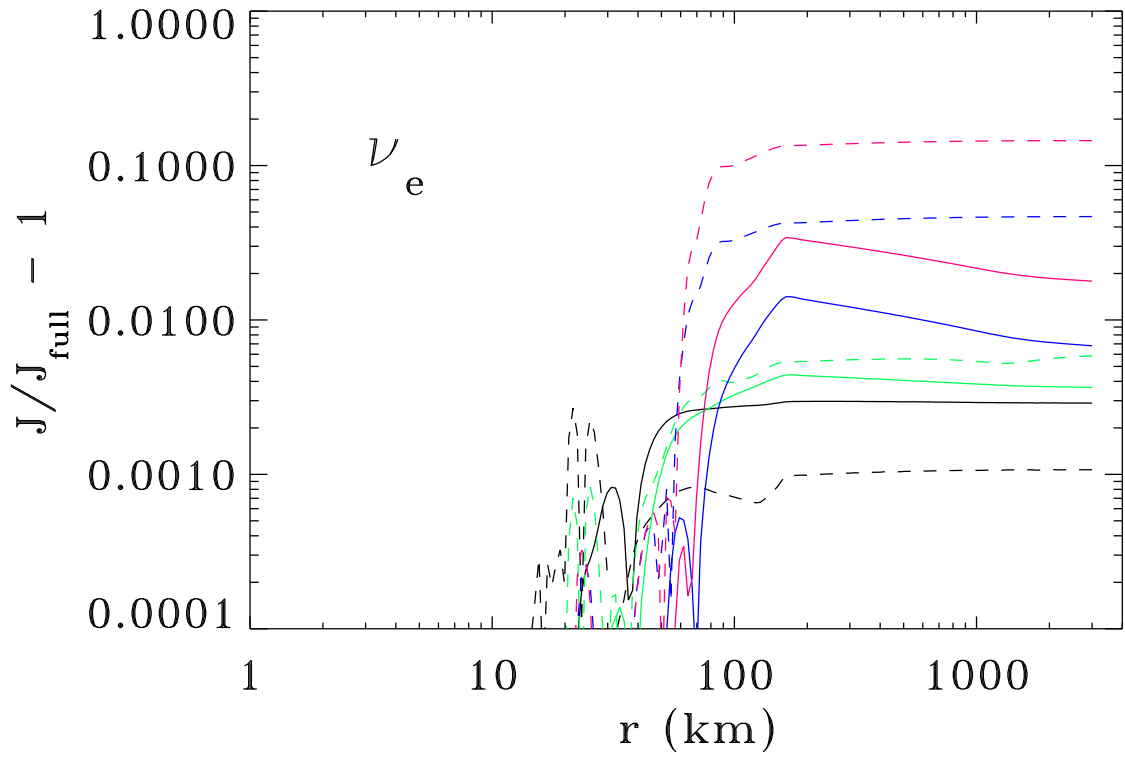}
\caption{A comparison of the relative differences of mean intensity 
for the full velocity-dependent, anisotropic scattering solution
with that setting $\delta=0$ (that is, assuming isotropic
scattering), but keeping the velocity-dependent terms (solid lines);
and with setting $v=0$ (but keeping $\delta\not=0$ -- dashed lines.
The color pattern is the same as in Fig. \ref{fig_sol2}.
}
\label{fig_vel1}
\end{figure}

%%%%%%%%%%%%%%%%%%%%%%%%%%%%%%%%%%%%%%%%%%%%%%%%%%%

\begin{figure}
\epsscale{1.0}
\plotone{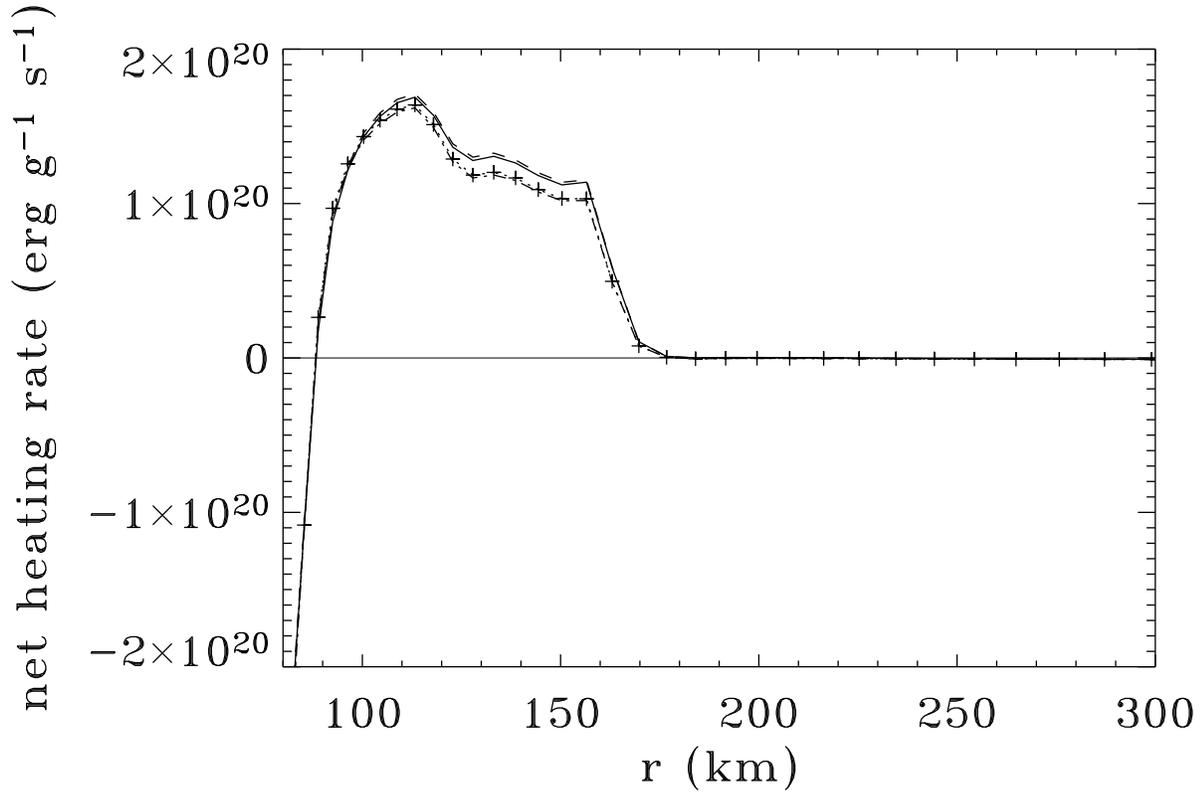}
\caption{A comparison of the net heating rate (``gain'') for the full 
velocity-dependent, anisotropic scattering solution (full lines)
with that setting $\delta=0$ (that is, assuming isotropic
scattering), but keeping the velocity-dependent terms (dashed lines);
and with setting $w=0$ (but keeping $\delta\not=0$) -- dotted lines.
Crosses (which lie indistinguishably close to the dotted line)
represent the solution with both $w=0$ and $\delta=0$.
Note that the sign of the $w$-correction to the net gain is different
from that found in the comoving-frame formalism (Buras et al. 2006).
See text for a discussion.
}
\label{fig10}
\end{figure}

%%%%%%%%%%%%%%%%%%%%%%%%%%%%%%%%%%%%%%%%%%%%%%%%%%%

\begin{figure}
\epsscale{1.0}
\plotone{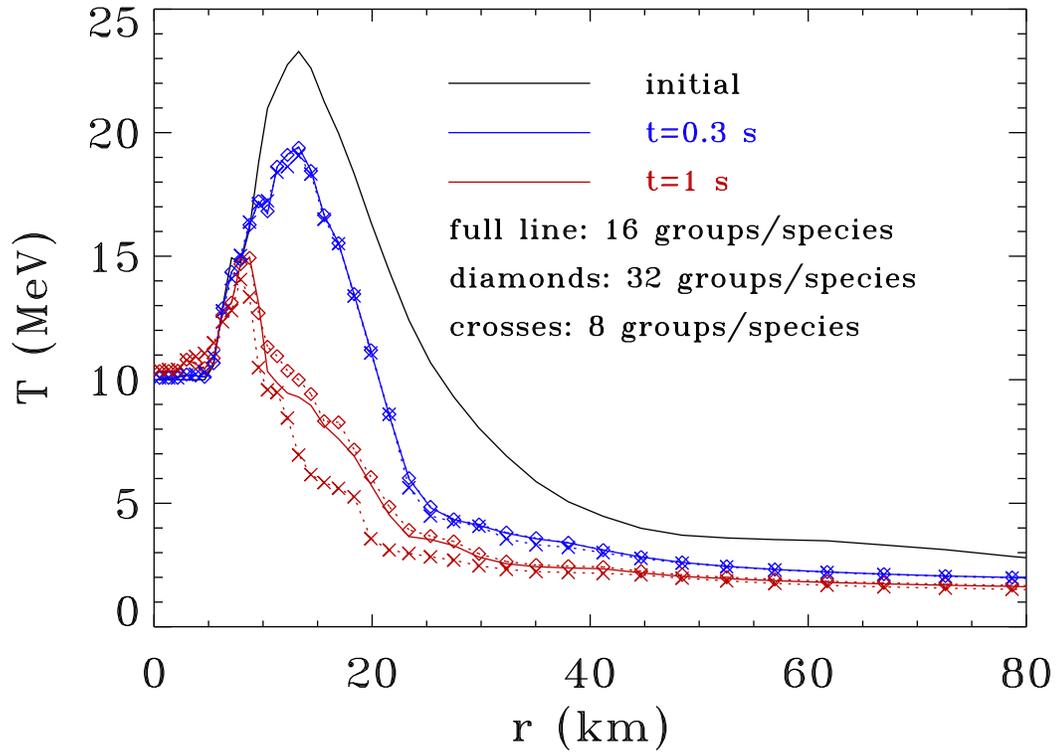}
\caption{Temperature as a function of radius, for the initial configuration
(full black line), and for 0.3 seconds (blue lines/symbols) and 1 second 
(red lines and symbols) of evolution. Solid lines display the evolution 
where we employ 16 energy groups per neutrino species;
diamonds display the evolution where we employ 32 energy groups per neutrino
species, and crosses display models where we employ 8 energy groups 
per neutrino species.
}
\label{fig_e1}
\end{figure}

%%%%%%%%%%%%%%%%%%%%%%%%%%%%%%%%%%%%%%%%%%%%%%%%%%%

\begin{figure}
\epsscale{1.0}
\plotone{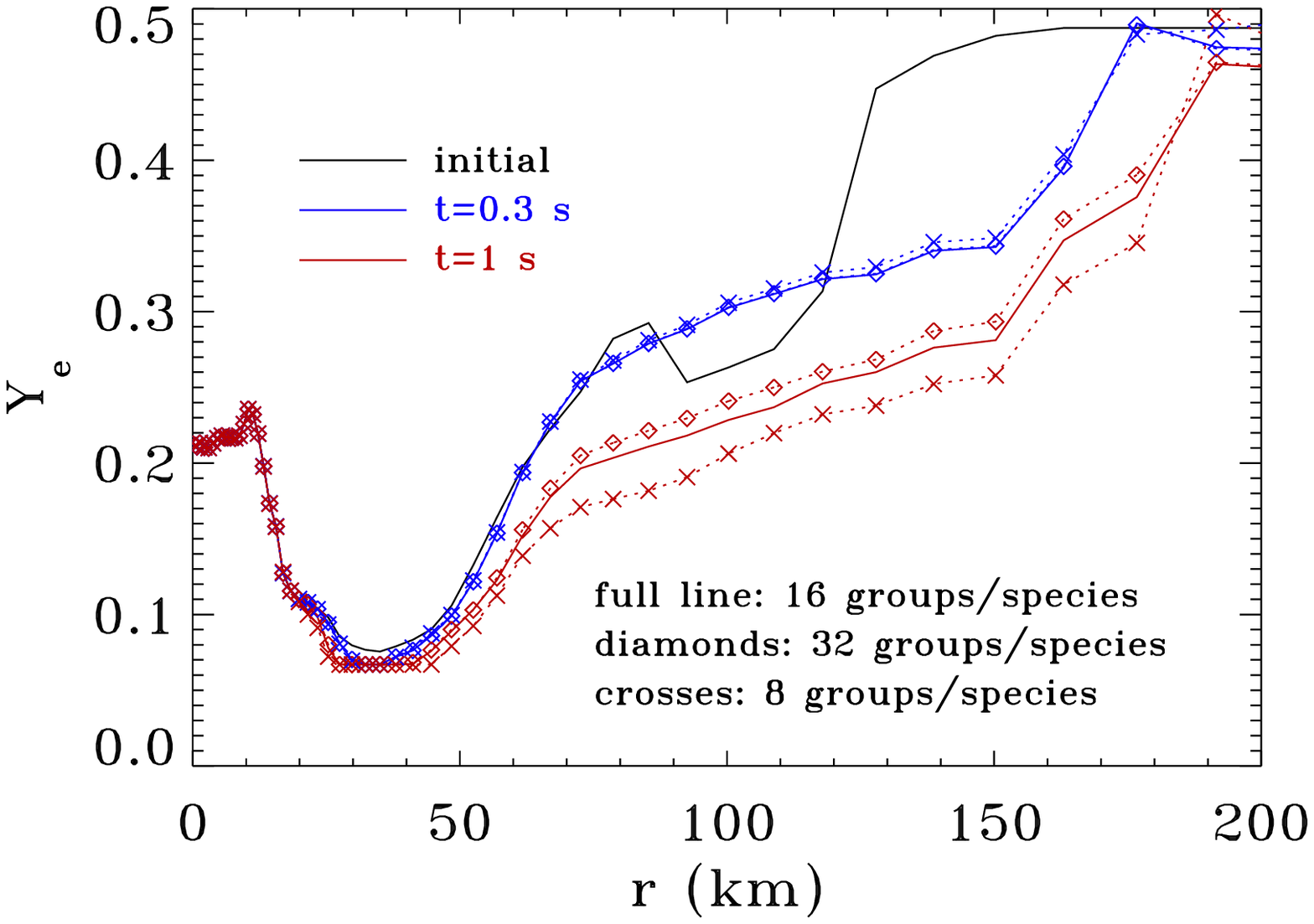}
\caption{The same as in Fig. \ref{fig_e1}, but for $Y_e$.
}
\label{fig_e2}
\end{figure}

%%%%%%%%%%%%%%%%%%%%%%%%%%%%%%%%%%%%%%%%%%%%%%%%%%%

\begin{figure}
\epsscale{1.0}
\plotone{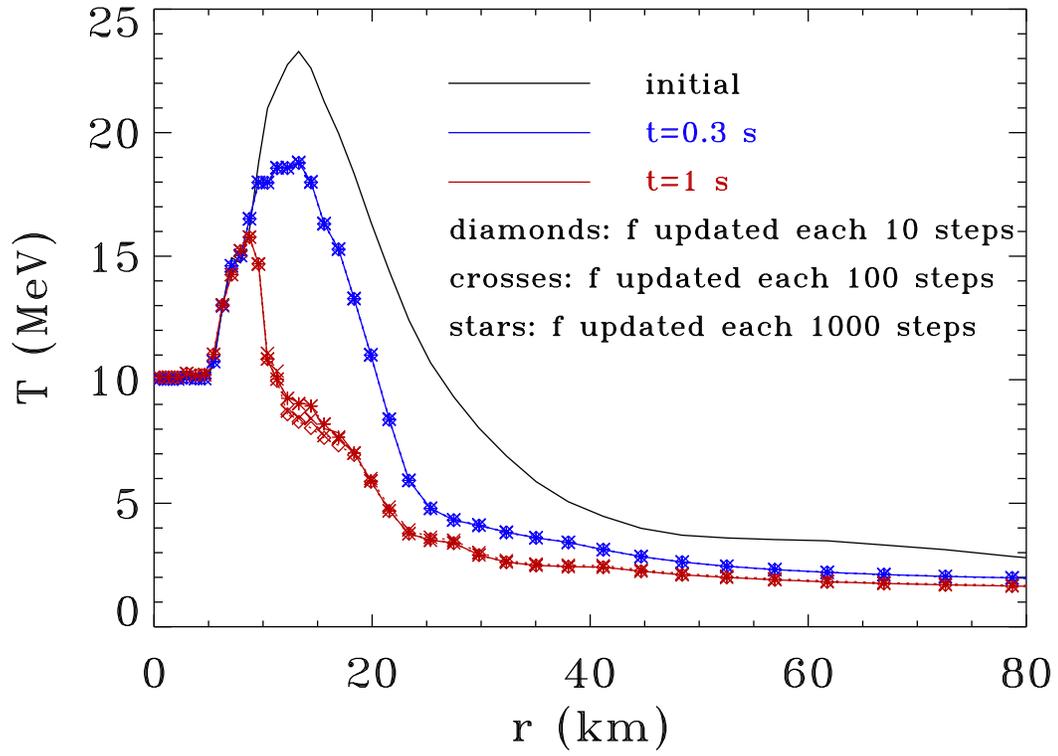}
\caption{Temperature as a function of radius, for the initial configuration
(full black line), and for 0.3 seconds (blue lines/symbols) and 1 second 
(red lines and symbols) of evolution. Diamonds display the evolution where
the Eddington factor ($f$) was updated every 10 timesteps; crosses display
models where the Eddington factor was updated every 100 timesteps,
while stars display models where the Eddington factor was updated every 1000
timesteps. Notice that in many instances all the symbols overlap.
}
\label{fig_e3}
\end{figure}

%%%%%%%%%%%%%%%%%%%%%%%%%%%%%%%%%%%%%%%%%%%%%%%%%%%

\begin{figure}
\epsscale{1.0}
\plotone{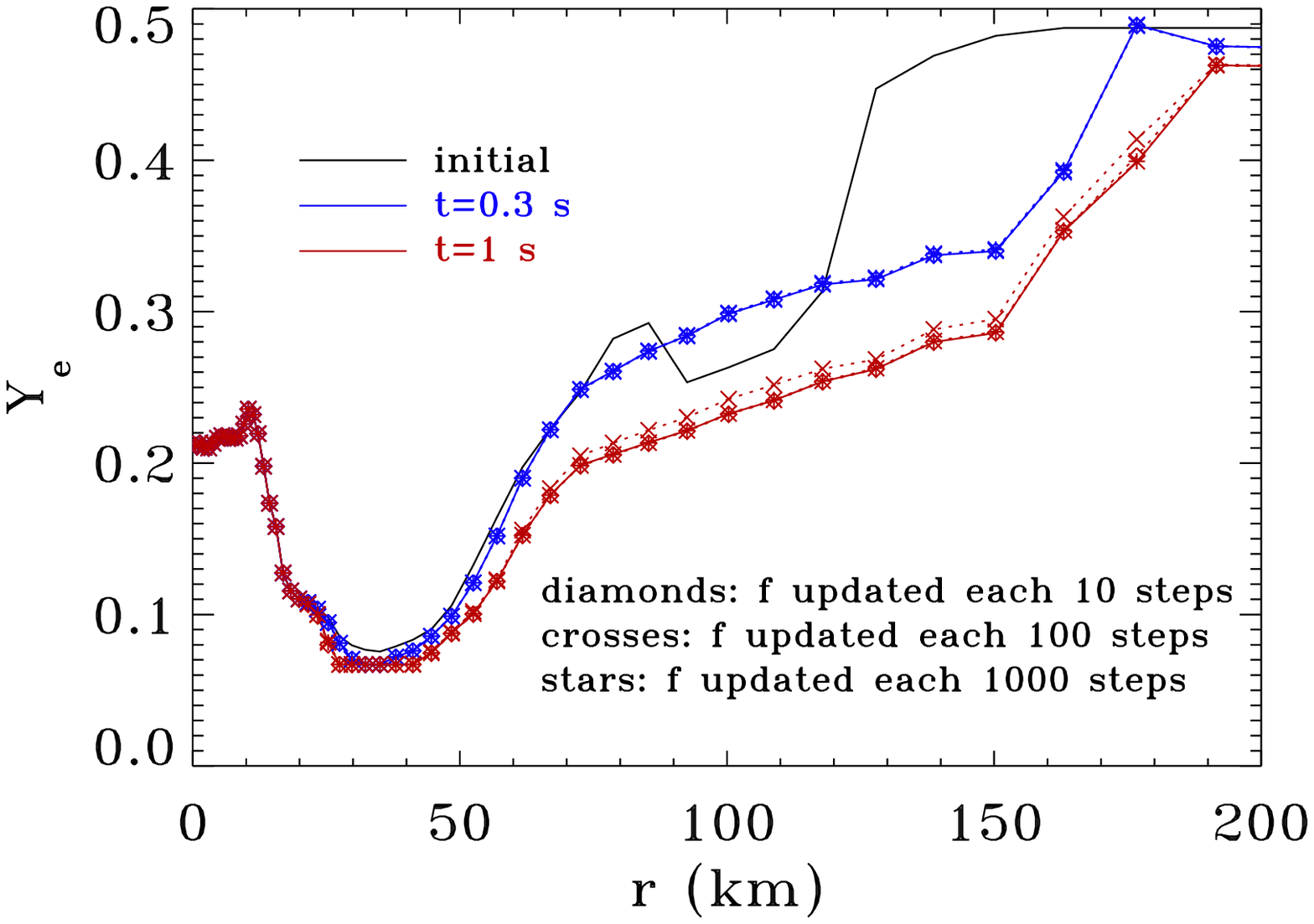}
\caption{The same as in Fig. \ref{fig_e3}, but for $Y_e$.
}
\label{fig_e4}
\end{figure}


\begin{thebibliography}{}

\bibitem[Auer 1971]{auer71} Auer, L.H., 1971, \jqsrt, 11, 573
\bibitem[Auer 1991]{auer91} Auer, L.H., 1991, in Stellar Atmospheres:
Beyond Classical Models, ed. by L. Crivellari, I. Hubeny, and D.G. Hummer,
NATO ASI Ser. C, Vol. 341, p.9
\bibitem[Bowers \& Wilson 1982]{bowers}
Bowers, R.L.~\& Wilson, J.R. 1982, \apjs, 50, 115
\bibitem[Bruenn 1985]{BR1} Bruenn, S.W. 1985, \apjs, 58, 771
\bibitem[Buras et al. 2006]{buras2006} Buras, R., Rampp, M., Janka, H.-Th., \&
Kifonidis, K.
2006,
\aap, 447, 1049
\bibitem[Burrows,~Hayes,~\&~Fryxell 1995]{bhf}
Burrows, A., Hayes, J., \& Fryxell, B.A.~1995, \apj, 450, 830
\bibitem[Burrows et al. 2006]{bur06} Burrows, A., Livne, E., Dessart, L., Ott, C.D., \&
Murphy,
J.
2006, \apj, 640, 878
\bibitem[Cardall \& Mezzacappa 2003]{card03} Cardall, C.Y. \& Mezzacappa, A. 2003,
\prd, 68,
023006
\bibitem[Cardall, Lentz, \& Mezzacappa 2005]{cardall} Cardall, C.Y.,
Lentz, E.J., \& Mezzacappa, A. 2005, \prd, 72, 043007
\bibitem[Castor, Dykema, \& Klein 1992]{castor} Castor, J.I., Dykema, P.G., \& Klein,
R.I,
1992, \apj, 387, 561 
\bibitem[Dessart et al. 2006a]{dessarta} Dessart, L., Burrows, A., Ott, C.D., Livne,
E.,
Yoon, S.-Y., \& Langer, N. 2006a, \apj, 644, 1063
\bibitem[Dessart et al. 2006b]{dessartb} Dessart, L., Burrows, A., Livne, E., \& Ott,
C.D.
2006b, astro-ph/0510229
%\bibitem[Fryer et al.~1999]{fryer}
%Fryer, C.L., Benz, W., Herant, M., \& Colgate, S. 1999, \apj, 516, 892
\bibitem[Hayes \& Norman 2003]{H1} Hayes, J.C. \& Norman, M.L. 2003, \apj,
147, 197
\bibitem[Herant et al. 1994]{herant}
Herant, M., Benz, W., Hix, W.R., Fryer, C.L., \& Colgate, S.A. 1994, \apj, 435, 339
\bibitem[Hubeny 2003]{hubeny03} Hubeny, I., 2003, in ASP Conf. Ser. 288, Stellar
Atmosphere
Modeling, ed. I. Hubeny, D. Mihalas, and K. Werner, 
Astronomical Society of the Pacific, San Francisco, p.17
\bibitem[Hubeny \& Lanz 1992]{hubeny92} Hubeny, I., \& Lanz, T., 1992, \aap, 262, 501
\bibitem[Hubeny \& Lanz 1995]{hubeny95} Hubeny, I., \& Lanz, T., 1995, \apj, 439, 875
\bibitem[Klein et al. 1989]{klein}
Klein, R.I, Castor, J.I., Dykema, P.G., Greenbaum, A., \& Taylor, D., 1989,
\jqsrt, 41, 199
\bibitem[Janka et al. 2005a]{janka05} Janka, H.-T., Buras, R., Kifonidis, K., Marek,
A.,
\& Rampp, M. 2005, in Cosmic Explosions, On the 10th Anniversary of SN1993J.
Proceedings of IAU Colloquium 192, edited by J.M. Marcaide and
Kurt W. Weiler, Springer Proceedings in Physics, vol. 99. (Berlin: Springer), p.253
(astro-ph/0401461)
\bibitem[Janka et al. 2005b]{janka05b} Janka, H.-Th., Buras, R., Kitaura Joyanes, F.S.,
Marek,
A.,
Rampp, M., \& Scheck, L. 2005, ``Neutrino-driven supernovae: An accretion instability
in a
nuclear physics controlled environment," in Proceedings of the 8th International
Symposium on
Nuclei in the
Cosmos, Vancouver, Canada, July 19--23, 2005, Nuclear Physics A, 758, 19--26
\bibitem[Lanz \& Hubeny 2003]{lh03} Lanz, T., \& Hubeny, I., 2003,
\apjs, 146, 417
\bibitem[LeBlanc \& Wilson 1970]{leblanc} LeBlanc, J.M. \& Wilson, J.R., 1970, \apj,
161,541
\bibitem[Liebend\"orfer et al. 2001a]{liebena} Liebend\"orfer, M. et al. 2001a, \prd,
63,
103004
\bibitem[Liebend\"{o}rfer et al.~2001b]{lieben20012}
Liebend\"{o}rfer, M., Mezzacappa, A., Thielemann, F.-K.~2001b, \prd, 63, 104003
\bibitem[Liebend\"{o}rfer et al.~2004]{lieben2002} Liebend\"{o}rfer, M., Messer,
O.E.B.,  Mezzacappa, A., Cardall, C.Y., \& Thielemann, F.-K.~2004, \apjs, 150, 263
\bibitem[Liebend\"{o}rfer et al. 2005]{lieben05} Liebend\"{o}rfer, M., Rampp,
M., Janka, H.-Th., \& Mezzacappa, A. 2005, \apj, 620, 840
\bibitem[Livne et al. 2004]{livne04}Livne, E., Burrows, A., Walder, R.,
Thompson, T.A., and Lichtenstadt, I. 2004, \apj, 609, 277
\bibitem[Mayle, Wilson, \& Schramm 1987]{mayle_wilson}
Mayle, R., Wilson, J.R., \& Schramm, D.N. 1987, \apj, 318, 288
\bibitem[Mezzacappa \& Bruenn 1993]{mezzbruenn} Mezzacappa, A. \& Bruenn, S.W.~1993,
\apj, 410,
669
%\bibitem[Mezzacappa et al. 1998]{mezz98} Mezzacappa, A., Calder, A.C, Bruenn,
%S.W., Blondin, J.M., Guidry, M.W., Strayer, M.R., Umar, A.S. 1998, \apj, 495, 911
\bibitem[Mezzacappa et al. 2001]{mezz01} Mezzacappa, A., Liebend\"{o}rfer, M., Messer,
O.E.B.,
Hix, W.R., Thielemann, F.-K., \& Bruenn, S.W.~2001, \prl, 86, 1935
\bibitem[Mihalas \& Klein 1982]{MK} Mihalas, D. \& Klein, R.I. 1982, Journal of
Computational
Physics, 46, 97
\bibitem[Mihalas, Kunasz, \& Hummer 1975]{mihalas75}
Mihalas, D., Kunasz, P.B., \& Hummer, D.G., 1975, \apj, 202, 465
\bibitem[Mihalas \& Mihalas 1984]{mihalas84}
Mihalas, D. \& Mihalas, B., {\it Foundations of Radiation Hydrodynamics},
New York, Oxford University Press, 1984
\bibitem[Ng 1974]{ng} Ng, K.C., 1974, J. Chem. Phys., 61, 2680 
\bibitem[Ott et al. 2006a]{ott06a} Ott, C.D., Burrows, A., Dessart, L., \& Livne, E.
2006a,
\apj~Suppl., 164, 130 %astro-ph/0508462
\bibitem[Ott et al. 2006b]{ott06b} Ott, C.D., Burrows, A., Dessart, L., \& Livne, E.
2006b,
\prl,  96, 201102  % astro-ph/0605493)
\bibitem[Rampp \& Janka 2000]{rampp2000} Rampp, M. \& Janka, H.-T. 2000, \apj, 539, L33
\bibitem[Rampp \& Janka (2002)]{rampp20022} Rampp, M. \& Janka, H.-Th. 2002, \aap, 396,
331
\bibitem[Stone, Mihalas, \& Norman 1992]{stone} Stone, J.M., Mihalas, D., \& Norman,
M.L. 1992,
\apjs, 80, 819
\bibitem[Rybicki \& Hummer]{rybicki91} Rybicki, G. \& Hummer, D.G., 1991, \aap, 245,
171
\bibitem[Swesty \& Myra 2005a]{swesty2005a} Swesty, F.D., \& Myra, E.S. 2005a,
astro-ph/0506178
\bibitem[Swesty \& Myra 2005b]{swesty2005b} Swesty, F.D., \& Myra, E.S. 2005b,
astro-ph/0507294
\bibitem[Swesty \& Myra 2006]{swesty2006} Swesty, F.D., \& Myra, E.S. 2006,
astro-ph/0607281
\bibitem[Thompson, Burrows, \& Pinto 2003]{TBP} Thompson, T.A., Burrows, A., \& Pinto,
P.A.
2003, \apj, 592, 434
\bibitem[Wilson 1985]{wilson1985} Wilson, J.R. 1985, in {\it Numerical Astrophysics},
ed. J.
Centrella,
J. M. LeBlanc, R. L. Bowers (Boston: Jones \& Bartlett), p. 422
\bibitem[Walder et al. 2005]{walder} Walder, R., Burrows, A., Ott,
C.D., Livne, E., Lichtenstadt, I., \& Jarrah, M. 2005, \apj, 626, 317

\end{thebibliography}
\end{document}